\documentclass[amsmath,amssymb,aps,prx, twocolumn,superscriptaddress, longbibliography]{revtex4-2}

\RequirePackage{epsfig}
\RequirePackage{graphicx}

\RequirePackage{xcolor}
\RequirePackage{siunitx}[=v2]
\RequirePackage{nicefrac}

\RequirePackage{booktabs}
\RequirePackage{makecell}

\RequirePackage{enumerate}
\RequirePackage{enumitem}

\RequirePackage[normalem]{ulem}
\RequirePackage{wrapfig}
\RequirePackage{verbatim}

\RequirePackage[colorlinks=true,linkcolor=blue,urlcolor=blue,hyperfootnotes=false,citecolor=black]{hyperref}
\RequirePackage{hypernat} 

\RequirePackage[capitalise]{cleveref}

\newcommand{\taup}{\tau_{\mathrm{p}}}
\newcommand{\taupe}{\tau_{\mathrm{p}}^{\mathrm{e}}}
\newcommand{\taupo}{\tau_{\mathrm{p}}^{\mathrm{o}}}

\newcommand{\Tone}{T_{1}}

\newcommand{\Ttwostar}{T_{2}^{\mathrm{*}}}
\newcommand{\Techo}{T_{2}^{\mathrm{echo}}}

\newcommand{\EJ}{E_{J}}
\newcommand{\EC}{E_{C}}

\newcommand{\JJa}{\mathrm{JJ}_{\mathrm{a}}}
\newcommand{\JJb}{\mathrm{JJ}_{\mathrm{b}}}
\newcommand{\EJa}{E_{\mathrm{Ja}}}
\newcommand{\EJb}{E_{\mathrm{Jb}}}
\newcommand{\EJone}{E_{\mathrm{J1}}}
\newcommand{\EJtwo}{E_{\mathrm{J2}}}
\newcommand{\EJthree}{E_{\mathrm{J3}}}

\newcommand{\EJm}{E_{\mathrm{J}m}}

\newcommand{\EJam}{E_{\mathrm{Ja,}m}}
\newcommand{\EJbm}{E_{\mathrm{Jb,}m}}

\newcommand{\Ica}{I_{\mathrm{c,a}}}
\newcommand{\Icb}{I_{\mathrm{c,b}}}
\newcommand{\DelL}{\Delta_\mathrm{B}}
\newcommand{\DelR}{\Delta_\mathrm{T}}
\newcommand{\delDel}{\delta\Delta}

\newcommand{\fLR}{f_{\delta\Delta}}

\def\bra#1{\mathinner{\langle{#1}|}}
\def\ket#1{\mathinner{|{#1}\rangle}}

\newcommand{\fzeroone}{f_{01}}
\newcommand{\fzerotwoovertwo}{f_{02}/2}
\newcommand{\fzerotwo}{f_{02}}
\newcommand{\fonetwo}{f_{12}}
\newcommand{\fzerothree}{f_{03}}

\newcommand{\fcav}{f_{\mathrm{cav}}}

\newcommand{\Vg}{V_{\mathrm{g}}}

\newcommand{\Bperp}{B_{\mathrm{\perp}}}
\newcommand{\Bparone}{B_{\mathrm{\parallel}, 1}}
\newcommand{\Bpartwo}{B_{\mathrm{\parallel}, 2}}
\newcommand{\Bpar}{B_{\mathrm{\parallel}}}

\newcommand{\Bcrit}{B_{\mathrm{c}}}
\newcommand{\BcritL}{B_{\mathrm{c}}^\mathrm{B}}
\newcommand{\BcritR}{B_{\mathrm{c}}^\mathrm{T}}

\newcommand{\Bphinaught}{B_{\Phi_{0}}}
\newcommand{\Bphinaughta}{B_{\Phi_{0},\mathrm{a}}}
\newcommand{\Bphinaughtb}{B_{\Phi_{0},\mathrm{b}}}

\newcommand{\sinc}{\mathrm{sinc}}

\newcommand{\xL}{x_B}
\newcommand{\xR}{x_T}
\newcommand{\xRp}{x_{T>}}
\newcommand{\xRm}{x_{T<}}

\begin{document}

\title{Quasiparticle effects in magnetic-field-resilient 3D transmons}

\author{J.~Krause}
\affiliation{Physics Institute II, University of Cologne, Zülpicher Str. 77, 50937 Köln, Germany}
\author{G.~Marchegiani}
\affiliation{Quantum Research Center, Technology Innovation Institute,  Abu Dhabi 9639, UAE}
\author{L.~M.~Janssen}
\affiliation{Physics Institute II, University of Cologne, Zülpicher Str. 77, 50937 Köln, Germany}
\author{G.~Catelani}
\affiliation{JARA Institute for Quantum Information (PGI-11), Forschungszentrum Jülich, 52425 Jülich, Germany}
\affiliation{Quantum Research Center, Technology Innovation Institute,  Abu Dhabi 9639, UAE}
\author{Yoichi~Ando}
\affiliation{Physics Institute II, University of Cologne, Zülpicher Str. 77, 50937 Köln, Germany}
\author{C.~Dickel}
\email[correspondence should be addressed to: \newline]{ando@ph2.uni-koeln.de, \newline dickel@ph2.uni-koeln.de}
\affiliation{Physics Institute II, University of Cologne, Zülpicher Str. 77, 50937 Köln, Germany}

\begin{abstract}
Recent research shows that quasiparticle-induced decoherence of superconducting qubits depends on the superconducting-gap asymmetry originating from the different thicknesses of the top and bottom films in Al/AlO$_x$/Al junctions.
Magnetic field is a key tuning knob to investigate this dependence as it can change the superconducting gaps in situ.
We present measurements of the parity-switching time of a field-resilient 3D transmon with in-plane field up to \SI{0.41}{\tesla}.
At low fields, small parity splitting requires qutrit pulse sequences for parity measurements.
We measure a non-monotonic evolution of the parity lifetime with in-plane magnetic field, increasing up to \SI{0.2}{\tesla}, followed by a decrease at higher fields.
We demonstrate that the superconducting-gap asymmetry plays a crucial role in the observed behavior. 
At zero field, the qubit frequency is nearly resonant with the superconducting-gap difference, favoring the energy exchange with the quasiparticles and so enhancing the parity-switching rate.
With a higher magnetic field, the qubit frequency decreases and gets detuned from the gap difference, causing the initial increase of the parity lifetime, while photon-assisted qubit transitions increase, producing the subsequent decrease at higher fields.
Besides giving a deeper insight into the parity-switching mechanism in conventional transmon qubits, we establish that Al-AlO$_x$-Al JJs could be used in architectures for the parity-readout and manipulation of topological qubits based on Majorana zero modes.
\end{abstract}

\maketitle

\section{Introduction}

Superconducting quantum circuits such as qubits, sensors, and amplifiers have drastically improved over the last two decades~\cite{Kjaergaard20}, mostly thanks to the reduction and mitigation of charge noise, a recent example being the transmon qubit~\cite{Koch07} based on tantalum capacitors~\cite{Place21}.
Consequently, non-equilibrium quasiparticles are slowly becoming a dominant source of losses, in addition to already being a main cause of residual excitations~\cite{Serniak18}.
Moreover, recent reports of chip-wide correlated decoherence due to quasiparticles~\cite{McEwen22} show that quasiparticle loss is a major obstacle to scaling up superconducting quantum processors.
A wide range of solutions to quasiparticle loss is being explored: quasiparticle traps~\cite{Riwar.2016}, pumping~\cite{Gustavsson16}, gap engineering~\cite{Riwar.2019,McEwen24}, phonon downconversion~\cite{Henriques19,Iaia22}, optimizing device geometry~\cite{Pan22}, shielding and filtering~\cite{Gordon22} or even shielding the device from cosmic rays~\cite{Vepsalainen20, Cardani21}.
Magnetic-field-resilient transmons made of thin-film Al/AlO$_x$/Al Josephson junctions (JJs)~\cite{Krause22} provide yet another angle to tackle quasiparticle loss:
the in-plane-magnetic field tunes in-situ both the transmon transition frequencies and the superconducting gaps of top and bottom aluminum electrodes, making these transmons an ideal system to study quasiparticle effects and optimize gap engineering.

In a parallel effort, magnetic-field-resilient transmons with aluminum tunnel JJs can be a key component for topological quantum-computation protocols based on precise readout and control of the fermion parity of Majorana zero modes (MZMs). 
Circuit QED measurements~\cite{Blais04} of offset-charge-sensitive transmon circuits allow for robust and fast detection of charge-parity switching~\cite{Riste13, Serniak18}, and hence transmons incorporating topological-superconductor nanowires provide an ideal platform to measure and manipulate MZMs~\cite{Hassler11, Hyart13}.
To host MZMs, the wires typically need to be threaded by large parallel magnetic fields on the order of at least \SI{0.5}{\tesla}, posing strong requirements on the magnetic-field resilience of the readout circuitry, too.
In this context hybrid JJs for field compatible transmons have also been explored~\cite{Luthi18,  Kroll18, Kringhoj21, Uilhoorn2021}.
With magnetic-field-resilient aluminum transmons, the advantages of highly coherent and reliable tunnel junctions can be exploited to achieve high-fidelity parity measurements even in large magnetic fields.

In this article, we present measurements of the quasiparticle-induced parity-switching rates in a 3D transmon~\cite{Paik11} as a function of in-plane magnetic field up to \SI{0.41}{\tesla}, promoting thin-film aluminum transmons for the parity readout of future topological qubits.
Using a comprehensive model, we distinguish the relevant mechanisms of parity switching, providing evidence for an interplay between Cooper-pair-breaking photons, the superconducting gaps in both sides of the JJs, and the Fraunhofer effect in the JJs.
We report signatures of near-resonantly enhanced quasiparticle tunneling due to the superconducting-gap difference approaching the transmon frequency at low in-plane magnetic fields; this resonance condition is gradually lifted with increasing fields until photon-assisted parity switching dominates.
Thus, counter-intuitively, the maximum parity lifetime is reached at a finite field of about \SI{0.2}{\tesla}.
Measurements of the temperature dependence of the parity-switching time for selected in-plane magnetic fields further support these findings.
We use different transmon transitions in the parity measurements~\cite{Kurter22, Tennant22}, as the Josephson energy ($\EJ$) changes with the magnetic field, mainly due to the Fraunhofer effect.
Our results complement recent experimental~\cite{Diamond22,connolly2023coexistence} and theoretical~\cite{Marchegiani22} works on the impact of gap asymmetry on the quasiparticle decoherence rate of transmons. 

This article is organized as follows. 
In Sec.~\ref{sec:experiment_design} we discuss the experiment design, while in Sec.~\ref{sec:spectrum_vs_Bpar} the magnetic-field dependent spectrum is shown and parameters of the transmon are determined by modeling it. 
Sec.~\ref{sec:parity_lifetime_vs_Bpar} is devoted to the in-plane magnetic field dependence of the parity-switching time and Sec.~\ref{sec:parity_lifetime_vs_temp} to its temperature dependence for selected in-plane magnetic fields. 
In the Conclusions [Sec.\ref{sec:conclusion}], we summarize the results and give an outlook for future research.

\section{Experiment design}
\label{sec:experiment_design}
\begin{figure}
  \centering
  \includegraphics[width=\columnwidth]{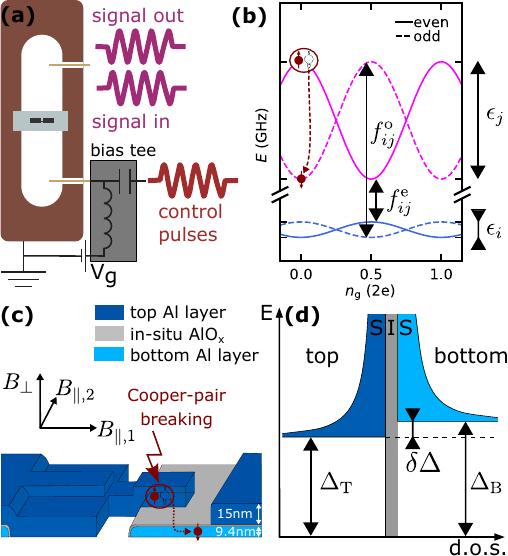}
  \caption{
    Schematic of the experiment design. 
    \textbf{(a)} Sketch of the 3D copper cavity with the transmon.
    A bias-tee at one of the cavity pins enables voltage biasing of the transmon.
    \textbf{(b)} Successive transmon energy levels $i$ and $j$ as a function of $n_\mathrm{g}$ (proportional to $\Vg$).
    There are two manifolds (``even'' and ``odd'') connected by the incoherent tunneling of single quasiparticles.
    Energy levels have a charge dispersion with peak-to-peak value $\epsilon_i$ and transition frequencies $f_{ij}^\mathrm{e}$ and $f_{ij}^\mathrm{o}$.
    \textbf{(c)} Sketch of a Dolan-bridge Josephson junction (JJ) relating the magnetic-field axes to the JJ geometry.
    A vector magnet is used to flux-bias the SQUID loop along $\Bperp$ and to apply in-plane magnetic fields along $\Bparone$ and $\Bpartwo$.
    Quasiparticles can be generated by Cooper-pair breaking photons; any event where single quasiparticles tunnel through the JJ changes the charge parity of both electrodes and makes the transmon transition frequencies jump from $f_{ij}^\mathrm{e}$ to $f_{ij}^\mathrm{o}$ or vice versa. The two electrodes in the JJ have different thicknesses for the top and bottom aluminum layers, leading to different superconducting gaps in the excitation spectrum. 
    \textbf{(d)} Density of states (d.o.s.) and superconducting gap for the two electrodes.
    }
  \label{fig:fig1}
\end{figure}

The experiment is conceived to measure the effect of magnetic fields on the parity-switching time $\taup$ of an offset-charge-sensitive transmon~\cite{Serniak19}. For dispersive readout, the transmon is coupled to a 3D copper cavity [\cref{fig:fig1} (a)], which is unaffected even by large magnetic fields.
We use a bias tee at one of the cavity pins to apply an offset voltage $\Vg$ to the two transmon islands (see \cref{app:setup} for details of the setup and \cref{app:device_fab_geometry} for the device fabrication and geometry). In this way, we can measure
the charge dispersion, the dependence of the spectrum on the offset charge $n_\mathrm{g}\propto \Vg$.

The charge dispersion is $2e$-periodic, where $e$ is the elementary charge (for transmon Hamiltonian see \cref{app:transmon_hamiltonian}). 
We define the peak-to-peak variation of transmon level $i$ as $\epsilon_i$.
However, the transmon energy spectrum splits into two separate manifolds that differ by one electron charge [see \cref{fig:fig1} (b)] and are labeled ``even'' and ``odd''.
Microwave transitions are only possible within each manifold with frequencies $f_{ij}^\mathrm{e}$ and $f_{ij}^\mathrm{o}$ respectively, but not between them, as this would require a term in the Hamiltonian that connects them~\cite{Ginossar14}.
The two manifolds are incoherently connected by the tunneling of single quasiparticle excitations across the junction~\cite{Joyez94, Aumentado04, Lutchyn06}.
This is equivalent to a $1/2$ shift in $n_\mathrm{g}$ (a shift by $1e$), which changes the transmon transition frequencies from $f_{ij}^\mathrm{e}$ to $f_{ij}^\mathrm{o}$ or vice versa and can be accompanied by qubit relaxation or excitation.
We define the transmon parity as the parity of the number of single quasiparticles that have tunneled across the junction~\cite{Diamond22} and the parity-switching time $\taup$ as the mean dwell time between two such events.
In addition to the discrete parity jumps caused by quasiparticles tunneling across the junction, the charge environment creates noise in $n_\mathrm{g}$ including slow drift, shifting by about $1e$ over a time scale of order \SI{10}{\minute}.
The transmon was introduced~\cite{Houck08} as a qubit design less sensitive to noise in $n_\mathrm{g}$, but charge noise in transmons continues to be a subject of research~\cite{Christensen19}.

Our transmon includes a Superconducting Quantum Interference Device (SQUID) comprising two Al/AlO$_x$/Al tunnel junctions.
Thin aluminum films yield relatively high critical fields $\Bcrit\sim\SI{1}{\tesla}$ for magnetic fields applied in-plane~\cite{Fulde1973, Meservey1994}, and narrow superconducting electrodes minimize vortex losses.
Previously, we showed that a device based on the same geometry can remain sufficiently coherent (with lifetimes of the order of $\SI{1}{\micro\second}$) even at high magnetic fields~\cite{Krause22}.
The two superconducting electrodes composing each junction are characterized by different superconducting gaps, $\DelL$ and $\DelR$, due to the different film thicknesses of the bottom ($\mathrm{B}$) and top ($\mathrm{T}$) aluminum layer [\cref{fig:fig1} (c) and (d)].
The gap asymmetry $\delDel = \DelL-\DelR$ estimated with a phenomenological law~\cite{Chubov1969, Meservey1971, Marchegiani22} is on the order of the transmon transition energy $h\fzeroone$, making it relevant for quasiparticle-tunneling processes~\cite{Marchegiani22}.
A SQUID transmon was chosen such that at low fields, the bottom sweet spot has a sufficient charge dispersion for charge-parity measurements, while at higher fields eventually, the top sweet spot charge dispersion becomes sufficient.
However, due to flux instabilities, we could not measure the spectrum for $\Bparone>\SI{0.41}{\tesla}$ for reasons that are currently not understood. 
Coherence times start to drop sharply slightly below \SI{0.40}{\tesla} (see supplementary information~\cite{SOM} for details). 
Due to flux noise, the qubit coherence is severely reduced away from the top and bottom sweet spots of the SQUID, so measurements of the parity-switching time were mainly performed at the bottom sweet spot.

A vector magnet is used to flux-bias the SQUID loop with an out-of-plane field $\Bperp$ and to apply in-plane magnetic fields along $\Bparone$ and $\Bpartwo$ [see \cref{fig:fig1} (c)].
The frequency modulation due to the SQUID is highly sensitive to $\Bperp$ which enables precise alignment of the magnetic field to the sample plane to within $\pm$\SI{0.05}{\degree}.
To measure the magnetic-field dependence of the parity-switching time, we apply the field along the $\Bparone$ axis. 
Previously we saw that the magnetic-field dependence along the $\Bpartwo$ axis was more erratic~\cite{Krause22}, which we hypothesize to be due to different spurious JJs in the two field directions.
In the following, we will use $\Bpar$ as a shorthand for $\Bparone$.

\section{In-plane magnetic field dependence of the transmon spectrum}
\label{sec:spectrum_vs_Bpar}

\begin{figure*}
  \centering
      \includegraphics[width=\textwidth]{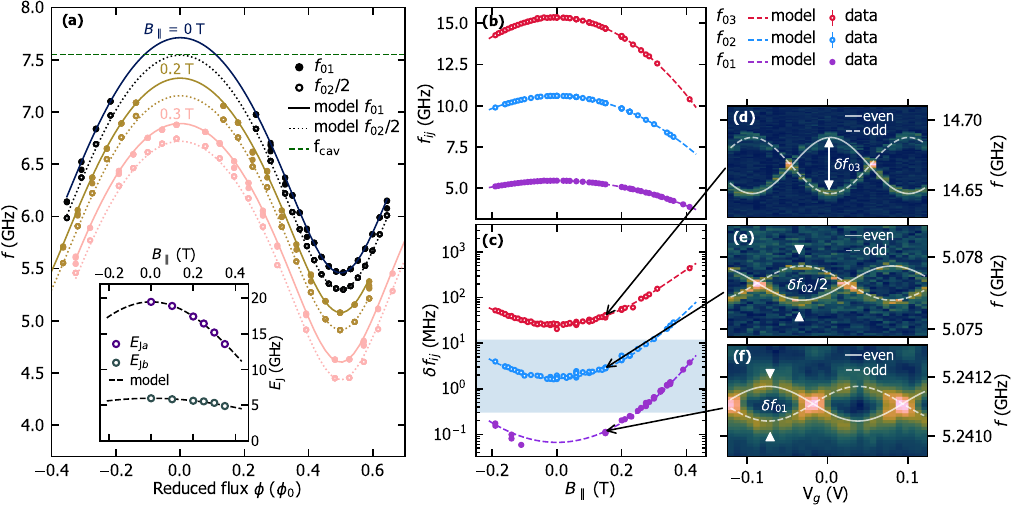}
  \caption{
      Transmon spectroscopy in a magnetic field.
      \textbf{(a)} Transmon frequencies $\fzeroone$, $\fzerotwoovertwo$ as a function of $\Bperp$ (here expressed as the reduced flux $\Phi/\Phi_0$) for selected $\Bpar$.
      The asymmetric SQUID oscillates between top and bottom flux sweet spots, corresponding to the sum and difference of the two junctions' Josephson energies $\EJa$ and $\EJb$.
      For $\Bpar\le\SI{0.2}{\tesla}$ the top sweet spot is close to $\fcav$ and therefore we do not perform dispersive two-tone spectroscopy for frequencies
      above $\SI{7.1}{\giga\hertz}$.
      The inset shows $\EJa$ and $\EJb$ as a function of $\Bpar$.
      Their monotonic decrease with $\Bpar$ is mainly due to a Fraunhofer-like pattern in the junctions' critical currents $\Ica$ and $\Icb$; the magnetic field further decreases $\EJ$ by suppressing the superconducting gap.
      However, in the magnetic-field range covered by quasiparticle-parity measurements, the effect is small, less than $3\%$ (see \cref{app:fraunhofer_and_critical_field}).
      \textbf{(b)} and \textbf{(c)} In-plane magnetic field dependence of the bottom-sweet spot transitions $\fzeroone$, $\fzerotwo$ and $\fzerothree$ and their respective parity-frequency splittings $\delta f_{ij} = \max_{n_\mathrm{g}}(f_{ij}^\mathrm{e}-f_{ij}^\mathrm{o})$.
      As $\EJ$ decreases all $f_{ij}$ decrease, while their $\delta f_{ij}$ increase; we assume $\EC$ to be unaffected by $\Bpar$.
      We obtain the $f_{ij}$ and $\delta f_{ij}$ in voltage-gate scans as shown in panels \textbf{(d)}-\textbf{(f)}.
      For each transition, two separate frequency branches differ by one missing or extra quasiparticle in each of the transmon's electrodes, making it possible to turn the transmon into a quasiparticle parity meter.
      The blue-shaded area in panel (d) indicates where $\delta f_{ij}$ is in a useful range for parity-state-mapping.
      For $\Bpar\le\SI{0.3}{\tesla}$ we employ the $\fonetwo$ transition, for $\Bpar\ge\SI{0.25}{\tesla}$ we can also use the $\fzeroone$ transition.
      The joint fit to all $f_{ij}$ and $\delta f_{ij}$ shown in this figure requires higher harmonics $E_\mathrm{J,m} \cos(m\phi)$ (with $m$ integer) in the Josephson energy term of the transmon Hamiltonian, for details see \cref{app:transmon_hamiltonian}.
      We find $\EC/h=\SI{327.5}{\mega\hertz}$ and the zero-field Josephson energies $\EJa^0/h=\SI{19.47}{\giga\hertz}$ and $\EJb^0/h=\SI{5.97}{\giga\hertz}$.
    }
  \label{fig:fig2}
\end{figure*}

First, we investigate the evolution of the transmon spectrum with the applied magnetic field to extract the parameters of our device, such as the Josephson energy ($\EJ$) and the charging energy ($\EC$). 
The in-plane magnetic field $\Bpar$ modulates $\EJ$, while $\EC$ remains unaffected.
Consequently, the transmon spectrum changes as a function of $\Bpar$; both the transition frequencies between different transmon levels and their respective charge dispersions are modified.

Two different mechanisms contribute to the suppression of the Josephson energy $\EJ$ in the presence of an in-plane magnetic field. First, the magnetic field weakens superconductivity in the two electrodes composing the SQUID~\cite{Tinkham04}. Specifically, the superconducting gaps $\DelL$, $\DelR$, and so $\EJ\propto \DelL\DelR/(\DelL+\DelR)$, decrease monotonically with increasing $\Bpar$. Second, the magnetic field directly affects the Josephson coupling at the junction by laterally penetrating the JJ's oxide barrier. 
This causes a Fraunhofer-like pattern in the dependence of the junction's critical current $I_c$ on $\Bpar$~\cite{Tinkham04} (c.f. \cref{fig:estimating_fraunhofer_fields}); this modulation is significant when the lateral flux associated with the field is of the order of the superconducting flux quantum $\Phi_0=h/(2e)$.
In our case, with JJs of width $l_2\sim\SI{200}{\nano\meter}$ and film thickness $t\sim \SI{10}{\nano\meter}$~\footnote{More generally, given the penetration depths $\lambda_T,\lambda_B$ and the film thicknesses of the two electrodes $t_{T}$, $t_{R}$, the effective thickness determining the Fraunhofer critical field amounts to $\lambda_T \tanh(t_T/2\lambda_T)+\lambda_B \tanh(t_B/2\lambda_B)+t_{\rm AlOx}$, where $t_{\rm AlOx}$ is the thickness of the insulating oxide-barrier of the Josephson junction.}, this field is of order $\Bphinaught(t,l_2)\sim\Phi_0/(l_2 t)\sim\SI{1}{\tesla}$.
We were able to explore this regime here due to the large in-plane critical field $B_c$ of thin-film superconducting electrodes~\cite{Meservey1994}.
In typical JJs, the electrodes are thicker, usually \SI{30}{\nano\meter} and above; as the critical field $\Bcrit$ decreases faster with thickness than $\Bphinaught$ ($t^{-3/2}$~\cite{Meservey1994} vs $t^{-1}$), the Fraunhofer effect becomes less important as thickness increases.

To quantify both contributions to $\EJ$, we measure the perpendicular flux dependence of the lowest energy transitions in a SQUID transmon for selected in-plane magnetic fields $\Bpar$ [Fig.~\ref{fig:fig2} (a)].
The SQUID is asymmetric; the two JJs have different dimensions, and hence different Josephson energies $\EJa$ and $\EJb$ proportional to the junction area.
Therefore the flux arcs shown in Fig.~\ref{fig:fig2} (a) display both a top and a bottom flux sweet spot, for which $E_J(\phi=0)=\EJa+\EJb$ and $E_J(\phi=0.5\Phi_0)=|\EJa-\EJb|$, respectively. The inset of Fig.~\ref{fig:fig2} (a) shows the extracted Josephson energies as a function of $\Bpar$. In general, the magnitude of gap suppression at a given field is determined by material properties and film thickness; since the two arms of the SQUID have been fabricated simultaneously, we take the two junctions to be equally affected by gap suppression. In contrast, the characteristic fields for the Fraunhofer contribution to $I_c$ are different ($\Bphinaughta\neq\Bphinaughtb$) due to the SQUID asymmetry.
Consequently, the parameter characterizing the SQUID asymmetry $\alpha_{\rm JJ}=\left|\EJa-\EJb\right|/(\EJa+\EJb)$ changes with $\Bpar$; this feature allows us to discriminate between the 
Fraunhofer contribution and the suppression of the superconducting gaps (for details see \cref{app:fraunhofer_and_critical_field}). 
In the magnetic-field range covered by quasiparticle-parity measurements, we estimate the contribution of the gap suppression to $\EJ\propto\DelL\DelR/(\DelR+\DelL)$ to be less than $3\%$, so the decrease in $\EJ$ is mainly due to the Fraunhofer contribution.

Measuring the parity-switching time requires the transmon to be in an offset-charge sensitive regime.
The charge dispersion's peak-to-peak values $\epsilon_i\propto \EC (\EJ/\EC)^{i/2+3/4}\exp(-\sqrt{8\EJ/\EC})$~\cite{Koch07} depend exponentially on the ratio of $\EJ/\EC$ and are larger for higher levels.
Experimentally we measure the transition frequencies $f_{ij}$ between two levels $\ket{i}$ and $\ket{j}$ and their parity-frequency splittings  $\delta f_{ij} = \max_{n_\mathrm{g}}(f_{ij}^\mathrm{e}-f_{ij}^\mathrm{o})$. Now, as $\EJ$ decreases with $\Bpar$, the $f_{ij}$ decrease and the $\delta f_{ij}$ increase.
Figure~\ref{fig:fig2} (b) and (c) show $\fzeroone$, $\fzerotwo$ and $\fzerothree$ as well as the corresponding $\delta f_{ij}$  measured at the bottom flux sweet spot as a function of $\Bpar$.  The data is determined from voltage-gate scans like the ones displayed in \cref{fig:fig2} (d)-(f); these examples were taken at $\Bpar=\SI{0.15}{\tesla}$ (see also \cref{app:gate_scans}). We observe both frequency branches $f_{ij}^\mathrm{e}$ and $f_{ij}^\mathrm{o}$ in two-tone spectroscopy as the measurement is slow compared to the characteristic parity-switching time $\taup$.
In successive single-shot measurements, however, the transmon is either in the ``odd'' parity state or in the ``even'' parity state, and parity switches come with a measurable frequency jump.
This jump is maximally resolved when the gate voltage $\Vg$ is set to the charge sweet spot, i.e., $n_\mathrm{g}=0,1/2,1\dots$ in \cref{fig:fig1} \textbf{(b)},
and it enables a charge-parity meter based on a frequency-dependent gate~\cite{Riste13}. To measure the parity-switching rates we use different transmon transitions depending on the field range:
For $\Bpar\ge\SI{0.25}{\tesla}$, $\delta\fzeroone\ge\SI{0.2}{\kilo\hertz}$, and we can use the $\fzeroone$ transition.
For smaller fields, however, the $\EJ/\EC$-ratio is on the order of \SIrange{45}{30}{} even for the bottom sweet spot, and $\delta\fzeroone$ is too small for charge parity measurements given our dephasing times.
With $\delta\fzerotwo$ being a factor $10$ larger with comparable dephasing times, we employ the $\fonetwo$ transitions instead. 
That way, we can measure the transmon parity also for higher $\EJ/\EC$-ratios and cover a wide range of magnetic fields.

Notably, a joint fit to all $f_{ij}$ and $\delta f_{ij}$ shown in \cref{fig:fig2} requires the inclusion of higher Josephson harmonics $E_\mathrm{J,m} \cos(m\phi)$ in the transmon Hamiltonian.
The data we collected in the process of pursuing the quasiparticle physics shows good evidence for the Josephson harmonics; it has been combined with results from other groups to make the case for a more complex current-phase relationship in conventional Al/AlO$_x$/Al tunnel JJs that had previously been overlooked, see Ref.~\cite{Willsch23}. 
Here we explicitly consider the SQUID nature of the transmon in analyzing our data, see \cref{app:transmon_hamiltonian} for details.
The parameters we extract from the spectroscopic data, in particular the charging and Josephson energies, also enter into the estimates of the parity-switching time that we consider in the next Section.

\section{In-plane magnetic field dependence of the parity-switching time}
\label{sec:parity_lifetime_vs_Bpar}

\begin{figure}
  \centering
  \includegraphics[width=\columnwidth]{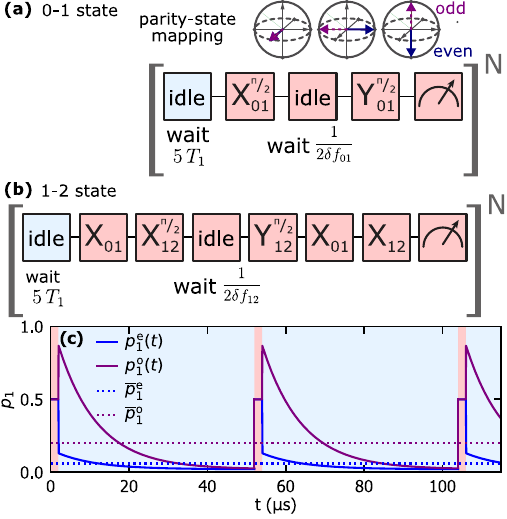}
  \caption{
    Parity-mapping scheme. \textbf{(a)} and \textbf{(b)} show the gate sequences for the parity measurements using superpositions $\vert 0 \rangle+\vert 1 \rangle$, and $\vert 1 \rangle+\vert 2 \rangle$ respectively.
    They are repeated $N=2^{18}$ times.
    For an initial $\vert 0 \rangle$ state, we define ``even'' (``odd'') parity as the parity that is ideally mapped on $\vert 0 \rangle$ ($\vert 1 \rangle$).
    During the parity measurement sequence, the transmon is out of equilibrium. 
    \textbf{(c)} estimated excited-state population $p_1$ for ``even'' and ``odd'' parity as a function of time during the parity-measurement cycle.
    From the hidden-Markov model, we obtain measurement outcomes for declared ``even'' and ``odd'' parities, which give the initial populations after the measurement.
    Finite parity-measurement fidelity leads to slight deviations from the ideal $p_1=0$ for ``even'' parity.
    Taking the average over the cycle, we can estimate the mean populations $\overline{p}_1^\mathrm{e}$, $\overline{p}_1^\mathrm{o}$ for the ``even'' and ``odd'' parities, which are input parameters for modeling the switching-time data.
    }
  \label{fig:parity_schemes}
\end{figure}

We now turn to the characterization of the transmon's parity-switching time as a function of $\Bpar$.
Measurements are performed at the bottom sweet spot in $\Bperp$, see discussion in Sec.~\ref{sec:experiment_design}.
We use a Ramsey-based parity measurement~\cite{Riste13} with superpositions of $\ket{0}$ and $\ket{1}$ as well as $\ket{1}$ and $\ket{2}$, as illustrated in in \cref{fig:parity_schemes}.
At a gate-voltage sweet spot ($n_\mathrm{g}=0,1/2,1\dots$), the parity-measurement protocol projects the lower frequency branch predominantly onto the transmon ground state $\ket{0}$, and the upper-frequency branch onto the first excited state $\ket{1}$.
We henceforth call ``even'' the parity branch that is projected onto $\ket{0}$ and denote its mean dwell time by $\taupe$.
Similarly, ``odd'' is the parity branch that is projected onto $\ket{1}$, with a mean dwell time $\taupo$.
As we do not actively reset the qubit state after an individual single-shot parity measurement, we include a waiting time of $5\Tone$ before taking the next single-shot parity measurement.
As a result, the qubit's first-excited-state population $p_1$ is on average higher for ``odd'' than for ``even'' parity mapping, as the transmon relaxes from the excited state to the ground state during the waiting time, see Fig.~\ref{fig:parity_schemes}~(c).
At the base temperature of the cryostat, $T\sim \SI{7}{\milli\kelvin}$, we estimate a mean excited-state population $\overline{p}_1^\mathrm{e}=4.3\%$ for ``even'' parity and $\overline{p}_1^\mathrm{o}=19.5\%$ for ``odd'' parity.
More details on the parity-measurement sequence and its analysis in terms of a Hidden Markov Model can be found in \cref{app:parity_schemes}.

\begin{figure}
  \centering
  \includegraphics[width=\columnwidth]{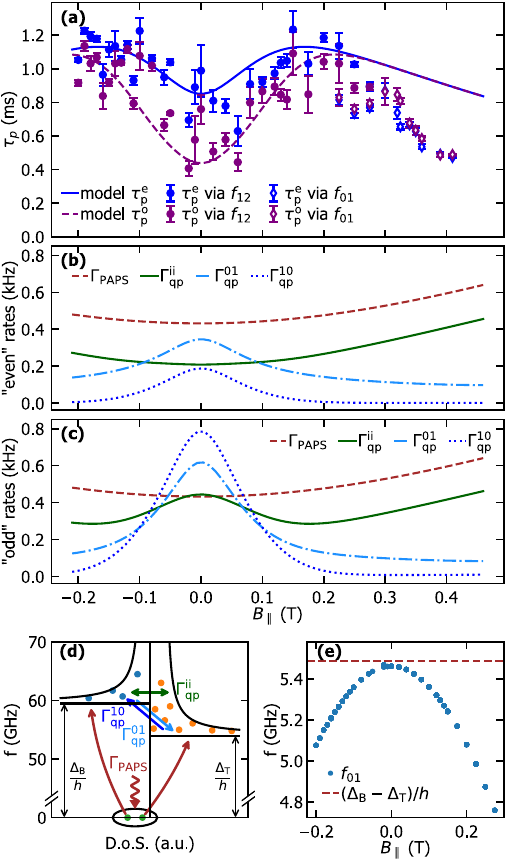}
  \caption{
    \textbf{(a)} Parity-switching times $\taupe$ and $\taupo$ for ``even'' and ``odd'' parity vs $\Bpar$
    measured at the flux bottom sweet spot.
    Model contributions to \textbf{(b)} $1/\taupe$ and \textbf{(c)} $1/\taupo$.
    Around $\Bpar=\SI{0}{\tesla}$ the population-weighted quasiparticle-tunneling rates $\Gamma_\textrm{qp}^{01}$ and $\Gamma_\textrm{qp}^{10}$ [Eqs.~\eqref{eq:gamma01QP} and \eqref{eq:gamma10QP}] show a peak.
    These two processes excite or relax the transmon while bridging the superconducting-gap difference and are resonantly enhanced when $\fzeroone\simeq\delDel/h$, explaining the low-field dip in $\taup$.
    A higher first-excited-state population $p_1$ further increases $\Gamma_\textrm{qp}^{10}$ for the ``odd'' case, which is why $\taupo<\taupe$ in this region. 
    Rate $\Gamma^\mathrm{ii}_\mathrm{qp}$ accounts for quasiparticle tunneling not accompanied by qubit logical state change.
    Photon-assisted quasiparticle tunneling $\Gamma_\mathrm{PAPS}$ dominates for higher fields.
    \textbf{(d)} Schematic density of states for the quasiparticle excitation spectra in high- and low-gap (bottom and top) electrodes.
    The quasiparticles are generated by pair-breaking photons; they can tunnel at constant quasiparticle and transmon energy or by exchanging energy with the transmon.
    \textbf{(e)} Qubit frequency $\fzeroone$ and estimated superconducting-gap difference  $\delDel/h = (\DelL-\DelR)/h$ vs $\Bpar$.
    While at zero field $\fzeroone\simeq \delDel/h$, the transmon is tuned further away from resonance for higher fields.
    Gaps used for the theoretical curves are $\DelR/h=\SI{54}{\giga\hertz}$ and $\delDel/h=\SI{5.49}{\giga\hertz}$.
   For the remaining model parameters see \cref{app:paritymodel}.
    }
  \label{fig:fig3}
\end{figure}

Figure~\ref{fig:fig3} (a) shows the two parity-switching times for ``even'' ($\taupe$) and ``odd'' ($\taupo$) parity as a function of $\Bpar$.
Most notably, the two parity-switching times evolve non-monotonically with $\Bpar$: for both $\taupe$ and $\taupo$ we observe an initial increase with applied magnetic field, reaching a maximum at $\Bpar \approx \SI{0.2}{\tesla}$, before decreasing again at higher fields.
Besides, we generally observe $\taupe\neq\taupo$; in particular for $|\Bpar|\le\SI{0.2}{\tesla}$, $\taupo$ is smaller than $\taupe$.
Theoretically, we expect $\taupe\approx\taupo$, since the asymmetry between ``even'' and ``odd'' parity-switching rates is at most of order $e^{-h\delta f_{01}/k_B T}$~\cite{Catelani14}. 
Even at the nominal base temperature of the cryostat $T_0\approx \SI{7}{\milli\kelvin}$, we have $k_B T_0\approx \SI{150}{\mega\hertz} \gg \delta f_{01}\sim 0.1-10$~MHz [cf. Fig.~\ref{fig:fig2} (c)].
As we will argue below, the difference between $\taupe$ and $\taupo$ is instead due to the measurement-induced $\overline{p}_1$.

\subsection{Modelling the parity switching time}
\label{sec:parityModelMain}

To understand the non-monotonic magnetic-field dependence of $\taup$ and the difference between ``even'' and ``odd'' parities we model different contributions to the parity-switching time. Adopting the notation of Ref.~\cite{Diamond22}, we distinguish photon-assisted parity switching (PAPS)~\cite{Houzet19}, from number-conserving parity switching (NUPS) events, expressing
\begin{equation}
\frac{1}{\taup}=\Gamma_{\rm PAPS}+\Gamma_{\rm NUPS} \,.
\end{equation}
For PAPS events, pair-breaking photons with energy larger than the gap sum $\DelL+\DelR$ are absorbed right at the JJ, leading to the generation of quasiparticles in the bottom and top electrode at a total rate 
\begin{equation}
\label{eq:PAPS}
\Gamma_{\rm PAPS}=p_0(\Gamma_{00}^{\rm ph}+\Gamma_{01}^{\rm ph})+p_1(\Gamma_{11}^{\rm ph}+\Gamma_{10}^{\rm ph}) \,.
\end{equation}
In Eq.~\eqref{eq:PAPS}, $p_1$ is the occupation probability of the first excited state of the qubit, $p_0=1-p_1$ [we disregard the occupation of higher excitation levels, see discussion in Appendix~\ref{app:paritymodel}], and
the subscripts in $\Gamma_{ij}^{\rm ph}$ denote the initial and final transmon logical states $i$ and $j$. 

In NUPS events, quasiparticles tunnel from one side of the JJ to the other, conserving the total quasiparticle number. 
We distinguish three terms in the total rate of these events,
\begin{equation}
\Gamma_{\rm NUPS}= \Gamma_\mathrm{qp}^\mathrm{ii} +\Gamma_\mathrm{qp}^\mathrm{01} +\Gamma_\mathrm{qp}^\mathrm{10}
\end{equation}
In general, both quasiparticles in the bottom and top electrodes  
contribute to each term. For instance, for the ``quasi-elastic'' events modifying the qubit parity only but not its logical state (i.e., with exchanged energy $|\epsilon_i| \ll hf_{01}$) their rate takes the form
\begin{align}
\Gamma_\mathrm{qp}^\mathrm{ii}=&\,p_0(\tilde\Gamma_{00}^{B}x_{B}+\tilde\Gamma_{00}^{T}x_{T})+ p_1(\tilde\Gamma_{11}^{B}x_{B}+\tilde\Gamma_{11}^{T}x_{T}) 
\nonumber\\
\approx&\,\tilde\Gamma_{00}^{B}x_{B}+\tilde\Gamma_{00}^{T}x_{T}
\label{eq:gammaiiQP}
\end{align}
where $\xL$, $\xR$ are the dimensionless quasiparticle densities in the top and bottom electrodes, respectively [$x_\alpha=N_\alpha/N_{\rm Cp}^\alpha$, with $N_\alpha$ and $N_{\rm Cp}^\alpha$ the numbers of quasiparticle and the Cooper pairs in electrode $\alpha$].
The rates $\tilde{\Gamma}_{ij}^\alpha$ denote tunneling of quasiparticles initially located in electrode $\alpha$ into the other electrode with initial and final transmon states $i$ and $j$.
The approximation in the last expression is valid at leading order in the ratio $\EJ/\EC$ for a transmon qubit. For the rates of events changing the transmon state by exchanging energy $\sim hf_{01}$ (weighted by the qubit state occupation probability) we consider 
 \begin{align}
 \label{eq:gamma01QP}
\Gamma_\mathrm{qp}^\mathrm{01}=&\,
p_0 \tilde\Gamma_{01}^{B}x_{B} \, ,
    \\  \label{eq:gamma10QP} \Gamma_\mathrm{qp}^\mathrm{10}=&\,p_1(\tilde\Gamma_{10}^{T}x_{T}+\tilde\Gamma_{10}^{B}x_{B})
 \end{align}
for events associated with qubit excitation and relaxation, respectively (see Appendix~\ref{app:paritymodel} about the contribution to $\Gamma^{01}_\mathrm{qp}$ from the top electrode's quasiparticles). 
Note that the excited state population $p_1$ is in general determined not just by quasiparticles, but by all the mechanisms affecting the qubit, including measurement as explained above; since $p_1$ can be determined experimentally, we treat it as an input parameter in analyzing the parity-switching time.

The steady-state values of the quasiparticle densities in the two electrodes $x_B$ and $x_T$ are computed by solving a system of coupled rate equations. Physically, these densities are determined by the balance between processes keeping the quasiparticles away from the junctions, such as recombination and trapping, and quasiparticle generation, which in our model originates from 
pair breaking by high-frequency photons with rate $\propto \Gamma_{\rm PAPS}$. 
Details on the approximate solution to the rate equations and on the temperature and parallel field dependencies of the rates $\tilde{\Gamma}^\alpha_{ij}$ and $\Gamma^{\rm ph}_{ij}$ are given in Appendix~\ref{app:paritymodel}.
The parallel magnetic field modulates these rates via three distinct effects: 
first, $\EJ$ changes due to a combination of gap suppression and Fraunhofer effect;
this also changes the qubit frequency and influences the matrix elements for quasiparticle transitions.
Second, similarly to the Fraunhofer modulation of the critical current, these matrix elements are also directly altered by phase-interference effects. Third, the two superconducting gaps get suppressed differently because the critical field depends on film thickness; in modeling the parity-switching time in the following section, we can neglect the field dependence of the superconducting gap as we only reach a small fraction of the critical field.

\subsection{Distinguishing parity switching mechanisms}
Figure~\ref{fig:fig3} (b) and (c) show the magnetic-field dependence of the estimated contributions $\Gamma_\mathrm{PAPS}$, $\Gamma_\mathrm{qp}^\mathrm{01}$, $\Gamma_\mathrm{qp}^\mathrm{10}$ and $\Gamma_\mathrm{qp}^\mathrm{ii}$ to $\taupe$ and $\taupo$, respectively.
We first address the nonmonotonic behavior of $\taup$, so we focus on the ``odd'' case, where this feature is more prominent.
At zero field, the quasiparticle tunneling rates $\Gamma_\mathrm{qp}^\mathrm{01}$ and $\Gamma_\mathrm{qp}^\mathrm{10}$ are maximum, decreasing monotonically with $|\Bpar|$.
The transitions described by the rates $\Gamma_\mathrm{qp}^\mathrm{01}$ and $\Gamma_\mathrm{qp}^\mathrm{10}$ excite and relax the transmon, respectively, exchanging an energy $\sim hf_{01}$.
This process is strongly enhanced when $\DelL-\DelR\simeq h\fzeroone$ due to the large density of states in the quasiparticle spectrum near the gap edge [see \cref{fig:fig3}~(d) and Ref.~\cite{Marchegiani22}].
At zero field, the gap difference estimated through fitting slightly exceeds the qubit frequency, i.e., $\delta\Delta\gtrsim f_{01}$.
With increasing magnetic field, $\fzeroone$ decreases and is progressively detuned from the gap difference [\cref{fig:fig3}~(e)], causing the initial increase of $\taup$ with applied magnetic field.
In contrast, photon-assisted pair-breaking increases as $\EJ$ decreases, producing the subsequent decay of $\taup$ at higher fields.
The competition between PAPS and NUPS is similar to the one reported in Ref.~\cite{Diamond22}, where the authors distinguish the two contributions by
exploiting the modulation of the frequency of a SQUID transmon with an out-of-plane magnetic field.
Here, instead, the interplay of PAPS and NUPS is tuned by $\Bpar$, causing the non-monotonic evolution of $\taup$: while PAPS dominates for higher fields, NUPS dominates at low fields.

The difference between ``even'' and ``odd'' parity-switching times $\taupe$ and $\taupo$ is related to the dependence of the average excited-state population $p_1$ on the parity measurements:
as discussed above, $p_1$ is higher for ``odd'' than for ``even'' parity, because the ``odd'' parity predominantly ends up in $\ket{1}$ at the end of the parity mapping and then relaxes during the idling, while the even parity predominantly ends up in $\ket{0}$ (see Fig.~\ref{fig:parity_schemes}).
Quasiparticle transitions with concomitant transmon relaxation at rate $\Gamma_\mathrm{qp}^\mathrm{10}$ require the transmon to be in the excited state and are therefore proportional to $p_1$.
Consequently, the contribution of $\Gamma_\mathrm{qp}^\mathrm{10}$ is enhanced for ``odd'' parity compared to ``even'' parity.
At the same time, quasiparticle transitions at rate $\Gamma_\mathrm{qp}^\mathrm{10}$ are dominantly from the low-gap (top) electrode to the high-gap (bottom) electrode, since $\xR\gg\xL$ for $\delta\Delta>\fzeroone$~\cite{Marchegiani22}.
These transitions further increase the quasiparticle density in the high-gap electrode. As a result, $\xL$ is larger for ``odd'' parity, thus enhancing the contribution of $\Gamma_\mathrm{qp}^\mathrm{01}\propto\xL$, despite the lower occupation of the ground state $p_0=1-p_1$.
Plots of the estimated quasiparticle densities $\xL$, $\xR$ can be found in the supplemental material~\cite{SOM}.

At the highest fields, i.e., $\SI{0.3}{\tesla}\lesssim\Bpar\lesssim\SI{0.41}{\tesla}$, $\taup$ decays more strongly than the model predicts.
This coincides with an overall deterioration of the transmon coherence:
For $\Bpar\gtrsim\SI{0.3}{\tesla}$, the cavity $Q$-factor starts to decrease and the transmon lifetime $\Tone$ and Ramsey-coherence time $\Ttwostar$ drop by an order of magnitude between $\Bpar=\SI{0.3}{\tesla}$ and $\SI{0.41}{\tesla}$ (see supplemental material~\cite{SOM}).
Above $\Bpar=\SI{0.41}{\tesla}$  we observe an extended magnetic-field range where the transmon frequency is very unstable in $\Bperp$ and in time and therefore largely not measurable.
While generally recovering for $\Bpar\ge\SI{0.6}{\tesla}$, the SQUID remains unstable around its top sweet spot hence thwarting our plan to use the top sweet spot for high-field measurements of $\taup$.
In Ref.~\cite{Krause22} we also observed an unstable region between $\Bpar=\SI{0.4}{\tesla}$ and $\SI{0.5}{\tesla}$ with a similar device and suspected spurious JJs in the leads to cause it.
Here, these effects may be connected to the reduction of $\taup$ in a way that we currently do not understand.
It is also possible that the theory overestimates $\taup$ for higher magnetic fields as we disregard the field dependence of the gaps.
As discussed in \cref{sec:spectrum_vs_Bpar}, we estimate the contribution of the gap suppression to $\EJ$ to be less than $3\%$ and therefore only account for the Fraunhofer effect on the quasiparticle-tunneling rates.
However, the model uses a simplified description of the pair-breaking photons in terms of a single frequency (monochromatic radiation, see Appendix~\ref{app:paritymodel}).
More generally, few or several modes can contribute to the photon-assisted pair-breaking rate, and considering for example black-body radiation originating from higher-temperature stages of the refrigerator, the photon-assisted switching rates could increase more strongly than predicted by our model.

\section{Temperature dependence of the parity-switching time}
\label{sec:parity_lifetime_vs_temp}
\begin{figure}
  \centering
  \includegraphics[width=\columnwidth]{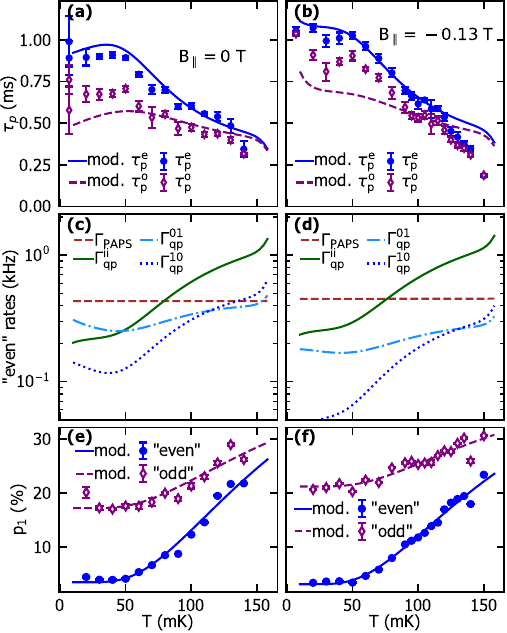}
  \caption{
    Temperature dependence of $\taup$ for \textbf{(a)} $\Bpar=\SI{0}{\milli\tesla}$ and \textbf{(b)} $\Bpar=\SI{-0.13}{\tesla}$.
    A cascading decay starting already around $T\sim$\SI{50}{\milli\kelvin} cannot be explained by quasiparticle generation due to thermal phonons.
    \textbf{(c)} and \textbf{(d)}: Model contributions to $1/\taupe$ for both fields. The increase in quasiparticle-tunneling with temperature is dominantly caused by enhanced ``quasi-elastic'' tunneling $\Gamma_\mathrm{qp}^\mathrm{ii}$ due to a change in the quasiparticle distributions.
    Two different processes contribute here: first, an increasing excited-state populations $p_1$ enhances $\Gamma_\mathrm{qp}^\mathrm{10}$, resulting in an increasing high-gap quasiparticle density.
    Additionally, a thermal broadening of the quasiparticle energy distributions in bottom and top electrode yields more quasiparticles that can tunnel at a given energy.
    \textbf{(e)} and \textbf{(f)}: Temperature dependencies of $p_1$. The semi-phenomenological expression for the fits in these panels is discussed in Appendix~\ref{app:p1}.
    }
  \label{fig:fig4}
\end{figure}

To complement the data on the magnetic-field dependence of $\taup$, we measured its temperature dependence for selected in-plane magnetic fields $\Bpar$.
Figure~\ref{fig:fig4} (a) and (b) show the temperature dependence of ``even'' and ``odd'' parity-switching times $\taupe$ and $\taupo$ for $\Bpar=\SI{0}{\tesla}$ and $\Bpar=\SI{-0.13}{\tesla}$ (which is close to the maximum observed $\taup$; measurements performed at additional values of $\Bpar$ can be found in Ref.~\cite{SOM}).
Interestingly, we observe a cascading decay of $\taup$ starting already at relatively low temperatures of order $T\sim\SI{50}{\milli\kelvin}$.
This behavior cannot be explained by quasiparticle generation due to thermal phonons alone; indeed, this contribution is exponentially suppressed as $\exp(-2\DelR/T)$ and typically dominates over pair-breaking photons only at higher temperatures (see \cref{app:T1temp} and Ref.~\cite{Fischer24}). A similar temperature dependence has been observed but not explained at finite field in a semiconducting nanowire-based transmon~\cite{Uilhoorn2021}. At zero field and in tunnel-junction-based transmons, a decrease in $\taup$ with temperature has been attributed to the excitation of quasiparticles out of superconducting traps by phonons~\cite{Pan22} and more recently related to the gap asymmetry~\cite{connolly2023coexistence}.

Our model can capture the cascading decay by taking into account, along with the gap asymmetry, two distinct effects: the thermal broadening of the quasiparticle energy distributions and an increase in the excited state population $p_1$.
To illustrate these points, we consider the same contributions determining $\taup$ in our model as in the previous section, now additionally accounting for the temperature variation [\cref{fig:fig4} (c) and (d)]. The contribution $\Gamma_{\rm PAPS}$ from photon-assisted parity switching is temperature-independent in the measured range~\footnote{The gap suppression with temperature can be safely neglected in this range, temperature being much smaller than the critical temperatures of the two electrodes.}.
At base temperature $T\sim\SI{7}{\milli\kelvin}\ll \delDel/k_B$ quasiparticle tunneling rates that leave the transmon state unchanged are slower, $\Gamma_\mathrm{qp}^\mathrm{ii} < \Gamma_{\rm PAPS}$; quasiparticles are mainly located in the lower-gap electrode with energy $\sim\DelR$, and rates for events where they tunnel from the bottom to the top electrode without changing the qubit logical state are suppressed as $e^{-\delDel/k_B T}$.
Starting from $T\gtrsim\SI{50}{\milli\kelvin}$, the excited state population $p_1$ increases [see \cref{fig:fig4} panels (e) and (f)], and so the (weighted) rate $\Gamma_\mathrm{qp}^\mathrm{10}$ increases. Since this process leads to a larger quasiparticle density in the high-gap (bottom) electrode (see plots in \cite{SOM}), the rates $\Gamma_{01}^{\rm qp}$ and $\Gamma_{ii}^{\rm qp}$ are enhanced, too.
The increase of $\Gamma_{ii}^{\rm qp}$ is also assisted by the thermal broadening of the quasiparticles distributions [see \cref{fig:fig3}(d)].
All these processes gradually suppress $\taup$ already for temperatures well below those at which quasiparticles generated by thermal phonons eventually become the limiting factor, $T\gtrsim \SI{150}{\milli\kelvin}$.

The robustness of our modeling is demonstrated by the fact that a fixed set of (few) parameters captures both the magnetic field and the temperature dependencies of $\taup$.
Moreover, the estimated
superconducting-gap difference $\delDel=\SI{5.49}{\giga\hertz}$ is in reasonable agreement with a simple estimate based on a phenomenological law for aluminum thin films~\cite{Marchegiani22}. Our estimate for the trapping rates $s_\mathrm{B}=s_\mathrm{T}=\SI{3.23}{\hertz}$ is smaller yet comparable to previously reported values for transmons of similar geometry, which are in the range $10$ to $30\,$Hz~\cite{Wang14, Diamond22}; we note that when fitting the data, the photon-assisted rate $\Gamma_{\rm PAPS}$ weakly correlates with $s$.
We independently quantify the gap $\DelR$, and so the onset of thermal quasiparticles, by fitting the temperature dependence of $\Tone$ measured for different in-plane magnetic fields $\Bpar$ (\cref{app:T1temp}).
The resulting $\DelR/h=\SI{54}{\giga\hertz}$ slightly overestimates the onset temperature for thermal quasiparticles to dominate the temperature dependence of $\taup$; accounting for the temperature dependence of the trapping dynamics could perhaps reduce the mismatch.
Moreover, the ``even'' case is generally better described than the ``odd'' case as the model considers a steady-state solution of coupled rate equations.
Our pulse sequence, however, maps ``odd'' parity to the transmon state $\ket{1}$, repeatedly driving the system away from its steady state. Hence, possible deviations due to the dynamics of the qubit during the measurement of $\taup$ affect more significantly the ``odd'' assignment.

\section{Conclusions and outlook}
\label{sec:conclusion}
We have explored the magnetic-field dependence of parity switching times in a 3D SQUID transmon with thin-film Al/AlO$_x$/Al Josephson junctions.
The magnetic field provides a tuning knob to explore the physics of quasiparticle dynamics in aluminum tunnel junctions; it suppresses the superconducting gaps $\DelL$, $\DelR$ of both sides of the JJ and it introduces a Fraunhofer-like pattern in the junction's critical current.
Both effects can influence the parity switching times.
We observe a non-monotonic evolution of the ``even'' and ``odd'' parity-switching times $\taupe$ and $\taupo$ with $\Bpar$:
a dip around $\Bpar=\SI{0}{\tesla}$ is followed by an initial increase reaching a maximum at $\Bpar\sim\SI{0.2}{\tesla}$, before decaying again at higher fields.
The observed maximum parity-switching time ($\taup$) of \SI{1.2}{\milli \second} is below the current state of the art on the order of a minute~\cite{vanWoerkom15, Iaia22}, but the shielding and filtering of our setup are not as advanced as those used in these works, and the 3D transmon has an antenna-like geometry that is more prone to absorption of pair-breaking photons~\cite{Rafferty21, Liu22}.

Modeling the quasiparticle dynamics between the two electrodes separated by the JJ with photon-induced pair-breaking as a generation mechanism,
we explain this behavior by a changing dominant contribution to the parity-switching rate as the magnetic field increases:
while photon-assisted parity switching (PAPS) dominates for higher fields, number-conserving parity switching (NUPS) dominates at lower fields.
More precisely, NUPS processes $\Gamma_\mathrm{qp}^{01}$ and $\Gamma_\mathrm{qp}^{10}$ that excite or relax the transmon to bridge the superconducting-gap difference $\delDel$ are resonantly enhanced around zero field as $h\fzeroone\simeq\delDel$.
The effect is more pronounced for $\taupo$ than for $\taupe$, which we attribute to the additional excited-state population $p_1$ introduced by our pulse sequence for ``odd'' parity, which further enhances $\Gamma_\mathrm{qp}^{10}$.
While photon-assisted parity switches could likely be reduced by better shielding and filtering, the superconducting-gap difference needs to be carefully engineered to avoid any resonance with transmon transition energies. 
As this work shows, the magnetic field lifts the resonance condition $h\fzeroone\simeq\delDel$ and can in principle be used post-fabrication as an analog to gap engineering.

To complement the magnetic-field data we also measured the temperature dependence of $\taup$ for selected $\Bpar$.
A cascading decay starting already at relatively low temperatures of  $T\sim\SI{50}{\milli\kelvin}$ is not due to quasiparticle generation by thermal phonons, as our model helps us understand.
Instead, the ``quasi-elastic'' tunneling rate $\Gamma_\mathrm{qp}^\mathrm{ii}$ rapidly increases above this temperature as the quasiparticle energy distributions in the bottom and top electrodes broaden,  yielding more available quasiparticles for tunneling at constant energy.
Additionally, $\Gamma_\mathrm{qp}^\mathrm{10}$ increases with temperature due to an increasing excited-state population $p_1$.
With the same model parameters, we consistently describe the magnetic-field and temperature dependencies of the measured parity-switching times.
The model thus helps to identify the relevant tunneling mechanisms and provides insight into how they contribute to the total parity switching rate: $\Gamma_\mathrm{qp}^{01}$ and $\Gamma_\mathrm{qp}^{10}$ dominantly capture the low-field behavior, $\Gamma_{\rm PAPS}$ the high field behavior, and $\Gamma_\mathrm{qp}^{ii}$ the temperature dependence.

Our results suggest that Al/AlO$_x$/Al JJ circuits are a viable option for parity readout of topological qubits~\cite{Hyart13} and that the required magnetic fields do not necessarily cause quasiparticle tunneling to limit the coherence of Al/AlO$_x$/Al JJ circuits.
The model we developed can help choose appropriate film thicknesses and junction geometries to approach operation up to \SI{1}{\tesla}.
Moreover, magnetic fields can help in understanding and optimizing gap engineering, as they provide a way to in-situ change the superconducting gap without heating the sample.
In samples with large gap asymmetry, the difference in critical fields will be more pronounced; hence, magnetic fields will strongly change the gap difference elucidating its impact on quasiparticle dynamics.

\begin{acknowledgments}
We acknowledge Ioan Pop, Dennis Willsch, and the other authors of Ref.~\cite{Willsch23}, for help in modeling and understanding the spectroscopic data. We also thank Dennis Willsch for his feedback on a draft of the manuscript. 
We would like to thank Philipp Janke for contributing to data analysis during a student internship.
We thank Kelvin Loh, M. Adriaan Rol, and Garrelt Alberts from Orange Quantum Systems for developing software with us to compile the time-domain sequences to the Zürich Instruments hardware.
We thank Michel Vielmetter for supporting the time-domain software development.
We thank Jurek Frey and Felix Motzoi for discussions on improving the parity-to-state mapping sequences.
We thank Lucie Hamdan and Timur Zent for their technical assistance.
This project has received funding from the European Research Council (ERC) under the European Union’s Horizon 2020 research and innovation program (grant agreement No 741121) and was also funded by the Deutsche Forschungsgemeinschaft (DFG, German Research Foundation) under CRC 1238 - 277146847 (Subproject B01) as well as under Germany’s Excellence Strategy - Cluster of Excellence Matter and Light for Quantum Computing (ML4Q) EXC 2004/1 - 390534769. 
\end{acknowledgments}

\section*{author contributions}

The project was conceived by C.D. and Y.A..
J.K. and C.D. fabricated the device and took the measurements with help from L.M.J..
The data was analyzed by J.K. and C.D. with help from L.M.J..
The theoretical model was extended to the magnetic field by G.M. and G.C. and fit to the data by J.K. and G.M..
The manuscript was written by J.K., G.M., G.C., and C.D. with input from all coauthors.

\section*{Software}

The setup was controlled based on \href{https://github.com/QCoDeS/Qcodes}{QCoDeS} drivers and logging~\cite{Qcodes}, while the measurements were run using \href{https://gitlab.com/quantify-os/quantify-core}{Quantify-core} and the time-domain experiments were defined and compiled to hardware using \href{https://gitlab.com/quantify-os/quantify-scheduler}{Quantify-scheduler}~\cite{rol2021quantify,crielaard2022quantify}.
In the data analysis, we used \href{https://github.com/qutip/qutip}{qutip} for modeling a qutrit lindblad equation~\cite{Johansson2013} and \href{https://github.com/hmmlearn/hmmlearn}{hmmlearn} for fitting a Gaussian hidden-Markov model to analyze the parity-switching datasets~\cite{hmmlearn}.

\section*{Data availability}

Datasets and analysis in the form of Jupyter notebooks that create the figures of this manuscript are available on Zenodo with doi \href{https://doi.org/10.5281/zenodo.10728469}{10.5281/zenodo.10728469}.

\appendix

\section{Experimental setup}
\label{app:setup}

\begin{figure*}
	\centering
		\includegraphics[width=\textwidth]{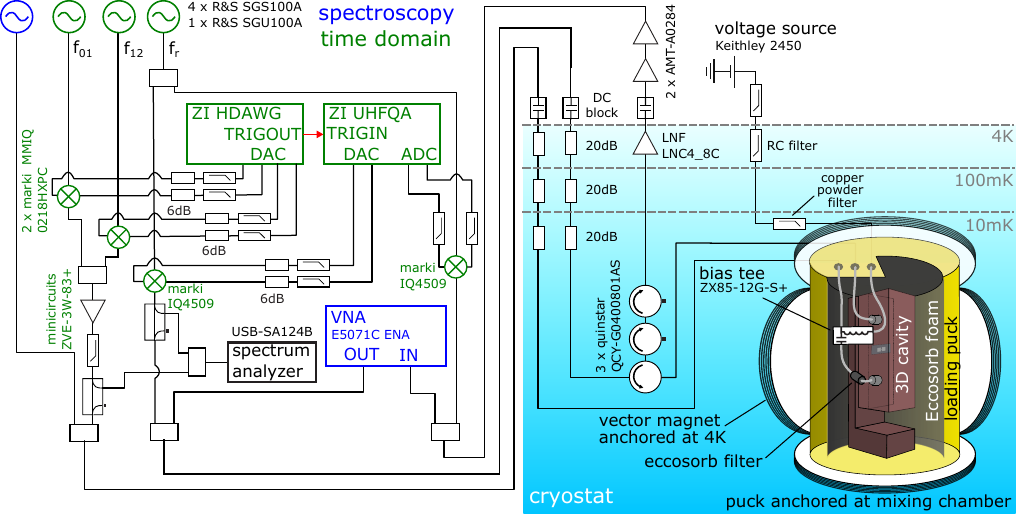}
	\caption{
		Wiring diagram of the experiment with the setup for spectroscopy using a two-port VNA and an additional microwave source and the time-domain setup using ADCs, DACs, and mixers.
		The transmon is inside a 3D cavity that is in the puck of the bottom loading dilution refrigerator.  
	}
	\label{fig:wiring_diagram}
\end{figure*}

The transmon is mounted in a 3D copper cavity which is loaded into a bottom-loading dilution refrigerator (Triton 500, Oxford Instruments) with a nominal base temperature of $\sim$\SI{7}{\milli\kelvin}. 
The cavity was wrapped in Eccosorb LS-26 foam to improve the radiation shielding.
This might be necessary because the outside of the puck sees the still plate environment, which is at about \SI{1}{\kelvin}.

A detailed wiring diagram of the experiment can be found in \cref{fig:wiring_diagram}.
An Eccosorb filter on the input line filters high-frequency radiation. 
However, the output line remained unfiltered to avoid signal loss.
Consequently, potential high energy photons coming from the LNF-LNC4\_8C HEMT amplifier at the \SI{4}{\kelvin} plate may leak back into the cavity to generate quasiparticles.
Likely filtering of the output line is currently the easiest way to reduce the Cooper-pair breaking radiation in the setup.

To voltage-bias the device we use a bias-tee at one of the cavity pins.
The voltage is applied relative to the ground of the dilution refrigerator, which is connected to the copper cavity, while the transmon islands are both floating.
The cavity pin couples asymmetrically to the two islands enabling efficient charge-bias (see \cref{app:gate_scans}).

The temperature control is done via a PID loop using the Lakeshore resistance bridge of the dilution refrigerator.
We took care to always stabilize for about \SI{10}{\minute} at every temperature to make sure the transmon reaches equilibrium.
All temperature readings are based on the mixing chamber $\mathrm{RuO_x}$ thermometer. 
To check the thermalization, we also looked at the transmon temperature based on ground-state occupancy measurements with the transmon nominally in the ground state (for details see Ref.~\cite{SOM}).
At the base temperature the transmon temperature is about \SI{50}{\milli\kelvin}. 

The transmon is controlled with standard DRAG pulses~\cite{Motzoi09, Chow10b}.
Predominantly the gate times for $\pi$ and $\nicefrac{\pi}{2}$ pulses were \SI{20}{\nano\second}.
The pulse amplitude is optimized based on Rabi sequences, the frequency based on Ramsey sequences (including beating) and the DRAG parameter is optimized based on an XY sequence~\cite{ReedPhD13}.
While the transmon control pulses and continuous-wave tones were routed to the input port of the cavity, we measured the cavity in reflection via the circulator.
The transmon readout was performed without a parametric amplifier. 
Nonetheless, typical assignment fidelities of \SI{>90}{\percent} could be achieved for the qubit subspace with readout durations around \SI{1}{\micro\second}.

\section{Device fabrication, geometry and film thickness}
\label{app:device_fab_geometry}

\begin{figure}
	\centering
		\includegraphics[width=\columnwidth]{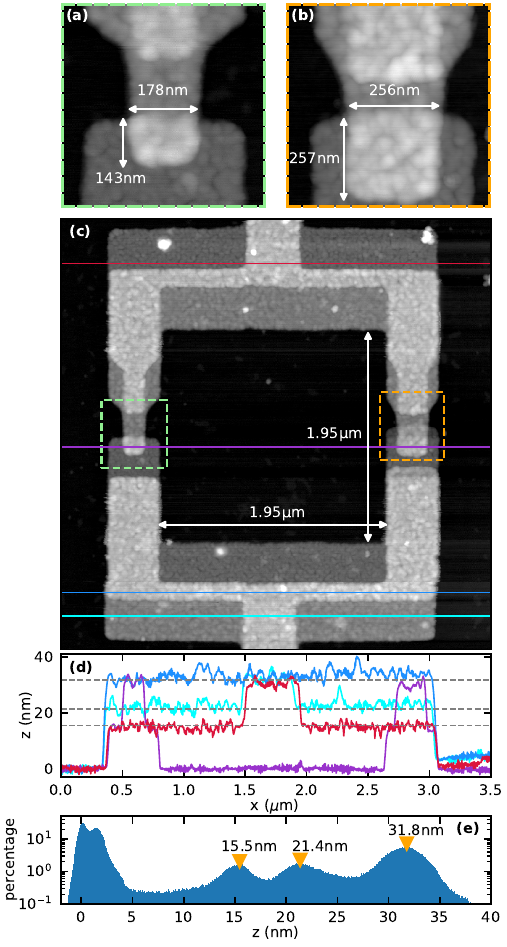}
	\caption{
		Atomic force microscope image showing in \textbf{(c)} the SQUID region of the transmon with the dimensions of the SQUID loop.
        The height is color-coded.
        Panels \textbf{(a)} and \textbf{(b)} are magnified images of the JJs and include their lateral dimensions.
		Aluminum grains with a typical size of \SI{50}{\nano\meter} are visible.
		\textbf{(d)} Height profiles along the lines indicated in \textbf{(c)}.
        \textbf{(e)} Histogram of the heights in the image with indicated peaks. 
        The three peaks at larger heights correspond to the bottom aluminum layer, the top layer, and the regions where they overlap.
	}
	\label{fig:device_figure}
\end{figure}

The 3D transmon used in this experiment was fabricated in the same batch and with the same capacitor geometry as the device in Ref.~\cite{Krause22}.
The aluminum film thickness plays a crucial role in making the transmon magnetic field resilient, but in the thin-film limit, it also has a strong effect on the magnitude of the superconducting gap~\cite{Chubov1969, Cherney1969, Meservey1971, Marchegiani22}. In Ref.~\cite{Krause22}, the nominal film thicknesses for the two aluminum layers according to the Plassys MEB 550S evaporator used for the film deposition were reported to be \SI{10}{\nano \meter} and \SI{18}{\nano \meter}.
The deposition rate was \SI{0.2}{\nano\meter\per\second}.
Between the two evaporations, the JJ barrier is formed by oxidation in  \SI{1}{\milli\bar} of pure oxygen in a static setting for \SI{6}{\minute}.
After the second layer, another oxidation step was added to grow the initial oxide in a more controlled way with \SI{1}{\milli\bar} of oxygen for \SI{10}{\minute}.

To measure the film thicknesses and JJ geometry, we perform Atomic force microscope imaging of the SQUID region of the device (\cref{fig:device_figure}).
The height profile [see~\cref{fig:device_figure} \textbf{(c)}, \textbf{(d)} and \textbf{(e)}] shows three characteristic thicknesses at
$t_1=\SI{15.5}{\nano \meter}$ for the bottom layer, $t_2=\SI{21.4}{\nano \meter}$ for the top layer, and $t_3=\SI{31.8}{\nano\meter}$ for the double layer. Since these measurements include either one or two additional oxide layers,
we estimate the thicknesses $t_\mathrm{B}$ and $t_\mathrm{T}$ of the bottom and top superconducting electrodes using the following equations:
\begin{align}
t_1&=t_\mathrm{B}+t_{\rm AlOx}^{\rm ext}\\
t_2&=t_\mathrm{T}+t_{\rm AlOx}^{\rm ext}\\
t_3&=t_\mathrm{B}+t_\mathrm{T}+t_{\rm AlOx}^{\rm ext}+t_{\rm AlOx}^{\rm int}
\end{align}
where $t_{\rm AlOx}^{\rm int}$ and $t_{\rm AlOx}^{\rm ext}$ are the thicknesses of the inter-electrode and external AlOx insulating barriers, respectively. 
We assume that, due to the exposure to ambient air and temperature, the oxide layers adding to $t_1$ and $t_2$ are the same, regardless of the extra-oxidation step that the bottom layer has faced during fabrication. 
Assuming that the thickness of the inter-layer barrier is approximately $t_{\rm AlOx}^{\rm int}\approx\SI{1}{\nano\meter}$, we obtain $t_\mathrm{B}\approx\SI{9.4}{\nano\meter}$, $t_\mathrm{T}\approx\SI{15.3}{\nano\meter}$ and an outer insulating oxide layer of $t_{\rm AlOx}^{\rm ext}\approx\SI{6.1}{\nano\meter}$. 
Notably, the gaps estimated with the phenomenological model $\Delta_{\mathrm{B},\mathrm{T}}\approx \Delta_{\rm Al}^{\rm bulk}+a/t_{\mathrm{B},\mathrm{T}}$ (with $a=600\,\mu$eV$/$nm and $\Delta_{\rm Al}^{\rm bulk}\approx 180\,\mu$eV, see Ref.~\cite{Marchegiani22} and references therein) are $\DelL=\SI{59}{\giga\hertz}\cdot h$ and $\DelR=\SI{53}{\giga\hertz}\cdot h$; these values differ by less than \SI{2}{\percent} (and their difference by less than \SI{10}{\percent}) from the gaps estimated in fitting the measured parity-switching and $T_1$ times.
Figure~\ref{fig:device_figure} also clearly shows that the aluminum film is polycrystalline with typical grain sizes on the order of \SI{50}{\nano \meter} and film thickness variations on the order of up to \SI{2}{\nano \meter}.

Additional information on the device stability, coherence times, and transmon temperature is provided in the supplementary information~\cite{SOM}.

\section{Josephson harmonics}
\label{app:transmon_hamiltonian}

The measurements of the frequencies $f_{ij}$ and their parity splittings $\delta f_{ij}$ of the three lowest logical transmon states unexpectedly cannot be described by the standard Cooper-pair box Hamiltonian
\begin{equation} \label{eq:transmon_hamiltonian}
	\begin{split}
		H = \; \; \;   4  \EC  & \sum_{n} (n-n_g)^2 \ket{n} \bra{n}  \\
		- \frac{1}{2} \EJ & \sum_{n}   \left( \ket{n}\bra{{n+1}} + \ket{n+1}\bra{n}\right),  
	\end{split}
\end{equation}
when trying to fit them simultaneously.
Here $\ket{n}$ represents the charge basis state with excess charge $2e n$.
Related observations by other groups have been combined with our data from this device into a joint publication that reports evidence of deviations from the sinusoidal current-phase relationship for conventional Al/AlO$_x$/Al JJs~\cite{Willsch23}. 
A more accurate description of the spectrum requires introducing higher harmonics $\EJtwo \cos(2\phi)$, $\EJthree \cos(3\phi)$, ... into the Hamiltonian.
In the charge basis, the modified transmon Hamiltonian reads
\begin{equation} \label{eq:transmon_hamiltonian_higher_harmonics_compact}
	\begin{split}
		H =  & \; \; \;   4  \EC \sum_{n} (n-n_g)^2 \ket{n} \bra{n}  \\
		- & \frac{1}{2}  \sum_{m=1}^\infty\sum_{n}\EJm\left(\ket{n}\bra{{n+m}} + \ket{n+m}\bra{n}\right)
	\end{split}
\end{equation}
where the index $m$ identifies the order of the Josephson harmonic.
Here we generalize this approach to a SQUID transmon with two junctions $a$ and $b$; then Eq.~\eqref{eq:transmon_hamiltonian_higher_harmonics_compact} becomes
\begin{equation} \label{eq:transmon_hamiltonian_higher_harmonics_compact_squid}
	\begin{split}
		H =  & \; \; \;   4  \EC \sum_{n} (n-n_g)^2 \ket{n} \bra{n}  \\
		- & \frac{1}{2}  \sum_{m=1}^\infty\left(\EJam+\EJbm e^{im\phi}\right)\sum_{n}\ket{n}\bra{{n+m}} \\
		- & \frac{1}{2}  \sum_{m=1}^\infty\left(\EJam+\EJbm e^{-im\phi}\right)\sum_{n}\ket{n+m}\bra{n} \, 
	\end{split}
\end{equation}
where the phase factors multiplying $\EJbm$ account for the (reduced) flux $\phi$ piercing the SQUID.

To start with, we fit \cref{eq:transmon_hamiltonian_higher_harmonics_compact_squid} to the zero-field data of \cref{fig:fig2}, including both the perpendicular flux scan for $\Bpar=0$ and the $f_{ij}$ and $\delta f_{ij}$ obtained in voltage-gate scans at bottom sweet spots of the flux arc.
We assume that JJs $a$ and $b$ have the same harmonics ratios $\EJam/E_\mathrm{Ja,1}= \EJbm/E_\mathrm{Jb,1} \equiv c_m$.
The voltage-gate scans of $\fzeroone$, $\fzerotwoovertwo$, $\fonetwo$, and $\fzerothree$ dominantly determine the Josephson energy at the bottom sweet spot, $\EJ(\phi=0.5)=E_\mathrm{Ja,1}-E_\mathrm{Jb,1}$, and the harmonics ratios at the bottom sweet spot,
\begin{equation}
\frac{\EJam+(-1)^m\EJbm}{E_\mathrm{Ja,1}-E_\mathrm{Jb,1}} = \begin{cases}
c_m &\text{for $m$ odd}\\
c_m/\alpha_\mathrm{JJ} &\text{for $m$ even\,.}
\end{cases}
\end{equation}
The flux scan in turn determines the SQUID asymmetry $\alpha_\mathrm{JJ}$.
As a result, we find the zero-field Josephson energies $\EJa(0)/h=\SI{19.47}{\giga\hertz}$, $\EJb(0)/h=\SI{5.97}{\giga\hertz}$ (corresponding to  $\alpha_\mathrm{JJ}(0)=0.53$), and the charging energy $\EC/h=\SI{327.5}{\mega\hertz}$.
We find that the higher harmonics decay rapidly with the index $m$, estimating $c_2 = -0.9\%$ and $c_3 = 0.03\%$.
To avoid overfitting, we set the harmonics of order $m\ge4$ to zero.

We note that our methodology is somewhat different from that employed in Ref.~\cite{Willsch23}: there, extensive scans of the sets $\left \{ E_{\mathrm{J}m}\right \}$ of harmonics compatible with the data are performed and the interaction of the transmon with the readout resonator or cavity is taken into account; here we instead just perform a fit to the data using a limited number of harmonics and do not include the effect of the cavity.
Thus, we find slightly different results based on the same data. 
For instance, for the charging energy we obtain a value about $1\%$ lower than the estimate in Ref.~\cite{Willsch23}. 
As for the values of the Josephson harmonics, our estimates become smaller with increasing order $m$ compared with those in that reference (by factors of order 1.4 and 13 for $c_2$ and $c_3$, respectively, when comparing to the model truncated to 4 rather than 3 harmonics).

Hereafter, we drop the label 1 to denote the first Josephson harmonic, i.e., $\EJ\equiv\EJone$, and $E_{Ja(b)}\equiv E_{Ja(b),1}$, for notational simplicity.

\section{Estimating Fraunhofer and critical fields}
\label{app:fraunhofer_and_critical_field}

The experimental data analysis of Appendix~\ref{app:transmon_hamiltonian} yields the zero-field Josephson energies $\EJa(0)$ and $\EJb(0)$ of the SQUID, and their harmonics.
The Josephson energies vary as a function of the magnetic field.
In this Appendix, we discuss a theoretical model for this dependence to quantify the relative contributions of i) field-suppression of the superconducting gaps $\DelL$ and $\DelR$, ii) Fraunhofer-like modulation of the Josephson energy, quantified in terms of two characteristic fields $\Bphinaughta$ and $\Bphinaughtb$ for $\JJa$ and $\JJb$, respectively.
We model the field evolution of $\EJa(\Bpar)$ and $\EJb(\Bpar)$ by the following expression, also used in Ref.~\cite{Krause22}:
\begin{equation}
    E_{Ja(b)}(\Bpar) = E_{Ja(b)}(0)\sqrt{1-\left(\frac{\Bpar}{\Bcrit}\right)^2}\left|\mathrm{sinc}\left(\frac{\Bpar}{B_{\Phi a(b)}}\right)\right|\, ,
    \label{eq:EJvsBpar}
\end{equation}
where $\sinc(x)\equiv\sin(\pi x)/(\pi x)$.
The factor in the absolute value in the right-hand side of Eq.~\eqref{eq:EJvsBpar} accounts for the Fraunhofer modulation of the Josephson energy for a rectangular junction~\cite{Barone}.
Even though thin aluminum films with different thicknesses and gaps display unequal critical fields $(\BcritL\neq \BcritR)$~\cite{Fulde1973}, we find that a single effective critical field $B_c$ can be used to capture the impact of the gaps' suppression on $\EJ$; the square root in the right-hand side of Eq.~\eqref{eq:EJvsBpar} models the gap suppression with a Ginzburg-Landau type formula $\Delta(\Bpar)/\Delta(0)=\sqrt{1-(\Bpar/B_c)^2}$~\cite{Tinkham04} (see \cite{Janssen2024} for the applicability of this approach). 
The Josephson energy of a JJ with asymmetric gaps is proportional to the harmonic mean of the two gaps, i.e., $\EJ\propto\DelL\DelR/(\DelR+\DelL)$, so at the leading order in $\Bpar\ll \BcritL,\BcritR$, Eq.~\eqref{eq:EJvsBpar} approximates the expression obtained retaining separately $\BcritL, \BcritR$, if we set $\Bcrit^{-2}= [\DelR(0)(\BcritL)^{-2}+\DelL(0)(\BcritR)^{-2}]/[\DelL(0)+\DelR(0)]$.
The validity of this approach is also supported by the analysis of the cavity frequency shown below.

\begin{figure}
	\centering
		\includegraphics[width=\columnwidth]{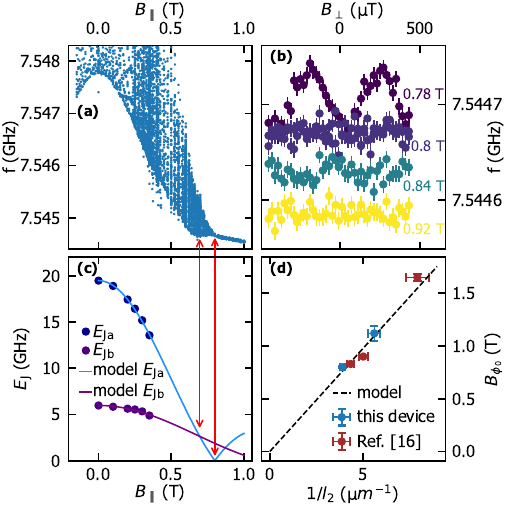}
	\caption{
        Estimation of the Fraunhofer modulation of the Josephson energy. \textbf{(a)}-\textbf{(b)} Measurements of the resonance frequency $\fcav$ of the cavity for different values of the out-of-plane ($\Bperp$) and in-plane ($\Bpar$) magnetic field plotted against \textbf{(a)} $\Bpar$ and \textbf{(b)} $\Bperp$ for selected values of $\Bpar$.
        The cavity is coupled to the transmon; the periodic oscillations of the transmon frequency with $\Bperp$ (due to the SQUID geometry) translate to oscillations in $\fcav$ via the dispersive shift.
        Due to the Fraunhofer effect, the oscillation vanishes at $\Bphinaughta=\SI{0.8}{\tesla}$ (see the text for the explanation).
        \textbf{(c)} The estimated Josephson energies $\EJa$ and $\EJb$ of the two JJs intersect at $\Bpar\simeq\SI{0.7}{\tesla}$, corresponding to the minimum $\fcav$ in \textbf{(a)}. That way we can estimate $\Bphinaughtb=\SI{1.12}{\tesla}$. \textbf{(d)} The estimated $\Bphinaughta, \Bphinaughtb$ are inversely proportional to the JJ width $l_2$ that is penetrated by $\Bpar$ (as expected for a rectangular junction~\cite{Barone}) and agree with more direct measurements performed in Ref.~\cite{Krause22} on other devices from the same batch.
	}
	\label{fig:estimating_fraunhofer_fields}
\end{figure}

To start with, we estimate the Fraunhofer fields for the two JJs.
While we could not systematically measure the transmon spectrum for $\Bpar>\SI{0.41}{\tesla}$, we did measure the cavity resonance frequency $\fcav$ up to \SI{1}{\tesla}, the limit of our magnet.
Sweeping $\Bperp$ by a few hundreds of \si{\micro\tesla} for every $\Bpar$, we could clearly observe the SQUID oscillations in $\fcav$ mediated by the dispersive shift up to $\SI{0.9}{\tesla}$ [see \cref{fig:estimating_fraunhofer_fields}(a) and (b)].
Around $\Bpar\simeq\SI{0.8}{\tesla}$ the oscillation collapses, before reviving again for fields $\SI{0.82}{\tesla}\le\Bpar\le\SI{0.9}{\tesla}$.
We attribute this behavior to one superconducting flux quantum ($\Phi_0= h/(2e)$) threading the larger of the two JJs, turning the SQUID effectively into a single-JJ transmon [right arrow in panel (c) of \cref{fig:estimating_fraunhofer_fields}]. Therefore, we estimate $\Bphinaughta =\SI{0.8}{\tesla}$.
Moreover, we observe a local minimum in $\fcav$ around \SI{0.7}{\tesla}.
As indicated by the left arrow in panel (c) of \cref{fig:estimating_fraunhofer_fields} this minimum 
corresponds to the condition $\EJa=\EJb$, 
where the SQUID is symmetric.
The resulting near-complete suppression of $\EJ$ for half a flux quantum through the SQUID leads to a reduction of the dispersive shift so that $\fcav$ approaches the bare cavity frequency.
The intersection of  $\EJa$ and $\EJb$ depends only on the zero-field Josephson energies $\EJa(0)$ and $\EJb(0)$, and on the characteristic Fraunhofer fields $\Bphinaughta$ and $\Bphinaughtb$, the latter being the only remaining unknown at this point.
Requiring $\EJa(\SI{0.7}{\tesla}) = \EJb(\SI{0.7}{\tesla})$ we estimate $\Bphinaughtb=\SI{1.12}{\tesla}$.
As shown in \cref{fig:estimating_fraunhofer_fields} (d) the estimated Fraunhofer fields $\Bphinaught$ of $\JJa$ and $\JJb$ are inversely proportional to the laterally penetrated junction width $l_2$, with a proportionality constant that agrees well with that for other devices from the same fabrication batch and that were measured in two-tone spectroscopy up to \SI{1}{\tesla} \cite{Krause22}.

To determine the effective critical field $\Bcrit$ we substitute Eq.~\eqref{eq:EJvsBpar} into the Hamiltonian that includes the higher Josephson harmonics, Eq.~\eqref{eq:transmon_hamiltonian_higher_harmonics_compact_squid}, and fit the spectroscopic data for the qubit transitions and parity-frequency splittings in \cref{fig:fig2}(b) and (c) [also reproduced in \cref{fig:transmon_transitions_and_charge_dispersions_vs_bpar}(a) and (e)].
For the higher harmonics, as in Appendix~\ref{app:transmon_hamiltonian} we keep only those with $m= 2$ and 3, and assume their ratios $c_m$ to be independent of parallel field (see end of this Appendix about this assumption).
Using this procedure, we estimate $\Bcrit=\SI{1.85}{\tesla}$.
The actual $\Bcrit$ is likely lower, given that we fit a Ginzburg-Landau dependence and for thin films at low temperatures one expects a faster suppression of $\Delta$ close to the critical field~\cite{Janssen2024}. Indeed, we do not observe SQUID oscillations in $\fcav$ for $\Bpar\ge\SI{0.92}{\tesla}$, and the critical field estimated for similar devices is about 1~T~\cite{Krause22}.
We also note that the fitted value of the critical field might in part account for the field-dependent effects, not included in the model, that we discuss at the end of this Appendix; in this context, one can regard $\Bcrit$ as an effective critical field that makes possible a more accurate modeling of the spectroscopic data.
Figure~\ref{fig:transmon_transitions_and_charge_dispersions_vs_bpar} shows the fit including the residuals.
As in \cref{app:transmon_hamiltonian}, the voltage-gate scans of $\fzeroone$, $\fzerotwoovertwo$, $\fonetwo$, and $\fzerothree$ dominantly determine the Josephson energy at the bottom sweet spot, $\EJ(\Bpar,\phi=0.5)=\EJa(\Bpar)-\EJb(\Bpar)$.
Fitting the remaining perpendicular flux scans for fields $\Bpar>0$, we can further determine $\alpha_\mathrm{JJ}(\Bpar)$, and hence the data points for $\EJa$ and $\EJb$ shown in the inset of \cref{fig:fig2} (a) [also in Fig.~\ref{fig:estimating_fraunhofer_fields} (c)].
We find their field dependence to be well described by \cref{eq:EJvsBpar} and the fit parameters obtained in the course of this section, showing the self-consistency of our modeling.
We note that, for our purposes, it is sufficient to quantify the gap suppression in the magnetic-field range covered by measurements of the parity-switching time, $|\Bpar|\le\SI{0.41}{\tesla}\ll\Bcrit$. In this field range the gaps are suppressed by less than $3\%$ compared to the zero-field values, based on Eq.~\eqref{eq:EJvsBpar} and the estimated $\Bcrit$. For this reason, we disregard the field-dependence of the gaps in modeling the parity-switching rates.

In closing this Appendix, we note that the assumption that the ratios $E_\mathrm{Jm}/E_\mathrm{J1}$ are independent of parallel field [implying $E_\mathrm{Jm} \propto \mathrm{sinc}(B_\parallel/B_\Phi)$]  is strictly speaking incorrect, since for rectangular junctions the argument~\cite{Barone} leading to the sinc modulation in Eq.~\eqref{eq:EJvsBpar} gives $E_\mathrm{Jm} \propto \mathrm{sinc}(m B_\parallel/B_\Phi)$. 
Similarly, a contribution to the second harmonic arising from an inductance in series to the junction would change as $[\mathrm{sinc}(B_\parallel/B_\Phi)]^2$ (at leading order in the ratio between Josephson and inductive energies, see supplementary to Ref.~\cite{Willsch23}). 
We disregard these effects for simplicity since an even more accurate modeling of the spectroscopic data is beyond the scope of the present work. 
Still, we note that the different dependencies of the junction harmonics and inductive corrections on the parallel field could in principle be used to distinguish these two contributions.

\begin{figure}
	\centering
		\includegraphics[width=\columnwidth]{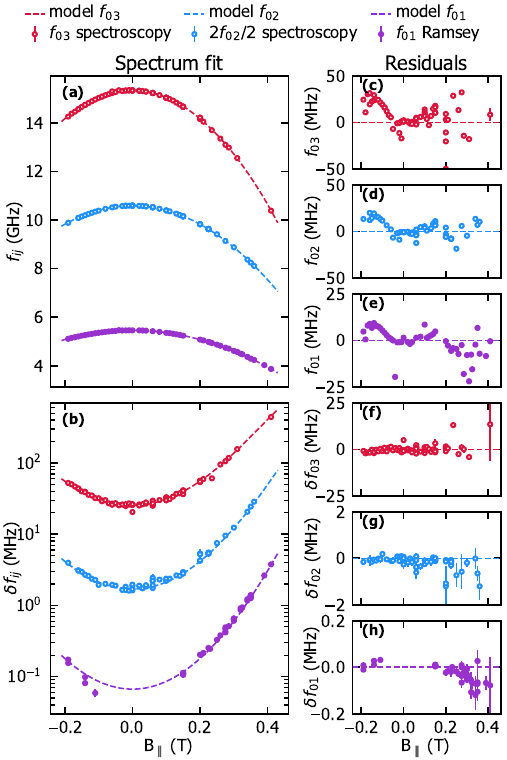}
	\caption{
        Transmon transitions frequencies $f_{ij}$ \textbf{(a)} and parity-frequency splittings $\delta f_{ij}$ \textbf{(b)} measured at the bottom sweet spot as a function of $\Bpar$.
        The data is obtained from voltage-gate scans like the ones shown in \cref{fig:fig2}.
        Substituting \cref{eq:EJvsBpar} into \cref{eq:transmon_hamiltonian_higher_harmonics_compact_squid} we estimate $\Bcrit=\SI{1.85}{\tesla}$ based on a fit to the data (see text).
        Fit residuals \textbf{(c)}-\textbf{(h)} are relatively flat for all $f_{ij}$ and $\delta f_{ij}$.
	}\label{fig:transmon_transitions_and_charge_dispersions_vs_bpar}
\end{figure}

\section{Gating the offset voltage}
\label{app:gate_scans}

\begin{figure*}
	\centering
		\includegraphics[width=\textwidth]{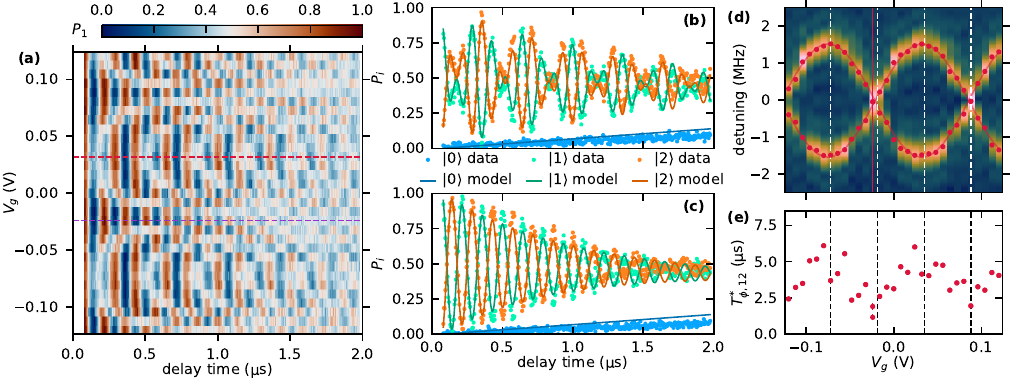}
	\caption{
		\textbf{(a)} Ramsey measurement for the 1-2 transition as a function of $V_\mathrm{g}$.
		The color plot gives the population of $\ket{1}$ as a function of waiting time between the $\nicefrac{\pi}{2}$ pulses and  $V_\mathrm{g}$.
		\textbf{(b)} and \textbf{(c)} linecuts of \textbf{(a)} close  the charge sweet spot and anti sweet spot. 
		In \textbf{(b)}, the characteristic beating for the two parities can be clearly seen.
		The three-state data is fitted with a qutip~\cite{Johansson2013} simulation.
		\textbf{(d)} Fourier transform of the temporal data in \textbf{(a)}, revealing the charge dispersion directly. 
		Data points are frequencies extracted from the fits to the linescans.
		\textbf{(e)} Ramsey pure dephasing time $T_{\phi,12}^*$ as a function of gate voltage. 
		While charge noise is not the dominant dephasing process, a slight sweet spot enhancement can be observed. 
	}
	\label{fig:Ramsey_gatescan}
\end{figure*}

In this Appendix, we show that we can control the charge bias of the transmon by applying a DC gate voltage $\Vg$  to the bias-tee on the cavity input port. 
Measuring the transmon transitions $\fzeroone$, $\fzerotwoovertwo$, $\fonetwo$, and $\fzerothree$ as a function of $\Vg$, we observe clear signature of charge-parity splitting for all the transitions (\cref{fig:fig2}).
This feature allows for gate-tuning of the transmon to a point of maximum charge-parity splitting, which is the ideal situation to measure the parity-switching time.

Some of the transitions shown in \cref{fig:fig2}~(d)-(f) were measured in time-domain, in a Ramsey-type measurement (see \cref{fig:Ramsey_gatescan} for an example measurement using a superposition of $\ket{1}$ and $\ket{2}$).
Depending on $\Vg$, the Ramsey data shows a beating of two frequency contributions or just a single frequency [see \cref{fig:Ramsey_gatescan}~(b) and (c)].
A Fourier transform of the Ramsey data [\cref{fig:Ramsey_gatescan} (d)] shows the detuning of the transition energies as plotted in \cref{fig:fig2}~(d)-(f); here the microwave frequency was chosen at the midpoint between the ``even'' and ``odd'' transition frequencies.
Accurately fitting the Ramsey data and tuning to this condition is essential for the subsequent parity measurements.
The Ramsey pure-dephasing time $T_{\phi,12}^*$ as a function of $\Vg$, \cref{fig:Ramsey_gatescan}~(e), is extracted from fitting the linecuts of \cref{fig:Ramsey_gatescan}~(a) using relaxation times from a preceding measurement of relaxation from the $\ket{2}$ state (see supplementary information~\cite{SOM} for an example). 
It shows an enhancement at the charge sweet spots but is not $\Tone$ limited as e.g. flux noise also contributes to $T_{\phi,12}^*$.
While Ramsey measurements are sensitive to $\Vg$, measurements of $\Tone$ and $\Techo$ do not show a $\Vg$ dependence~\cite{SOM}, suggesting that the dominant charge noise is slow.

We have also explored scanning $\Vg$ over many periods and see few jumps and no anomalies~\cite{SOM}. 
Generally, we find that the $\Vg$ scans show the expected double-sinusoidal dependence with few jumps and distortions, but we do find that $\Vg$ shows a slight hysteresis when reversing the scan direction.
We observe drift in the charge offset at fixed $\Vg$ that can cover a period on a timescale of \SI{10}{\minute}~\cite{Christensen19}.

\section{Parity measurement schemes}
\label{app:parity_schemes}

A single parity measurement is realized by a parity-dependent gate on the transmon that flips the qubit state for ``odd'' parity and leaves it unchanged for ``even'' parity, followed by a measurement of the transmon state.
The parity-dependent gate we use is Ramsey-based~\cite{Riste15}; its basic scheme and illustration on the Bloch sphere can be found in \cref{fig:parity_schemes}~(a).
The gate requires a careful tune-up: the carrier frequency of the qubit-control pulses needs to be well-centered between ``even'' and ``odd'' transmon frequencies $f_\mathrm{e}, f_\mathrm{o}$ and the voltage gate on the bias-tee, $\Vg$, is chosen such that $|f_\mathrm{e}-f_\mathrm{o}|$ is maximized.
Generally, for $\delta f_{ij}<\SI{10}{\mega\hertz}$, our usual \SI{20}{\nano\second} control pulses are not parity selective, meaning a $\pi/2$ pulse will map both parities onto the same point on the Bloch sphere [see \cref{fig:parity_schemes} (a)]. 
The microwave carrier defines a rotating frame, such that the detuning of each parity from the carrier gives the rotation frequency on the Bloch-sphere equator.
The detunings have opposite signs, and the parities precess in opposite directions.
After a waiting time $t_w=(2\, \delta f_{ij})^{-1}$, the two parities will have rotated to orthogonal states on the Bloch sphere that, with the right rotation, can be mapped to the ground and first-excited states $\vert 0 \rangle$ and $\vert 1 \rangle$, respectively. 
The same method can be used with a superposition of $\vert 1 \rangle$ and $\vert 2 \rangle$ [see \cref{fig:parity_schemes} (b)].
As explained in Sec.~\ref{sec:spectrum_vs_Bpar} of the main text, we explore a range of $\EJ/\EC$ ratios while sweeping $\Bpar$, such that the sequences involving higher transitions are useful when the charge dispersion for the 0-1 transition is too small:
the charge dispersion increases roughly by a factor of 10 with the level index, while $\Tone$ and $\Ttwostar$ decrease for the higher levels typically by less than 50\%.

\begin{figure}
	\centering
		\includegraphics[width=\columnwidth]{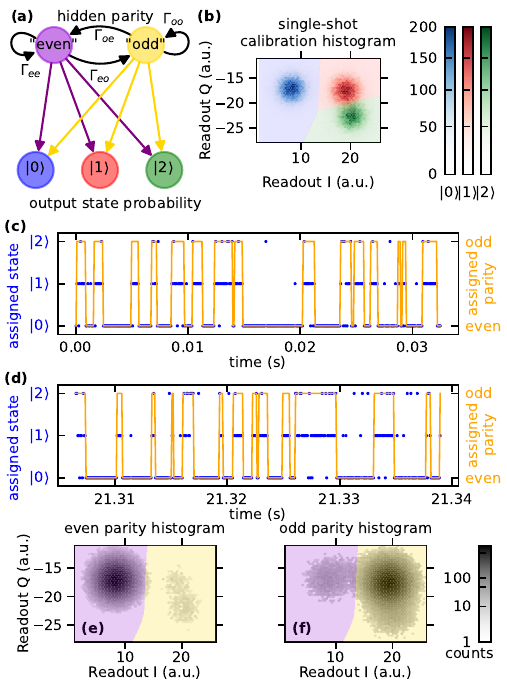}
	\caption{
		\textbf{(a)} Diagram illustrating the Gaussian mixture hidden Markov model. The parity state changes with certain (fixed) probabilities between runs. Both parities have different state probabilities, which correspond to different weights in a Gaussian mixture for the output probabilities.
		\textbf{(b)} Histogram of calibration points for the different prepared states directly preceding the parity measurement run. 
        The colors show the most likely state in each region based on a Gaussian classifier.
		\textbf{(c)} and \textbf{(d)} Assigned states from the Gaussian classifier and assigned parities from the HMM fit for the first 400 measurements at the beginning and the end of the parity measurement run comprising 262144 measurements.
  		\textbf{(e)} and \textbf{(f)} Histograms of the output voltages conditioned on the assigned parity.
        The colors show the most likely parity in each region based on a Gaussian classifier.
	   }
	\label{fig:parity_measurement_explanation_hmm}
\end{figure}

In addition to the parity-selective gate, the transmon needs to start in a known state for each run, and a good single-shot readout fidelity for the transmon is required to resolve the parity dynamics.
We generally achieve high single-shot fidelities $>90\%$ that can distinguish $\vert 0 \rangle$, $\vert 1 \rangle$ and $\vert 2 \rangle$ [see \cref{fig:parity_measurement_explanation_hmm} (b)].
We let the qubit relax by waiting $5\Tone$ between individual runs; at temperatures $\lesssim 50\,$mK, our qubit has a residual excitation of up to \SI{5}{\percent}.
In principle, one can also initialize by measurement~\cite{Riste13} or just look if the transmon state flips or stays the same compared to the previous measurement.

We then perform the parity-mapping sequence consisting of idling, parity gate, and readout $N=2^{18}$ times.
The measurement results form a sequence of $N$ points in the IQ plane, from which the $\taupe$ and $\taupo$ are determined by fitting a Gaussian hidden-Markov model (HMM)~\cite{Vool14}.
The hidden parameter in our Markov model is the ``true'' parity, while the measured parameters are the IQ Voltages which have different probability distributions for the $\ket{0}$, $\ket{1}$ and $\ket{2}$ states.
Depending on the parity, there are different probabilities of the transmon being mapped on the states indicated by colored arrows in \cref{fig:parity_measurement_explanation_hmm} (a), which we call emission probabilities.
Between two consecutive runs, the hidden parity can change or stay the same with fixed probabilities $t_\mathrm{rep}\Gamma_{ij}$ where $t_\mathrm{rep}$ is the repetition time and $\Gamma_{ij}$ [with $i,j$=\{``even'' (e), ``odd'' (o)\}] are the transition rates illustrated
by black arrows in \cref{fig:parity_measurement_explanation_hmm} (a).
In the model, we allow for different $\Gamma_\mathrm{eo}$ and $\Gamma_\mathrm{oe}$, which in turn fix $\Gamma_\mathrm{oo}$ and $\Gamma_\mathrm{ee}$.
The calibration points for the $\vert 0 \rangle$, $\vert 1 \rangle$ and $\vert 2 \rangle$ states are obtained by preparing those states followed by a measurement [see \cref{fig:parity_measurement_explanation_hmm} (b)].
Each IQ histogram for the prepared states ($\ket{0}$, $\ket{1}$ and $\ket{2}$) is well described by sums of three Gaussian probability distributions (one of the three Gaussians clearly dominates for each state, but we consider that our calibration points are slightly mixed due to residual excitation as well as relaxation).
The means and covariances of the dominant Gaussians corresponding to $\ket{0}$, $\ket{1}$ and $\ket{2}$, respectively, fix the probability distributions of the Gaussian HMM.
The free parameters of the HMM are the emission probabilities of the ``even'' and ``odd'' parities (the probabilities to measure each state given a parity) as well as $\Gamma_\mathrm{eo}$ and $\Gamma_\mathrm{oe}$.
These parameters are determined by a likelihood fit of the HMM to the data using the \emph{hmmlearn} package~\cite{hmmlearn}.

As outcomes of the HMM fit, we get the assigned parities for each run and the transition probabilities, from which $\taupe=(\Gamma_\mathrm{eo})^{-1}$ and $\taupo=(\Gamma_\mathrm{oe})^{-1}$ can be calculated with the known repetition time of the sequence.
\cref{fig:parity_measurement_explanation_hmm} (c) and (d) show the assigned parities as a function of time; here the assigned states are based on a Gaussian classifier yielding the most likely state for any point in the IQ plane [see colors in \cref{fig:parity_measurement_explanation_hmm} (b)].
We find that generally the ``even'' (``odd'') parity is associated with the $\vert 0 \rangle$ ($\vert 1 \rangle$) state. 
The histograms of the IQ data selected on the ``even'' and ``odd'' parity are shown in \cref{fig:parity_measurement_explanation_hmm} (e) and (f).
The overlap of the emission probability distributions of the two parities is $1-F_\mathrm{p}$ with $F_\mathrm{p}$ being the parity-measurement fidelity. 
To check the HMM methodology for consistency, we compared it with the results for the parity-switching time based on a fit to the power spectral density of the assigned states (see supplementary information~\cite{SOM}).

We filter our data based on Ramsey measurements before and after each parity measurement sequence:
The splitting $\delta f_{ij}$ between ``even'' and ``odd'' transmon frequencies has to match $t_w=(2\delta f_{ij})^{-1}$, and we reject data in which the transmon frequencies drifted during the time of the parity-measurement sequence.
Moreover, we ensured that $t_w$ is small compared to $\Ttwostar$ and $\Tone$.

\section{Modeling the contributions to the parity-switching time}
\label{app:paritymodel}

We discussed the theoretical model for the parity switching time in Sec.~\ref{sec:parityModelMain} of the main text. Here, we present explicit expressions for the rates associated with PAPS and NUPS contributions in Eqs.~\eqref{eq:PAPS}-\eqref{eq:gamma10QP}. The parity-switching time measurements have been performed at the lower sweet spot of the split-transmon device ($\Phi=\Phi_0/2$), where the gap difference is larger than the qubit energy transition, \textit{i.e.}, $\fLR>\fzeroone$ (denoted as case II in Ref.~\cite{Marchegiani22}) with $\fLR=(\DelL-\DelR)/h$. 
In this situation, quasiparticles accumulate in the lower-gap electrode at energies close to its gap due to relaxation (via phonon emission) and suppressed tunneling into the higher-gap electrode (because of the energy difference $\fLR-\fzeroone > 0$), see Ref.~\cite{Marchegiani22}.

As in the reference, we describe the quasiparticle dynamics in each superconducting electrode via rate equations for their density, see also Appendix~E of Ref.~\cite{Diamond22} and Sec.~III.D of the supplement to Ref.~\cite{connolly2023coexistence}. The equations are written in terms of the dimensionless quantities $\xL$ and $\xR$, obtained normalizing the quasiparticle density in electrode $\alpha=\{T,B\}$ by the Cooper pair density $n_{\rm Cp}^\alpha =2\nu_0 \Delta_\alpha$ (with $\nu_0\approx 0.73\times 10^{47}{\rm J}^{-1}{\rm m}^{-3}$ the single spin density of states of aluminum at the Fermi level and $\Delta_\alpha$ the superconducting gap),
\begin{align}
\frac{d \xL}{dt}&=g^B -s_B \xL -r^B \xL^2 + \delta \bar\Gamma^T \xR -\delta {\bar\Gamma}^B \xL \, ,
\label{eq:xLdot}\\
\frac{d \xR}{dt}&=g^T -s_T \xR -r^T \xR^2  - \bar\Gamma^T \xR + {\bar\Gamma}^B \xL \, . 
\label{eq:xRdot}
\end{align}
where $\delta=\DelR/\DelL$. In the right-hand sides of Eqs.~\eqref{eq:xLdot} and \eqref{eq:xRdot}, $g^{\alpha}$, $\alpha=\{T,B\}$, describe quasiparticle generation, and the terms proportional to $s_{\alpha}$ and $r^{\alpha}$ account for quasiparticle trapping and recombination, respectively; for the recombination rate, we use the values $r^T = r^B = 1/(160\,\mathrm{ns})$~\cite{Wang14,Marchegiani22}.
Finally, the terms proportional to
\begin{equation}
\bar{\Gamma}^\alpha = (1-p_1) (\bar{\Gamma}_{00}^\alpha + \bar{\Gamma}_{01}^\alpha)+ p_1 (\bar{\Gamma}_{11}^\alpha + \bar{\Gamma}_{10}^\alpha) 
\end{equation}
account for the tunneling of quasiparticles initially located in electrode $\alpha$ ($\alpha=\{T,B\})$ into the other electrode, where $p_1$ is the occupation probability of the excited state of the qubit. Note that in writing the first factor on the right-hand side in the form $(1-p_1)=p_0$ we implicitly assume that the occupations of higher levels can be ignored; it was estimated in the supplement to Ref.~\cite{connolly2023coexistence} that for high $p_1$ taking the occupation of second excited state into consideration could lead to corrections of the order of several percent, so we neglect this possibility here. In our notation for NUPS [see, for instance, Eqs.~\eqref{eq:gammaiiQP}-\eqref{eq:gamma10QP} in the main text], $\tilde{\Gamma}_{ij}^\alpha x_\alpha $ represents the quasiparticle-tunneling induced transition rate for the qubit from the initial logical state $i$ to the final state $f$ ($i,f =\{0,1\}$). The barred rates $\bar{\Gamma}_{ij}$ are obtained dividing $\tilde{\Gamma}_{ij}$ by the Cooper pair number in the low-gap electrode, \textit{i.e.}, $\bar{\Gamma}^\alpha=\tilde{\Gamma}^\alpha/(n_{Cp}^{T} {\mathcal V}_T)$, where $\mathcal{V}_T\simeq 3400\,\mu$m$^3$ is the volume of the low-gap electrode; this normalization is at the origin of the factors $\delta$ appearing in Eq.~\eqref{eq:xLdot}. 
In comparing Eqs.~\eqref{eq:xLdot} and \eqref{eq:xRdot} to the model of Ref.~\cite{Marchegiani22} [see Eqs.~(11)-(13) there], we note that here the terms $-s_\alpha x_\alpha$ in the right-hand side account for trapping (cf. Sec.~IV.D in Ref.~\cite{Marchegiani22}) and we additionally assume quasi-equilibrium quasiparticle distributions in both electrodes, having temperature coinciding with the fridge temperature, but allowing for different effective chemical potentials in the two electrodes. This assumption enables us to
use a single equation for $\xR=x_{T<}+x_{T>}$ obtained by summing Eqs.~(12) and (13) of Ref.~\cite{Marchegiani22} (the rates from the top electrode are defined so to obtain $\bar{\Gamma}_{ij}^T \xR = \bar{\Gamma}_{ij}^{T<} \xRm +\bar{\Gamma}_{ij}^{T>} \xRp $).

For the modeling of the parity-switching time, we are interested in the steady-state values of the quasiparticle densities in the two electrodes ($d\xL/dt=d\xR/dt=0$). Assuming $\delta\bar\Gamma_B+s_B\gg r^B\xL$, we can neglect recombination in the bottom electrode in Eq.~\eqref{eq:xLdot}, and then we can express the steady-state value of $\xL$ in terms of $\xR$,
\begin{align}
 \xL& =
     \,\frac{g^B/\delta+ [p_0(\bar\Gamma_{00}^T+\bar\Gamma_{01}^T)+p_1(\bar\Gamma_{11}^T+\bar\Gamma_{10}^T)]\xR}{p_0(\bar\Gamma_{00}^B+\bar\Gamma_{01}^B)+p_1(\bar\Gamma_{11}^B+\bar\Gamma_{10}^B)+s_B/\delta} \,
     \nonumber\\
     & \approx \,\frac{g^B/\delta+ (\bar\Gamma_{00}^T+p_1\bar\Gamma_{10}^T)\xR}{\bar\Gamma_{00}^B+p_0\bar\Gamma_{01}^B+p_1\bar\Gamma_{10}^B+s_B/\delta} 
     \, .\label{eq:xlss-II-Ref} 
\end{align}
The approximate expression in the second line of Eq.~\eqref{eq:xlss-II-Ref} is obtained noticing that parity switching rates for a transmon qubit are independent of the logical state of the qubit at the leading order in $E_J/E_C$~\cite{glazman21, Marchegiani22}, \textit{i.e.}, $\bar{\Gamma}_{00}^\alpha\approx \bar{\Gamma}_{11}^\alpha$, and neglecting the contribution of transitions exciting the qubit for quasiparticles initially located in the low-gap (top) electrode, since $\bar{\Gamma}_{01}^{T}\ll \bar{\Gamma}_{00}^{T},\bar{\Gamma}_{10}^{T}$ (the rates are discussed in more detail later in this Appendix). This strong inequality also justifies neglecting terms proportional to $\xR$ in qubit's excitation rates [see Eq.~\eqref{eq:gamma01QP} in the main text].
Substituting Eq.~\eqref{eq:xlss-II-Ref} into Eq.~\eqref{eq:xRdot}, the steady-state density in the top electrode $\xR$ is obtained by solving a quadratic equation, yielding the result 
\begin{align}
\xR=\frac{\sqrt{\tilde s^2+4\tilde g r^{T}}-\tilde s}{2 r^{T}} \, ,
\label{eq:xrss-II-Ref} 
\end{align}
where for notational convenience, we introduced an effective trapping rate $\tilde s$ 
\begin{align}
\tilde s&= s_T
+s_B\frac{p_0(\bar\Gamma_{00}^T+\bar\Gamma_{01}^T)+p_1(\bar\Gamma_{11}^T+\bar\Gamma_{10}^T)}{s_B+\delta[p_0(\bar\Gamma_{00}^B+\bar\Gamma_{01}^B)+p_1(\bar\Gamma_{11}^B+\bar\Gamma_{10}^B)]}
\nonumber\\
 &\approx s_T+s_B\frac{\bar\Gamma_{00}^T+p_1\bar\Gamma_{10}^T}
{s_B+\delta[\bar\Gamma_{00}^B+p_0\bar\Gamma_{01}^B+p_1\bar\Gamma_{10}^B]} \, ,
\end{align}
and generation rate $\tilde g$
\begin{align}
\tilde g &= g^T+ \frac{g^B}{\delta} \left(1-\frac{s_B}{\delta[p_0(\bar\Gamma_{00}^B+\bar\Gamma_{01}^B)+p_1(\bar\Gamma_{11}^B+\bar\Gamma_{10}^B)]+s_B}\right)
\nonumber\\
&\approx g^T+ \frac{g^B}{\delta} \left(1-\frac{s_B}{\delta[\bar\Gamma_{00}^B+p_0\bar\Gamma_{01}^B+p_1\bar\Gamma_{10}^B]+s_B}\right) \, .
\end{align}
The generation rates in the two electrodes originate from pair breaking by high-frequency photons with rate $g^{\rm PAPS}$ and by thermal phonons,
\begin{equation}\label{eq:grates}
g^B/\delta=g^{\rm PAPS} \, , \,   
g_T=g^{\rm PAPS} + 2\pi r^{T} \frac{k_B T}{\DelR} e^{-2\DelR/k_B T}
\end{equation}
where thermal phonons generation in the high-gap (bottom) electrode is neglected for consistency with the approximation ignoring recombination there (alternatively, once the parameters are determined from the experimental data, one can check if the thermal phonon generation term is small compared the the photon pair-breaking one, which independently justifies ignoring the former). The photon-assisted generation rate is computed by dividing the photon-assisted tunneling rates $\Gamma^{\rm PAPS}$~\cite{Houzet19,Marchegiani22} by the Cooper pair number in the low-gap electrode,
\begin{align}\label{eq:gPAPS}
g^{\rm PAPS}=\frac{\Gamma^{\rm PAPS}}{n_{Cp}^{T}\mathcal V_T}
&\approx \frac{\Gamma_{00}^{\rm ph}+p_0\Gamma_{01}^{\rm ph}+p_1\Gamma_{10}^{\rm ph}}{2\nu_0\mathcal V_T\DelR} \, .
\end{align}

The (inverse) parity-switching time can be expressed summing all the rates associated with quasiparticle tunneling, reading  
\begin{align}
\tau_P^{-1}=&\,\Gamma^{\rm PAPS}+[p_0(\tilde{\Gamma}_{00}^B+\tilde{\Gamma}_{01}^B)+p_1 (\tilde{\Gamma}_{11}^B+\tilde{\Gamma}_{10}^B)] \xL +
\nonumber\\
&\,[p_0 (\tilde{\Gamma}_{00}^{T}+\tilde{\Gamma}_{01}^{T})+ p_1 (\tilde{\Gamma}_{11}^{T}+\tilde{\Gamma}_{10}^{T})]\xR
\nonumber\\
\approx & \,\Gamma^{\rm PAPS}+(\tilde{\Gamma}_{00}^{B}+p_1 \tilde{\Gamma}_{10}^{B}+p_0 \tilde{\Gamma}_{01}^{B})\xL
+(\tilde{\Gamma}_{00}^{T}+p_1 \tilde{\Gamma}_{10}^{T})\xR
\nonumber\\
= & \left(2-\frac{s_B}{\delta[\bar\Gamma_{00}^B+p_0\bar\Gamma_{01}^B+p_1\bar\Gamma_{10}^B]+s_B}\right)
\nonumber \\
&\times
[\Gamma^{\rm PAPS}+(\tilde{\Gamma}_{00}^{T}+p_1 \tilde{\Gamma}_{10}^{T})\xR]
\, .
\label{eq:tauP}
\end{align}

The calculation of the quasiparticle rates $\tilde{\Gamma}_{ij}^\alpha$ and $\Gamma_{ij}^{\rm ph}$ can be performed using Fermi's Golden rule, as extensively discussed in the literature, see, for instance, Refs.~\cite{Catelani11,glazman21, Marchegiani22,connolly2023coexistence}. Below, we report the results for the rates entering $\tau_P^{-1}$ under the assumption of quasiequilibrium distributions in the two electrodes; the detailed derivation of the quasiparticle rates in the presence of Fraunhofer effect will be given elsewhere. First, we consider the relaxation and parity switching tunneling rates for quasiparticles initially located in the top electrode:
\begin{align}
\label{eq:Gamma10R}
&\tilde{\Gamma}_{10}^{T}=\frac{8E_{J\Sigma}^0}{h}\sqrt{\frac{E_C}{8E_J}}\sqrt{\frac{2\DelR}{\pi k_B T}}\,
{\rm Exp}\left[-\frac{h(\fLR-f_{01})}{2k_B T}\right]
\nonumber\\
&\bigg\{\gamma_+ K_0\left[\frac{h|\fLR-f_{01}|}{2 k_B T}\right] \nonumber \\ & +
\left(\gamma_-+\frac{\gamma_+}{2}\right)\frac{ h|\fLR-f_{01}|}{2\bar\Delta}K_1\left[\frac{h|\fLR-f_{01}|}{2 k_B T}\right]\bigg\} ,\\
\label{eq:Gamma00R}
&\tilde{\Gamma}_{00}^{T}=\frac{8E_{J\Sigma}^0}{h}\sqrt{\frac{2\DelR}{\pi k_B T}}\,
{\rm Exp}\left[-\frac{h\fLR}{2k_B T}\right] 
\nonumber\\
&\left\{\gamma_- K_0\left[\frac{h|\fLR|}{2 k_B T}\right]+\left(\gamma_++\frac{\gamma_-}{2}\right)\frac{h|\fLR|}{2\bar\Delta}K_1\left[\frac{h|\fLR|}{2 k_B T}\right]\right\} ,
\end{align}
where $E_J$ and $f_{01}$ are the Josephson energy and the frequency of the qubit at the lower sweet spot in the presence of the parallel field, $K_n$ are modified Bessel functions of the second kind, and $E_{J\Sigma}^0=E_{Ja}(\Bpar=0)+E_{Jb}(\Bpar=0)$.
The remaining rates are obtained from the ones given in Eqs.~\eqref{eq:Gamma10R} and \eqref{eq:Gamma00R} with the following substitutions:
\begin{align}
&\tilde{\Gamma}_{01}^{T}= \tilde{\Gamma}_{10}^{T} (f_{01}\to-f_{01})
\\
\label{eq:Gamma10L}
&\tilde{\Gamma}_{10}^{B}= \sqrt{\frac{\DelL}{\DelR}}\tilde{\Gamma}_{10}^{T} (\fLR\to-\fLR)
\, ,
\\
\label{eq:Gamma00L}
&\tilde{\Gamma}_{00}^{B}=\sqrt{\frac{\DelL}{\DelR}}{\rm Exp}\left[\frac{h\fLR}{k_B T}\right]\tilde{\Gamma}_{00}^{T}
\, ,
\\
\label{eq:Gamma01L}
&\tilde{\Gamma}_{01}^{B}=
\sqrt{\frac{\DelL}{\DelR}}{\rm Exp}\left[\frac{-h(f_{01}-\fLR)}{k_B T}\right]
\tilde{\Gamma}_{10}^{T}
\, .
\end{align}
The transformations in Eqs.~\eqref{eq:Gamma10L} and \eqref{eq:Gamma00L} exploit the symmetry of the rates and are equivalent to exchanging the role of the two electrodes $B\leftrightarrow T$, while ~\eqref{eq:Gamma01L} follows from the detailed balance principle.
The weights in the curly brackets of Eqs.~\eqref{eq:Gamma10R} and \eqref{eq:Gamma00R}
\begin{equation}
\gamma_\pm = \frac{2\pm(z_- + \alpha_{JJ,0} z_+)}{4}.
\label{eq:weightsFraunLowersweet spot}
\end{equation}
accounts for the interference effects related to the SQUID interferometer as well as the Fraunhofer effect: $\alpha_{\rm JJ,0}=\alpha_{\rm JJ}(\Bpar=0)=(E_{Ja}-E_{Jb})/(E_{Ja}+E_{Jb})$ is the split-transmon asymmetry parameter at zero parallel field, while $z_\pm ={\rm Sinc}(\Bpar/B_{\Phi a})\pm {\rm Sinc}(\Bpar/B_{\Phi b})$. For comparison, the rates for a single-junction transmon correspond to the case $\alpha_{\rm JJ,0}=1$; if we further consider zero parallel magnetic field, we have $\{\gamma_+,\gamma_-\}=\{1,0\}$. 

We can present the photon-assisted tunneling rates $\Gamma_{ij}^{\rm ph}$ in a similar way.
For simplicity, we consider monochromatic radiation with frequency $f_\nu >(\DelL+\DelR)/h$ to allow for Cooper-pair breaking.
The rates can be expressed as 
\begin{align}
\Gamma_{00}^\mathrm{ph} &=\Gamma_\nu\frac{g_\Sigma\DelL}{8e^2}
\left[
\gamma_-
S_{\rm ph}^+\left(\frac{hf_\nu}{\DelL},\delta\right)+\gamma_+S_{\rm ph}^-\left(\frac{hf_\nu}{\DelL},\delta\right)
\right]
\\
\Gamma_{10}^\mathrm{ph} &=\Gamma_\nu\frac{g_\Sigma\DelL}{8e^2}\sqrt{\frac{E_C}{8E_J}}
\bigg[
\gamma_+
S_{\rm ph}^+\left(\frac{f_\nu+f_{01}}{\DelL/h},\delta\right) \nonumber \\ & \qquad +\gamma_-S_{\rm ph}^-\left(\frac{f_\nu+f_{01}}{\DelL/h},\delta\right)
\bigg]
\\
\Gamma_{01}^\mathrm{ph} &=\Gamma_\nu\frac{g_\Sigma\DelL}{8e^2}\sqrt{\frac{E_C}{8E_J}}
\bigg[
\gamma_+
S_{\rm ph}^+\left(\frac{f_\nu-f_{01}}{\DelL/h},\delta\right) \nonumber \\ & \qquad +\gamma_-S_{\rm ph}^-\left(\frac{f_\nu-f_{01}}{\DelL/h},\delta\right)
\bigg]
\end{align}
where $g_\Sigma\approx 8g_k E_{J\Sigma}^0(\DelL+\DelR)/(2\DelL\DelR)$ is the total tunnel conductance of the SQUID in the normal state, $\Gamma_\nu$ is the dimensionless photon rate that accounts for the coupling strength between transmon and pair-breaking photons, and the photon spectral densities read~\cite{Marchegiani22} 
\begin{align}
& S_{\rm ph}^\pm(x,z)=\theta(x-1-z)\int\limits_1^{x-z} dy\frac{y(x-y)\pm z}{\sqrt{y^2-1}\sqrt{(x-y)^2-z^2}} 
\, \nonumber\\
& \ =\theta(x-1-z)\Bigg\{\sqrt{x^2-(z-1)^2}E\left[\sqrt{\frac{x^2-(z+1)^2}{x^2-(z-1)^2}}\right]
\nonumber\\
& \ -2z\frac{1\mp 1}{\sqrt{x^2-(z-1)^2}}K\left[\sqrt{\frac{x^2-(z+1)^2}{x^2-(z-1)^2}}\right]\Bigg\}
\end{align}
with $E$ and $K$ complete elliptic integrals of the second and first kind, respectively.

We note that all the rates, being expressed in terms of the conductance $g_\Sigma$, are calculated at leading order in the tunneling transmission probability.
This means in particular that we are ignoring effects comparable in magnitude to those of the Josephson harmonics, which can introduce corrections at most at the percent level, see Appendix~\ref{app:transmon_hamiltonian} and Ref.~\cite{Willsch23}. Given the limited accuracy of time-domain measurements, this approach is sufficient for our purposes.

Finally, let us comment on the data fitting procedure. We base values for $\EJ$, $\EC$ and $f_{01}$ on the fitting of the spectroscopic data (Appendices~\ref{app:transmon_hamiltonian} and \ref{app:fraunhofer_and_critical_field}); the occupation probability $p_1$ are estimated based on the single-shot parity outcomes and the measured relaxation times (see \cref{fig:parity_schemes} and \cref{app:p1}).
We estimate the smaller gap $\DelR$ based on modeling the temperature dependence of $\Tone$ (see \cref{app:T1temp}). 
That leaves as unknown parameters the gap difference, the photon frequency, the trapping rates, and the dimensionless photon rate. 
The values used in calculating the theoretical curves in Figs.~\ref{fig:fig3} and \ref{fig:fig4} are $\delDel/h=\SI{5.49}{\giga\hertz}$, $f_\nu = \SI{119}{\giga\hertz}$, $s_\mathrm{B}=s_\mathrm{T}=\SI{3.23}{\hertz}$, and $\Gamma_\nu = 1.69\times 10^{-8}$.

\section{Modeling the temperature dependence of the excited state population}
\label{app:p1}

In our model, the parity switching time is parametrically expressed in terms of the populations of the ground and first excited state of the qubit (see \cref{app:paritymodel}).
The population of the excited state of the qubit is given by
\begin{equation}
p_1=\frac{\Gamma_{01}^\mathrm{ee}+\Gamma_{01}^\mathrm{eo}}{\Gamma_{01}^\mathrm{ee}+\Gamma_{01}^\mathrm{eo}+\Gamma_{10}^\mathrm{ee}+\Gamma_{10}^\mathrm{eo}}=(\Gamma_{01}^\mathrm{ee}+\Gamma_{01}^\mathrm{eo})T_1
\label{eq:p1steady}
\end{equation}
where $\Gamma_{10}^\mathrm{ee}$, and $\Gamma_{01}^\mathrm{ee}$ are the parity-preserving (thus not associated with quasiparticles) qubit' relaxation and excitation rates, respectively. For terms changing the parity, we identity the rates adding NUPS and PAPS contributions, i.e., $\Gamma_{10}^\mathrm{eo}=\Gamma_{10}^{\rm ph}+\Gamma_{10}^{\rm qp}/p_1$ and $\Gamma_{01}^\mathrm{eo}=\Gamma_{01}^{\rm ph}+\Gamma_{01}^{\rm qp}/p_0$ [cf. Eqs.~\eqref{eq:gamma01QP} and \eqref{eq:gamma10QP}].
In the temperature regime for the parity-switching time measurements, the lifetime of the qubit is mainly limited by dielectric losses $T_1\approx 1/(\Gamma_{10}^\mathrm{ee}+\Gamma_{01}^\mathrm{ee})$. Indeed, the maximum qubit's lifetime is of the order of 10 $\mu$s (see Appendix \ref{app:T1temp}), which is more than an order of magnitude shorter than the parity lifetime. Assuming that the parity-preserving rates satisfy the detailed balance principle, i.e., $\Gamma_{01}^\mathrm{ee}=\Gamma_{10}^\mathrm{ee}e^{-h\fzeroone/k_BT}$, we can write 
\begin{equation}
p_1\approx T_1\Gamma_{01}^\mathrm{eo} + \frac{\exp(-h f_{01}/k_B T)}{1+\exp(-h f_{01}/k_B T)} \, .
\label{eq:p1steadyApp}
\end{equation}
In the experimental protocol for determining the parity-switching time, the qubit's parity is mapped onto the qubit's logical state, with the convention $\mathrm{e}\to 0$ and $\mathrm{o}\to 1$. After the mapping, the qubit's state is not reset to zero, rather there is a waiting time equal to five times $T_1$. As a result, the qubit's excited state population relaxes exponentially from $p_1=0$ (for even assignment) or $p_1=1$ (for odd assignment) to the steady-state value $p_1$ of Eq.~\eqref{eq:p1steady}. Integrating the rate equation for the qubit's excited state population over the waiting time $5T_1$, we obtain the average excited state population for the two parity assignments 
\begin{equation}
\overline{p}_1^\mathrm{e}\approx 0.8 \,p_1 \, \, , \, \, \overline{p}_1^\mathrm{o}\approx 0.2+ 0.8\,p_1 \, . 
\label{eq:assignmentp1}
\end{equation}

The analysis of the hidden Markov model, combined with the considerations on the parity assignment, allows us to determine an average population of the excited state during the measurements of the parity-switching rates [points in Figs.~\ref{fig:fig4}(e) and \ref{fig:fig4}(f)].  Motivated by the considerations made above, we use the following semi-phenomenological expression 
\begin{equation}
p_1^{\rm fit}=\frac{a}{1+{\rm Exp}[-h f_{01}/k_B T]} + b \,\frac{{\rm Exp}[-h f_{01}/k_B T]}{1+{\rm Exp}[-h f_{01}/k_B T]}
\label{eq:p1FitExpression}
\end{equation}
and we perform a two-parameter fit ($a,b$) to the data [solid curves in Figs.~\ref{fig:fig4}(e) and \ref{fig:fig4}(f)]
The parameter $a$ quantifies the average excited's state population at low temperatures $T\ll hf_{01}/k_B$; for even assignment, it is in the range of a few percent while for odd assignment is around 20\% [see the considerations leading to Eq.~\eqref{eq:assignmentp1}]. 
The prefactor of the last term in Eq.~\eqref{eq:p1FitExpression} turns out to be in the range of 0.8 to 1.3, in reasonable agreement with our interpretation.

\section{Temperature dependence of the qubit relaxation time}
\label{app:T1temp}

As mentioned in Appendix~\ref{app:p1}, the qubit relaxation time $T_1$ is much shorter than the parity-switching time at the base temperature of the dilution refrigerator, suggesting that $T_1$ is limited by decay processes unrelated to quasiparticle tunneling, such as dielectric losses; we model the latter as a two-level-system bath. At low temperatures, $T\ll h f_{01}/k_B$, the qubit's relaxation time is approximately the inverse of the parity-conserving relaxation rate, $T_1\approx 1/\Gamma_{10}^\mathrm{ee}$, see Appendix~\ref{app:p1}; this quantity is an unknown function of the parallel magnetic field. Upon increasing the temperature $T\lesssim hf_{01}/k_B$, the qubit lifetime decreases due to the non-negligible qubit excitation rate caused by the TLS bath. Moreover, the quasiparticle density increases exponentially with the phonon temperature; thus, the qubit's relaxation time is eventually limited by quasiparticle tunneling above a crossover temperature; we assume this temperature to be sufficiently high so that deviations of the quasiparticle distribution from thermal equilibrium due to \textit{e.g.} photon pair-breaking can be ignored
(see the discussion at the end of this Appendix).

\begin{figure}[!t]
	\centering
		\includegraphics[width=\columnwidth]{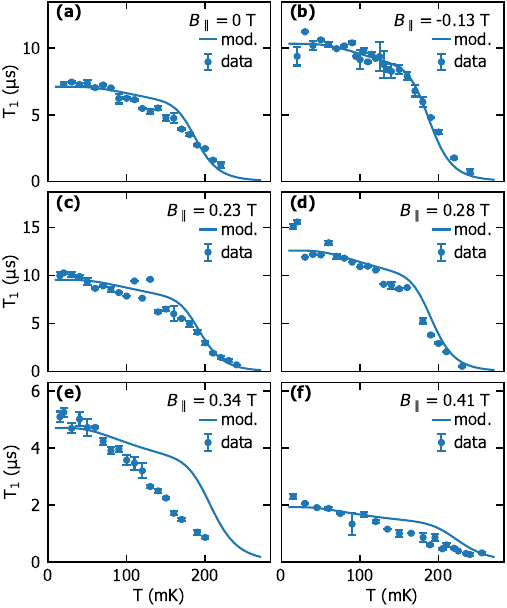}
	\caption{
    Qubit lifetime vs. temperature for different values of the in-plane magnetic field. 
    Points are the experimental data, continuous curves are given by Eq.~\eqref{eq:T1temp} and are obtained using $\DelL/h=59.49$ GHz, $\DelR/h=54\,$GHz, and the spectroscopic parameters estimated in Appendices~\ref{app:transmon_hamiltonian} and \ref{app:fraunhofer_and_critical_field}. 
    The parity preserving relaxation rate used in the plot is $\Gamma_{10}^\mathrm{ee}=$  \textbf{(a)} \SI{140}{\kilo\hertz}, \textbf{(b)} \SI{97}{\kilo\hertz}, \textbf{(c)} \SI{105}{\kilo\hertz}, \textbf{(d)} \SI{79}{\kilo\hertz}, \textbf{(e)} \SI{213}{\kilo\hertz}, \textbf{(f)} \SI{520}{\kilo\hertz}.
    The dependence of the superconducting gaps on the magnetic field has not been included.
    } 
	\label{fig:t1TempNew}
\end{figure}

As in previous works~\cite{Catelani11, glazman21}, to capture the temperature evolution of the qubit's relaxation time we sum the thermal quasiparticle decoherence rate ($T_{\rm 1,qp}^{-1}$) and the contribution of the dielectric losses,
\begin{equation}
\label{eq:T1temp}
T_1(T)=\left\{\Gamma_{10}^{\rm ee}\left[1+\exp\left(-\frac{h f_{01}}{k_B T}\right)\right]+T_{\rm 1,qp}^{-1}(T)\right\}^{-1}.
\end{equation}
The relaxation rate due to thermal quasiparticles alone is
\begin{equation}
T^{-1}_{1,\rm qp}=(\tilde{\Gamma}_{10}^{B}\xL^{\rm th}+\tilde{\Gamma}_{10}^{T}\xR^{\rm th})[1+\exp(-h f_{01}/k_BT)]
\label{eq:T1QPs}
\end{equation}
where 
\begin{equation}
x_\alpha^{\rm th}=\sqrt{\frac{2\pi k_B T}{\Delta_\alpha}}\exp\left[-\frac{\Delta_\alpha}{k_B T}\right]
\end{equation}
is the thermal quasiparticle density in the electrode $\alpha=\{T,B\}$. In Eq.~\eqref{eq:T1QPs}, the relaxation rate for quasiparticles tunneling from the top to the bottom electrode $\tilde{\Gamma}_{10}^{T}$ is given by Eq.~\eqref{eq:Gamma10R}, the corresponding rate for quasiparticles initially located in the bottom electrode is simply obtained exchanging $\DelL\leftrightarrow \DelR$ in Eq.~\eqref{eq:Gamma10R}.
The exponential factor in the square bracket of Eq.~\eqref{eq:T1QPs} accounts for qubit's excitation rates, according to the detailed balance principle.

To check that the thermal equilibrium assumption for quasiparticles is justified in fitting the $T_1$ data, we proceed as follows: we start by noticing that the thermal quasiparticle contribution becomes relevant at temperature $\sim170\,$mK above which $T_1$ decreases markedly faster with temperature. 
At that temperature, the thermal phonon generation rate is given by $r^\alpha (x_\alpha^\mathrm{th})^2$, cf. the last term in Eq.~\eqref{eq:grates}; for the low-gap (top) electrode, this rate is about $1.5\times 10^{-7}\,$Hz. 
The thermal phonon generation rate should be compared with $g^\mathrm{PAPS}$ of Eq.~\eqref{eq:gPAPS}, which we estimate to be at most of order $3.7\times10^{-8}\,$Hz (cf. Fig.~\ref{fig:fig3}); therefore, non-equilibrium generation by pair-breaking photons can be ignored when analyzing the $T_1$ data. 
Conversely, we conclude that in our experiment at temperatures below about 150~mK the generation by thermal phonons plays essentially no role in the temperature dependence of $\tau_p$; in fact, in Fig.~\ref{fig:fig4}~(a) and (b) we see a steeper decline in $\tau_p$ beginning at that temperature.

%

\setcounter{section}{0}
\renewcommand{\appendixname}{Supplementary}
\renewcommand{\appendixesname}{Supplementary}
\renewcommand{\thesection}{\Roman{section}} 
\setcounter{table}{0}
\renewcommand{\thetable}{S\arabic{table}}
\setcounter{figure}{0}
\renewcommand{\thefigure}{S\arabic{figure}}
\setcounter{equation}{0}
\renewcommand{\theequation}{S\arabic{equation}}

\clearpage

\onecolumngrid


\begin{center}
	\textbf{\large{Supplementary Information for \\\smallskip
		``Magnetic-field dependence of quasiparticle parity dynamics in an Al/AlO$_x$/Al Josephson junction transmon''}}\\
\end{center}

This supplement provides experimental details and additional data supporting the claims in the main text. Supplementary~\ref{som:full datasets} contains plots of the qubit lifetime, estimated mean excited-state populations for the ``even'' and ``odd'' parity outcomes, and parity lifetimes with corresponding model contributions for all temperature dependence datasets that were measured. Supplementary~\ref{som:parity_fidelities} gives the estimated single-shot fidelities of the parity measurements as a function of the parallel field $\Bpar$ and the temperature $T$. 
Supplementary~\ref{som:vg_control} contains data on the gate voltage $\Vg$ control of the offset charge of the transmon over a wider range compared to the main text [cf. Fig.\ref{fig:fig2} (d)-(f) in Sec.\ref{sec:spectrum_vs_Bpar}]; moreover we show that $V_g$ has no influence on qutrit measurements for $\Tone$ and $\Techo$.
The coherence times as a function of $\Bpar$ are presented in Supplementary~\ref{som:coherence_vs_Bpar}.
They show a clear decrease towards the edge of our measurement range $\vert \Bpar \vert > \SI{0.4}{\tesla}$ that also can be seen in the cavity quality factor.
Supplementary~\ref{som:hmm_psd} contains additional information on the extraction of the parity lifetimes from the data comparing the fit to the power spectral density \cite{Riste13} to the hidden-Markov model.
Supplementary~\ref{som:transmon_temperature} shows the transmon temperature extracted from single-shot measurements as a function of fridge temperature at different fields.
Supplementary~\ref{som:squid_stability} contains details on the SQUID instability at high fields.

\section{Full temperature dependence datasets}
\label{som:full datasets}

\begin{figure*}
  \centering
    \includegraphics{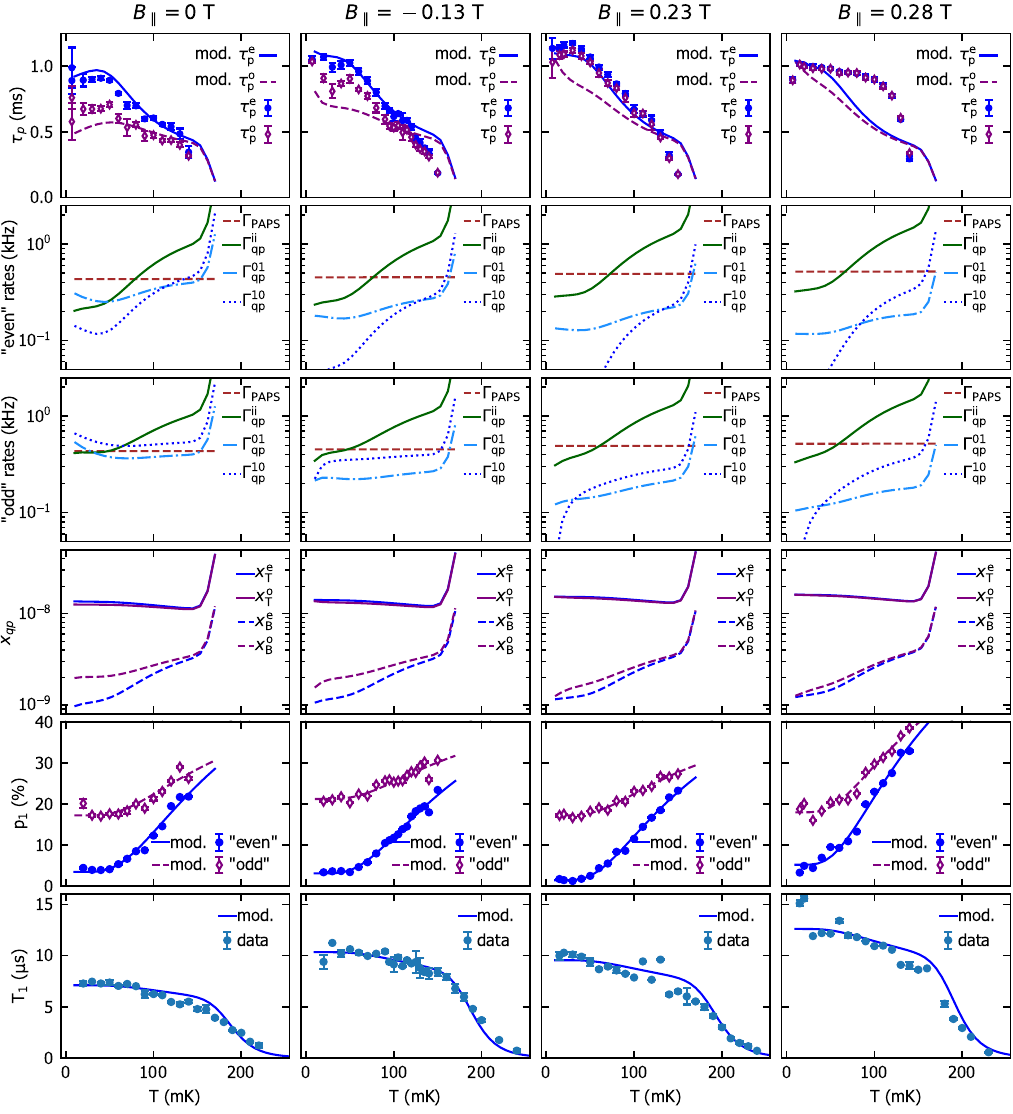}
  \caption{
    Temperature dependence of $\taup$, $p_1$ and $\Tone$ for different in-plane magnetic fields $\Bpar$. 
    Here, the entangled transition during parity measurements is $\fonetwo$.
    Columns one and two have been discussed extensively in the main text.
    For the ``even'' and ``odd'' parity-switching times $\taupe$ and $\taupo$, the agreement with model decreases with increasing magnetic field.
    For the highest field, $\Bpar=\SI{0.28}{\tesla}$, the predicted waterfall-like decay of $\taup$ is not observed, possibly due to a neglected evolution of the superconducting-gap difference $\delDel$ and/or altered quasiparticle-trapping dynamics. Parameters for the fit 
    of the excited state population are given in \cref{tab:p1_vs_T_params}, $\Gamma_{10}^\mathrm{ee}$ for the different fields are the same as in \cref{fig:t1TempNew}. The remaining parameters are given in the main text.
  }
  \label{fig:full_T_data_12}
\end{figure*}

\begin{figure*}
  \centering
    \includegraphics{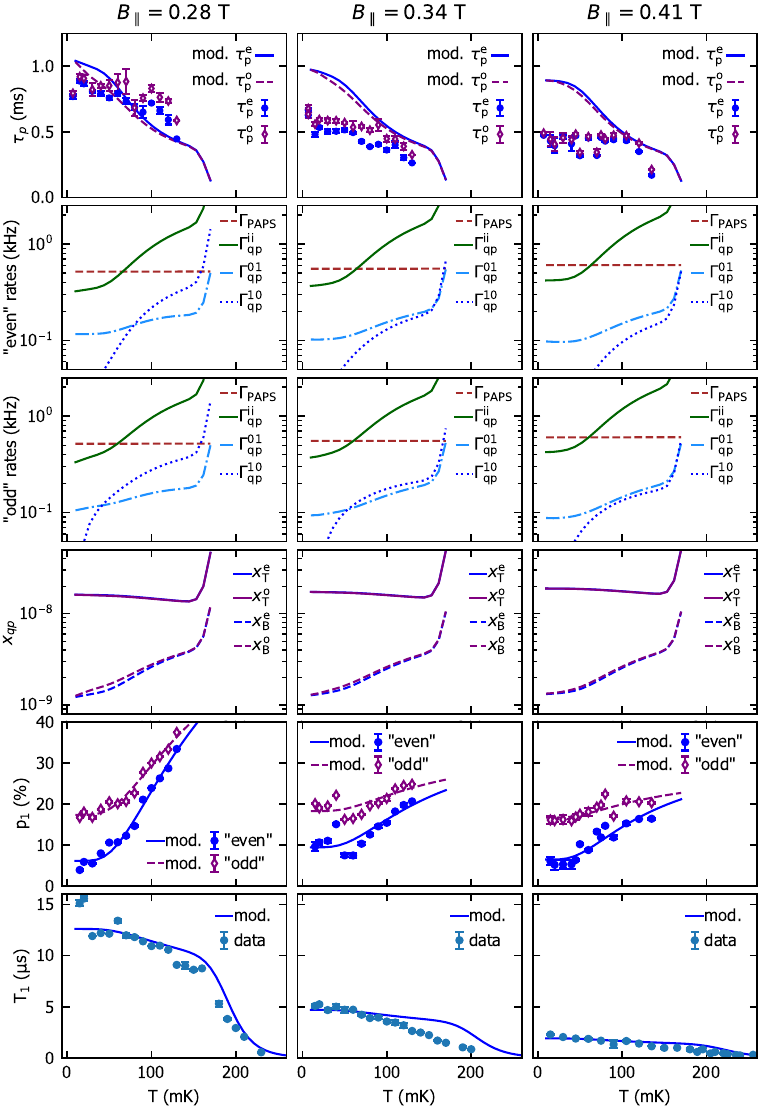}
  \caption{
    Temperature dependence of $\taup$, $p_1$ and $\Tone$ for different in-plane magnetic fields $\Bpar$. Here, the entangled transition during parity measurements is $\fzeroone$.
    The increasing mismatch with $\Bpar$ for low temperatures is the same as in \cref{fig:fig3}; its origin is unclear.
    Notably, some of the data suggests $\taupe<\taupo$ unlike for lower fields.
    However, the single-shot parity fidelity is significantly reduced in these measurements, $0.2\le F_{\mathrm{p},01}\le0.7$ (with fidelity generally decreasing with increasing temperature), compared to the parity measurements using the $\fonetwo$ transition shown above, where $0.5\le F_{\mathrm{p},12}\le0.9$.
    Parameters for the fit 
    of the excited state population are given in \cref{tab:p1_vs_T_params}, $\Gamma_{10}^\mathrm{ee}$ for the different fields are the same as in \cref{fig:t1TempNew}. The remaining parameters are given in the main text.
    }
\label{fig:full_T_data_01}
\end{figure*}

In the main text, we report the temperature evolution of the parity switching time only for a few in-plane magnetic fields, for the sake of brevity. Figures~\ref{fig:full_T_data_12} and \ref{fig:full_T_data_01} display the parity switching times $\taupe$ and $\taupo$, and the corresponding model contributions obtained through fitting for all the measured temperature-dependent datasets using the first-to-second excited (cf. Fig.~\ref{fig:fig3}b) and the ground to-first excited protocol (cf. Fig.~\ref{fig:fig3}a), respectively. Moreover, for each $\Bpar$, we conveniently compare with the corresponding temperature evolution of $\Tone$ already shown in Appendix~\ref{app:T1temp}. 

The model describes the $\taupe$ data quite accurately up to \SI{0.23}{\tesla}, while a noticeable mismatch can be observed at larger fields. For the highest field values, $\Bpar=0.34$~T and $\Bpar=0.41$~T, the $\taup$ is shorter (roughly half the value at the base temperature of the cryostat) than expected from the model, suggesting that there may be additional mechanisms that we do not account for; another possibility is that our model parameters depend more drastically on the magnetic field, even though  
we do not believe that $\delta\Delta$ change significantly in the measured range,
as the estimated critical field is above $1$T. The most puzzling behavior, which we currently cannot explain, concerns the \SI{0.28}{\tesla} dataset, where the model notably overestimates the decay of $\taupe$ and $\taupo$ with the temperature. 
Furthermore, we note that for this particular field, we measured the parity switching time using both the protocols (cf. Figs.\ref{fig:fig3}a and b) reporting similar values; however we point out that the protocol involving the ground-to-first excited transition is less trustworthy than the other (cf. Supplementary~\ref{som:parity_fidelities}), given the small charge dispersion of $\fzeroone$ transition, which implies that the waiting times of the Ramsey protocol are of the order of the coherence time of the transmon $\Ttwostar$.

\begin{table*}[h!]
\centering
\begin{tabular}{c @{\hskip 0.3in} | @{\hskip 0.3in} c @{\hskip 0.3in} c @{\hskip 0.3in} c @{\hskip 0.3in} c @{\hskip 0.3in} c @{\hskip 0.3in} c @{\hskip 0.3in} c} 
 \hline
    $\Bpar$ [\SI{}{\tesla}] & 0.  & -0.13 & 0.23 & 0.28 ($f_{12}$) & 0.28 ($f_{01}$) & 0.34 & 0.41 \\  
 \hline
    \makecell{$\bar{p}_1^{\mathrm{e}}$ vs $T$ \\ parameters } &
    \makecell{$a=0.034$\\ $b=1.461$ } &
    \makecell{$a=0.031$\\ $b=1.261$ } &
    \makecell{$a=0.014$\\ $b=1.280$ } &
    \makecell{$a=0.052$\\ $b=1.801$ } &
    \makecell{$a=0.061$\\ $b=1.859$ } &
    \makecell{$a=0.094$\\ $b=0.712$ } & 
    \makecell{$a=0.064$\\ $b=0.649$ } \\
 \hline
    \makecell{$\bar{p}_1^{\mathrm{o}}$ vs $T$ \\ parameters } & 
    \makecell{$a=0.172$\\ $b=0.932$ } &
    \makecell{$a=0.212$\\ $b=0.788$ } &
    \makecell{$a=0.172$\\ $b=0.788$ } &
    \makecell{$a=0.180$\\ $b=1.382$ } &
    \makecell{$a=0.173$\\ $b=1.448$ } &
    \makecell{$a=0.183$\\ $b=0.522$ } &
    \makecell{$a=0.167$\\ $b=0.407$ } \\
 \hline
\end{tabular}
\caption{
Fit Parameters for the temperature dependence of the excited state population of the qubit. 
}
\label{tab:p1_vs_T_params}
\end{table*}

\clearpage
\clearpage

\section{Parity measurement fidelities}
\label{som:parity_fidelities}

\begin{figure*}[h]
	\centering
    \includegraphics{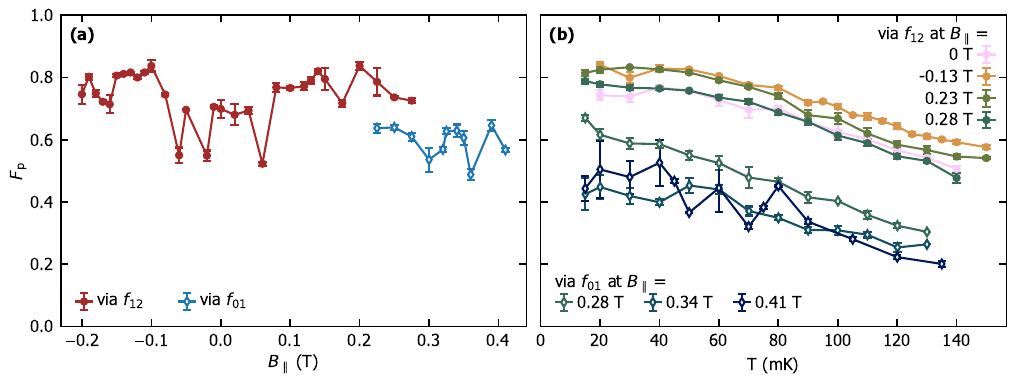}
	\caption{
		Single-shot parity measurement fidelities $F_\mathrm{p}$ based on the HMM as a function of \textbf{(a)} $\Bpar$ and \textbf{(b)} temperature.
	}
	\label{fig:parity_fidelities}
\end{figure*}

As shown in \cref{app:parity_schemes}, we can estimate a single-shot parity measurement fidelity based on the assigned parity histograms.  
In \cref{fig:parity_fidelities} we show the fidelities as a function of $\Bpar$ [panel (a)] and $T$ [panel (b)].
Generally, the fidelities of the parity mapping based on the first and the second excited states of the transmon are higher, because $\delta f_{12}>\delta f_{01}$, while the coherence times are comparable. 
Moreover, the measurements based on the ground-to-first-excited state transition are performed at higher fields, where $\Ttwostar$ is typically lower [see \cref{fig:coherence_times_vs_bpar}]. 

Fidelities decrease monotonically with $T$ mainly due to three distinct effects: firstly, the coherence time $\Ttwostar$ is reduced due to thermally populated photons in the cavity; secondly, parity switching becomes faster, which makes it harder to measure, in turn making the assignment less accurate, which reduces the fidelity (it also reduces $\Ttwostar$ as well);  thirdly, the transmon becomes hotter which also leads to assignment errors.

\section{Gate-voltage control of the excess charge of the transmon and potential effects on the transmon's lifetime and coherence time.}
\label{som:vg_control}

\begin{figure*}[t]
	\centering
		\includegraphics[width=\textwidth]{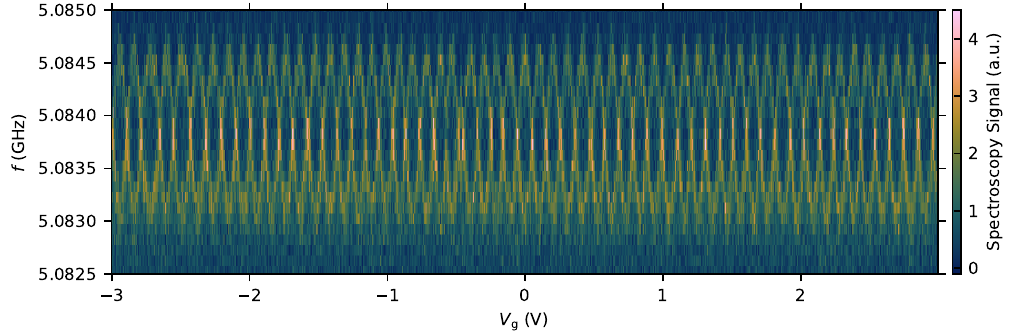}
	\caption{
		Wide range scan of the charge dispersion for $\fzerotwoovertwo$ as a function of $V_\mathrm{g}$.
		More than 25 periods can be observed with one jump around \SI{-0.5}{\volt}.
        The data was taken at the bottom sweet spot at $B_\parallel=\SI{150}{\milli\tesla}$.
	}
	\label{fig:wide_range_gatescan}
\end{figure*}

\begin{figure*}
	\centering
		\includegraphics[width=\textwidth]{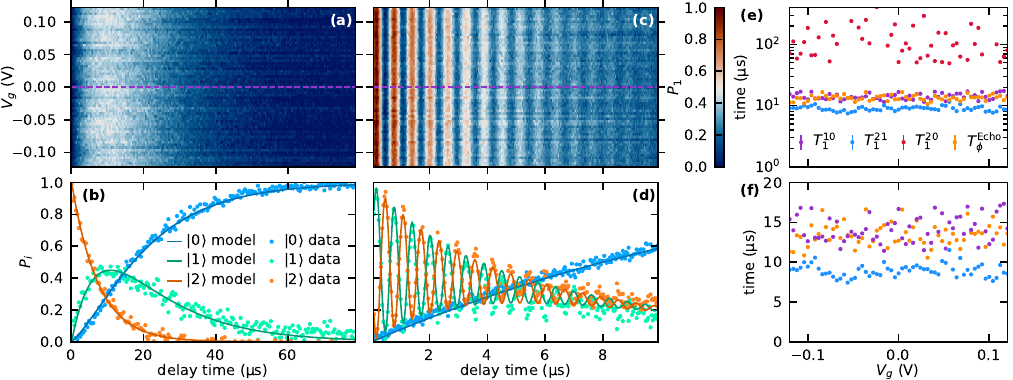}
	\caption{
	   Transmon qutrit lifetimes and echo coherence time $\Techo$ as a function of the gate voltage $\Vg$. Color plots of the $\ket{1}$ state population $P_1$ vs delay time and $V_g$ for (a) a $\Tone$ sequence where the transmon is initialized in the $\vert 2 \rangle$ state and left to relax to the steady state, and (c) for an echo coherence time experiment where the qubit is initialized in the superposition of $\vert 1 \rangle$, and $\vert 2 \rangle$. 
       The cuts at $V_g=0$ show the dynamics of the population of $\vert 0 \rangle$, $\vert 1 \rangle$, and $\vert 2 \rangle$ states (points) for (b) the $T_1$ sequence and (d) the echo experiment. Experimental data are compared with a fitted model (solid) based on Eqs.~\eqref{eq:P0qutrit}-\eqref{eq:P2qutrit} for panel (b) and a master equation for panel (d) [see text]. (e)-(f) Decay time and echo coherence times for the first two excitations of the transmon, as obtained from the fitted models, vs gate voltage. 
	}
	\label{fig:T1_Echo_gatescan}
\end{figure*}

We explored scanning the gate voltage $\Vg$ over a wide range of many of the charge dispersion periods  to  verify that the transmon levels follow the expected periodicity [see 
\cref{fig:wide_range_gatescan} for the two-photon ground to second excited transition].
Overall, we observe the expected periodicity with one jump around $\Vg=\SI{-0.5}{\volt}$. 
Our measurements suggest that charge jumps are relatively rare but we do observe drift in $\Vg$ that can cover up to a period on a~\SI{10}{\minute} timescale.
For this reason, we usually sandwich the parity measurement sequences between two Ramsey measurements to be able to post-select on runs where the drift was not severe.

We have also investigated if the gate voltage has measurable effects on transmon $\Tone$ and $\Techo$, concluding that they are not strongly affected by $\Vg$ [see \cref{fig:T1_Echo_gatescan} (a) and (c)].
As the majority of our measurements are done with qutrit pulse sequences, we also measured the relaxation of the transmon after preparing the state $\vert 2 \rangle$ and show an example of the estimated populations $P_i$ as a function of time [see \cref{fig:T1_Echo_gatescan}(b)].
We model the dynamics of the populations using standard rate equations, and we disregard transmon excitation processes, since at low temperatures we expect $\Gamma_{01},\Gamma_{12},\Gamma_{02}\ll \Gamma_{10}, \Gamma_{21},\Gamma_{20}$. Within this approximation, the time evolution of the populations can be readily obtained by solving the linear system of rate equations [with initial condition $P_0(0)=P_1(0)=1-P_2(0)=0$], yielding
\begin{align}
\label{eq:P0qutrit}
P_0(t)&=1-\frac{\Gamma_{21}}{\Gamma_{21}+\Gamma_{20}-\Gamma_{10}}{\rm Exp}[-\Gamma_{10}t]\{
1-{\rm Exp}[-(\Gamma_{21}+\Gamma_{20}-\Gamma_{10})t]
\}-{\rm Exp}[-(\Gamma_{21}+\Gamma_{20})t]\, , 
\\
P_1(t)&=\frac{\Gamma_{21}}{\Gamma_{21}+\Gamma_{20}-\Gamma_{10}}{\rm Exp}[-\Gamma_{10}t]
\{
1-{\rm Exp}[-(\Gamma_{21}+\Gamma_{20}-\Gamma_{10})t]
\} \, , \label{eq:P1qutrit} \\
P_2(t)&={\rm Exp}[-(\Gamma_{21}+\Gamma_{20})t]\label{eq:P2qutrit} 
\, .
\end{align}

The relaxation somehow cannot be accurately modeled by including only sequential decay $T_1^{21}=\Gamma_{21}^{-1}$ and $T_1^{10}=\Gamma_{10}^{-1}$, but we also need to take into account the non-sequential decay time $T_1^{20}=\Gamma_{20}^{-1}$ which should have a small transition matrix element in the transmon but has also been previously observed~\cite{peterer12}. 

We also show an example of a $\Techo$ experiment for a superposition of $\ket{1}$ and $\ket{2}$ [see \cref{fig:T1_Echo_gatescan} \textbf{(c)} and \textbf{(d)}]. In this case, the populations are modeled using a standard master equation for the density matrix to account for dephasing processes (see, for instance, Ref.~\cite{Morvan2021}).
In the modeling of the Echo experiment, we fix the relaxation rates based on a preceding relaxation experiment from the $\ket{2}$ state.
The echo pure-dephasing rate $T_{\phi}^{\mathrm{Echo}}$ is fitted using a qutrit lindblad-master equation that includes the wait-time-dependent phase in the second $\nicefrac{\pi}{2}$ pulse responsible for the oscillations.
The estimated $T_1^{10}$, $T_1^{21}$, $T_1^{20}$ and $T_\phi^\mathrm{Echo}$ as a function of $\Vg$ are shown in \cref{fig:T1_Echo_gatescan} \textbf{(e)} and \textbf{(f)}, showing no clear dependence on $\Vg$.
As the period of $\Vg$ is known to be \SI{0.206}{\volt}, one could take $\Tone$ vs $\Vg$ datasets over many periods and look for small effects in principle.
Combined with SQUID or Fraunhofer tunability, one could try to identify effects of stable two-level fluctuators and understand their dependence on $\Vg$ but this is beyond the scope of this work.

\section{Qutrit coherence times as a function of the in-plane magnetic field}
\label{som:coherence_vs_Bpar}

\begin{figure}
	\centering
	\includegraphics{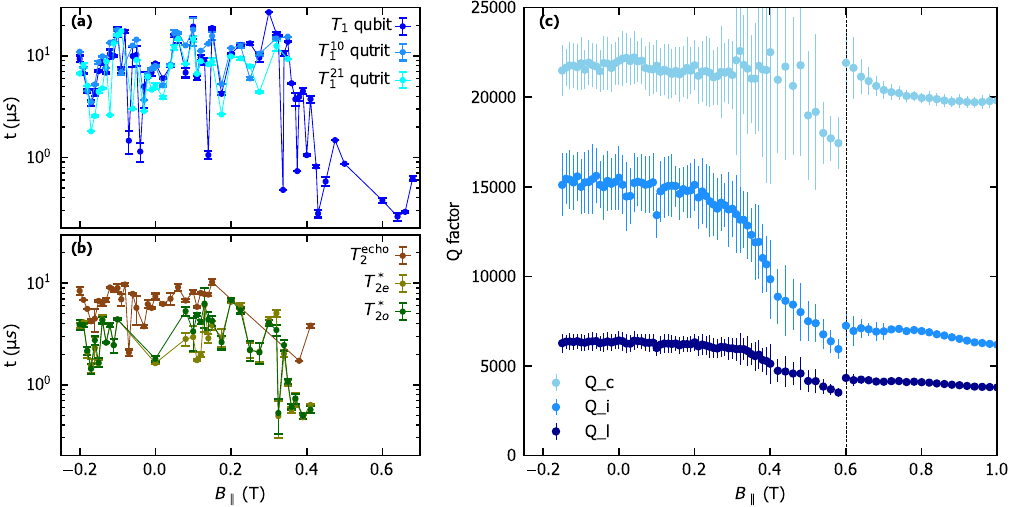}
	\caption{
        \textbf{(a)} relaxation times vs $\Bpar$.
        There are two curves for the lifetime $\Tone$, one for qubit and one for qutrit measurements of relaxation from $\ket{2}$.
        With the latter approach, also the decay rate from second to first excited states has been measured.
        \textbf{(b)} coherence times vs $\Bpar$.
        The two curves for $\Ttwostar$, in contrast, are obtained in a single measurement and correspond to the two frequency branches of ``even'' and ``odd'' parity ($T^*_{2e}$ and $T^*_{2o}$). 
        Notably, the coherence times drop by an order of magnitude when approaching the region of SQUID instability (see Supplementary~\ref{som:squid_stability}).
        \textbf{(c)} Loaded ($Q_l$), internal ($Q_i$), and coupling ($Q_c$) quality factors of the cavity vs $\Bpar$.
        There is a drop in cavity $Q_i$ setting in from \SI{0.2}{\tesla}, which was also reported in Ref.~\cite{Krause22}, described there as a step-like drop around \SI{0.45}{\tesla}, as only $Q_l$ was extracted there.
        At \SI{0.6}{\tesla}, there is a jump in the data, as the power was slightly increased due to reduced non-linearity in the cavity.
	}
\label{fig:coherence_times_vs_bpar}
\end{figure}

We generally find that the transmon lifetime $\Tone$ fluctuates with $\Bpar$ and small changes in $\Bperp$ (for strong $\Bperp$ it is of course strongly suppressed as seen in Ref.~\cite{Krause22}); nevertheless, for a given $\Bpar$ one can find a high-coherence range in $\Bperp$.
However, the $\Tone$ times of the transmon show a sharp drop for in-plane fields above $\Bpar=\SI{0.4}{\tesla}$ [see \cref{fig:coherence_times_vs_bpar} \textbf{(a)}], a similar behavior was observed in Ref.~\cite{Krause22}.
While in Ref.~\cite{Krause22} the coherence times show a revival, reaching values above \SI{10}{\micro\second} around $\Bpar=\SI{0.5}{\tesla}$, 
in this device, above $\Bpar=\SI{0.4}{\tesla}$ we did not find coherence times exceeding \SI{2}{\micro\second}. 
The decoherence mechanism causing this behavior is presently not understood, although we observe a strong drop of the cavity Q-factor in the same range of $\Bpar$ [see \cref{fig:coherence_times_vs_bpar} \textbf{(c)}].

For completeness, we also investigate the relaxation ($T_1^{10}$, $T_1^{21}$) of the first and second excited states of the transmon as well as the Ramsey and Echo coherence times for the qubit transition $\Ttwostar$ and $\Techo$, respectively [examples of qutrit coherence-time measurements can be found in \cref{fig:Ramsey_gatescan} and \cref{fig:T1_Echo_gatescan}].
We find that the $\Tone$ measurements based on measurements preparing the $\vert 1 \rangle$ and $\vert 2 \rangle$ states generally give consistent results for $T_1^{10}$. 
$\Techo$ is generally on the order of $\Tone$, while $\Ttwostar$ is typically around $\nicefrac{\Tone}{2}$ [see \cref{fig:coherence_times_vs_bpar} \textbf{(b)}].
We fit the $\Ttwostar$ Ramsey data with a model with two frequencies due to the two parities, but find that they generally have similar decay time constants, the difference is generally within the fit uncertainty.
Our data for the ratio of $T_1^{10}/T_1^{21}$ scatters significantly; typically  $T_1^{10}>T_1^{21}$ (theory expectation would be $T_1^{10}=2T_1^{21}$~\cite{peterer12}) but there are outliers for which $T_1^{10}<T_1^{21}$.
This observation shows that losses are possibly due to a complicated environment with resonant two-level fluctuators, higher cavity modes, and the nearly-resonant quasiparticle tunneling. 

\section{Hidden Markov Model vs power-spectral-density based parity extraction}
\label{som:hmm_psd}

In early experiments using transmons to detect quasiparticle tunneling, the parity lifetime estimation from the Ramsey-based parity meter was done by assigning qubit states to the measurement outcomes and taking the power spectral density ($PSD$) of the resulting bitstrings~\cite{Riste13},
while hidden-Markov model (HMM) methods were introduced slightly later~\cite{Vool14}.
The advantage of the Gaussian HMM method we use is that it explicitly takes into account the Gaussian distributions of the single-shot measurement outcomes instead of thresholding the data.
Moreover, the HMM method makes it possible to extract separate switching times $\taupe$ and $\taupo$ for the two parities in a straightforward way.
In the following, we confirm that both methods of data analysis give roughly consistent results: 
the HMM method is presented in \cref{app:parity_schemes}, while here we describe the $PSD$ method and compare the results.

\begin{figure}
	\centering
    \includegraphics{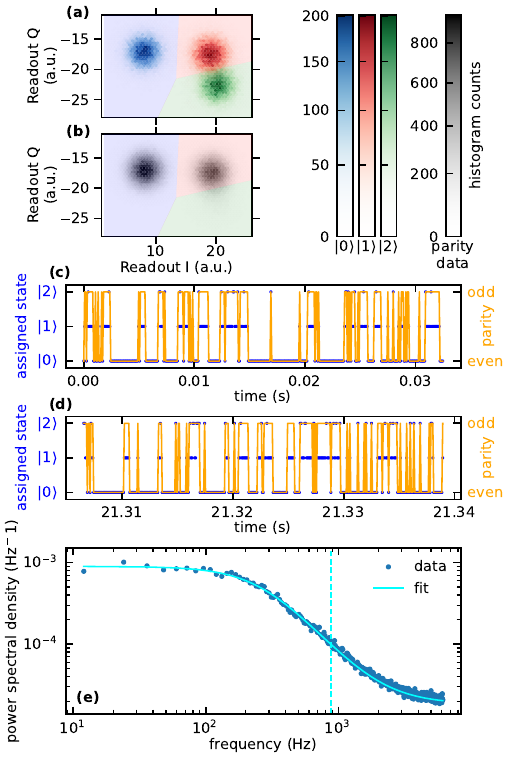}    
	\caption{
		\textbf{(a)} Calibration points for the classifier.
		The transmon is prepared in the $\ket{0}$, $\ket{1}$ and $\ket{2}$ state and measured.
		Histograms for the readout voltage in I and Q are shown. 
		The raw shots are fitted with Gaussian probability distributions.
		Thus a classifier can be constructed that assigns a most-likely state to every point in the IQ plane (overlayed colors).
		\textbf{(b)} Histogram of the data points from the parity measurements. 
		States can be assigned based on the classifier.
		As expected from the sequence, the main outcomes are $\ket{0}$ and $\ket{1}$. 
		\textbf{(c)} and \textbf{(d)} assigned state as a function of time for the beginning and end of the parity run. 
		As the $\ket{2}$ outcomes are most likely due to the finite overlap with the $\ket{1}$ state, even parity is assigned to the $\ket{0}$ state, while odd parity is assigned to $\ket{1}$ and $\ket{2}$ outcomes.
		\textbf({d}) Power spectral density calculated from the parity outcomes as a function of time.
		A fit of a ``noisy'' Lorentzian, Eq.~\eqref{eq:PSD}, is used to determine the parity-switching time $\tau_\mathrm{p}$ and fidelity $F$.
	}
	\label{fig:parity_measurement_explaination_psd}
\end{figure}

To extract $\taup$ and a parity fidelity from the raw IQ data using the $PSD$ method, there are several steps: 
\begin{enumerate}
    \item assign the transmon states to the IQ measurement outcomes
    \item assign a parity to each possible state 
    \item compute the $PSD$ of the parity-vs-time bitstring
    \item fit a Lorentzian-switching model to the $PSD$ to get $\taup$ as well as the parity-measurement fidelity 
\end{enumerate}
The different steps are illustrated in \cref{fig:parity_measurement_explaination_psd}.
We use our Gaussian classifier trained on calibration points to assign a state to each IQ value [\cref{fig:parity_measurement_explaination_psd} \textbf{(b)}].
Based on this calibration, different regions of the IQ plane are colored in according to the most likely state.
The raw IQ data of a calibration run and of a subsequent parity measurement is reported in \cref{fig:parity_measurement_explaination_psd} \textbf{(b)}. 
For this particular measurement, the parity outcomes are mostly $\vert 0 \rangle$ and $\vert 1 \rangle$ and any outcomes classified as $\vert 2 \rangle$ are likely due to the finite overlap between $\vert 1 \rangle$ and $\vert 2 \rangle$.
We therefore identify the ``even'' parity for the $\vert 0 \rangle$ outcomes and ``odd'' for $\vert 1 \rangle$ or $\vert 2 \rangle$.
In \cref{fig:parity_measurement_explaination_psd} \textbf{(c)} and \textbf{(d)}, we display the initial and the final section of the resulting parity-vs-time and state-vs-time sequences with a total duration of 21 s.
We model the parity sequence as a random telegraph signal, characterized by a Lorentzian $PSD$~\cite{Riste13}
\begin{equation}\label{eq:PSD}
    PSD(f) = F^2 \frac{4\taup^{-1}}{(2\taup^{-1})^2+(2\pi f)^2}+(1-F^2)\Delta t_\mathrm{exp},
\end{equation}
where $\taup$ and the parity-measurement fidelity $F$ are free parameters and $\Delta t_\mathrm{exp}$ is the repetition time of the runs.
The experimental $PSD$ and a fit are shown in \cref{fig:parity_measurement_explaination_psd} \textbf{(e)}.
Notably, with this specific method, one can only get a mean $\taup$, as one would have to modify the model for an asymmetric dwell time to extract $\taupe$ and $\taupo$. 
This particular dataset is the same that was shown in \cref{fig:parity_measurement_explanation_hmm} to illustrate the HMM method.
When comparing the assigned parities as a function of time for both figures, the HMM fit already filters out some wrong parity assignments due to ``readout'' errors (e.g. cases where there are relaxation events during the sequence or trivial readout errors). 
For this example, the fit yields $\taup=\SI{1.36}{\milli\second}$ and $F=0.88$, while the HMM gives $\taupe=\SI{1.21}{\milli\second}$ and $\taupo=\SI{1.09}{\milli\second}$ and $F=0.89$.
Alternatively, as done in Ref.~\cite{Riste13} or Ref.~\cite{Serniak18}, one can initialize the transmon as well as the parity by measurements and with post-selection extract the different individual rates; since we did not use a parametric amplifier, the measurements are not quantum non-demolition strictly-speaking, such that this was not attempted.

\begin{figure}
	\centering
	\includegraphics{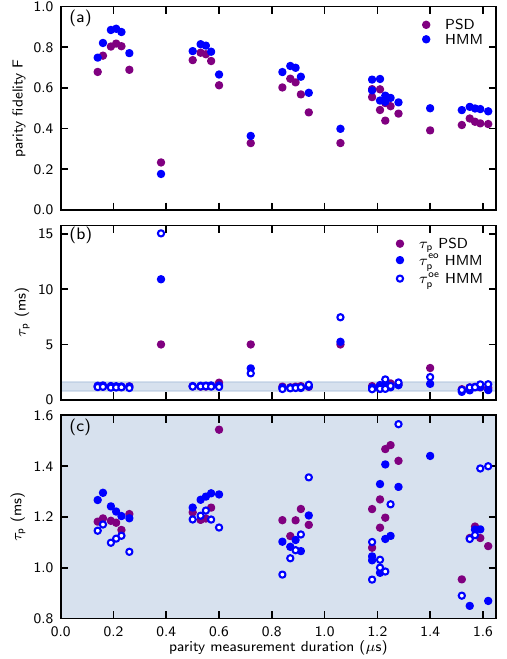}
	\caption{
		\textbf{(a)} Parity measurement fidelity $F$ extracted from the PSD fit and the HMM fit as a function of parity measurement duration.
        The duration comprises the \SI{20}{\nano\second} pulses as well as the Ramsey wait time.
        Oscillations in $F$ are expected and similar fidelities are extracted for both methods. 
        \textbf{(b)} and \textbf{(c)} (zoomed in) $\taup$ extracted from the PSD fit and $\taupe$ and $\taupo$ extracted from the HMM fit. 
        PSD and HMM fit generally give similar $\taup$ for a range of fidelities, showing that the measurements are relatively robust and that the two methods don't contradict each other. 
        For long parity-measurement times, decoherence leads to very noisy $\taup$.
        At the parity-insensitive wait time, the extracted $\taup$ and $F$ are dominated by other low-frequency noise processes. 
        }
	\label{fig:fidelity_taup_vs_waittime}
\end{figure}

We now compare the outcomes for the HMM and $PSD$-fit extraction of $\taup$ and $F$ for more datasets.
We checked the results for all of our data and generally find the deviations to be within $20\%$.
Only around zero field, where we observe a notable difference between $\taupe$ and $\taupo$ due to different $p_1$, $\taupo\sim 0.5\,\taup^\mathrm{PSD}$.
Here we compare the two approaches for a set of data where we performed the Ramsey-based parity measurement using the superposition of $\vert 1 \rangle$ and $\vert 2 \rangle$ for different wait times.
Sweeping the wait time ideally would see $F$ oscillate between 0 and 1, as the azimuthal angles for the even and odd parities would generally not be separated by 180 degrees on the Bloch sphere when the measurement is performed.
For the optimum wait time and at odd integer multiples of it, the parities reach orthogonal states on the Bloch sphere but they periodically converge as well.
The results are shown in \cref{fig:fidelity_taup_vs_waittime} \textbf{(a)}.
We see the expected oscillations in $F$ and a decay for long wait time, as the parity meter becomes limited by the coherence time.
The fidelity estimate for the two methods only differs slightly and follows the same trend.
Around the minimum $F$, the extracted $\taup$ do not reflect parity dynamics but other low-frequency noise processes in the setup [see outliers in  \cref{fig:fidelity_taup_vs_waittime} \textbf{(b)}].
This is likely the reason why $F$ does not drop to zero and that $F$ actually increases for the higher multiples of the wait times for the minimum.
However, in the high-fidelity region we find that the extracted $\taup$ are generally similar and are not too sensitive to imperfect tune-up of the parity measurement [see \cref{fig:fidelity_taup_vs_waittime} \textbf{(c)}]. 
At low fidelity, the HMM can not find a stable difference between $\taupe$ and $\taupo$.
The results presented in the main text are always measured around the first fidelity maximum.

\twocolumngrid

\clearpage

\onecolumngrid 

\section{Primary thermometry with the transmon}
\label{som:transmon_temperature}

\begin{figure}[h]
	\centering \includegraphics{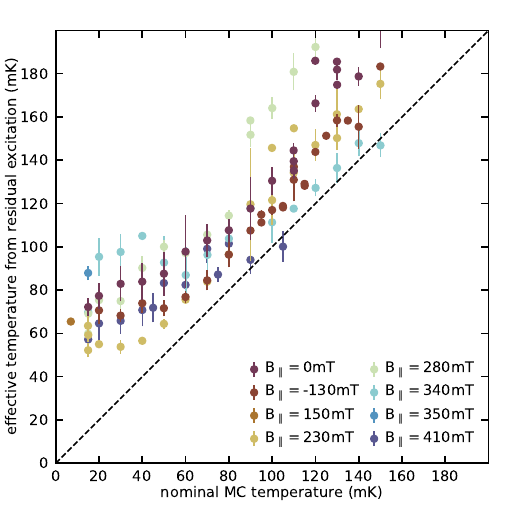}
	\caption{
		Transmon effective temperature extracted from the ground-state population as a function of mixing chamber (MC) temperature (see text). 
        Data suggest a saturation of effective temperature taking place between a mixing chamber temperature of \SI{50}{\milli\kelvin} and \SI{60}{\milli\kelvin}.
        }
	\label{fig:transmon_vs_fridge_temperature}
\end{figure}

As we measure the parity-switching time as a function of the nominal mixing-chamber temperature, it is instructive to compare the latter with the effective transmon temperature. 
If we assume that the transmon populations are Boltzmann distributed, we can approximately infer an effective temperature from the $\ket{0}$ state population $P_0$ and transition frequencies.
The ground-state population can be extracted from a Gaussian fit to the single-shot measurement data for the nominal $\ket{0}$ state - we can extract the fraction of the single-shot IQ values that are assigned as zero.
In principle, we can also extract $P_1$ and $P_2$ from the single-shot measurements to test and corroborate the hypothesis of the Boltzmann distribution (see for example Ref.~\cite{morstedt2024}); for the sake of simplicity, here we just consider the ground state population.
We have explicitly measured the first 3 transition frequencies and populations in the higher levels can be neglected, since we estimate that even at the highest field and temperature we consider, their inclusion would alter the results at most by a few percent. Then, treating the transmon as a ququart we can infer a temperature from $P_0$ by finding the root of the equation
\begin{equation}
P_0 = \frac{1}{ 1 + \exp(- \frac{hf_{01}}{k_B T}) + \exp(- \frac{hf_{02}}{k_B T}) + \exp(- \frac{hf_{03}}{k_B T})}.
\end{equation}
Alternatively, one could use the temperature estimation method described in Ref.~\cite{Sultanov21}, which does not rely on single-shot readout but rather on the contrast of Rabi oscillations between different states. 

The estimated transmon temperature as a function of mixing-chamber temperature at different in-plane fields is shown in  \cref{fig:transmon_vs_fridge_temperature}.
Generally, there is a roughly linear dependence between fridge and transmon temperature starting from about \SI{60}{\milli\kelvin}, in line with values reported in literature~\cite{Sultanov21}.
Below \SI{60}{\milli\kelvin} of mixing chamber temperature the transmon temperature decouples from the fridge temperature and has different saturation values at different magnetic fields.
Parity-switching qubit transitions are likely a main source of heating in this regime [cf. first term in the right-hand-side of Eq.~\eqref{eq:p1steadyApp}].
We find that the transmon temperature at the base temperature of the fridge depends on $\Bpar$, with estimated values in the range \SI{50}{\milli\kelvin}-\SI{80}{\milli\kelvin}.

\section{SQUID oscillation stability and anomalies at high field}
\label{som:squid_stability}

\begin{figure}[h]
  \centering
    \includegraphics{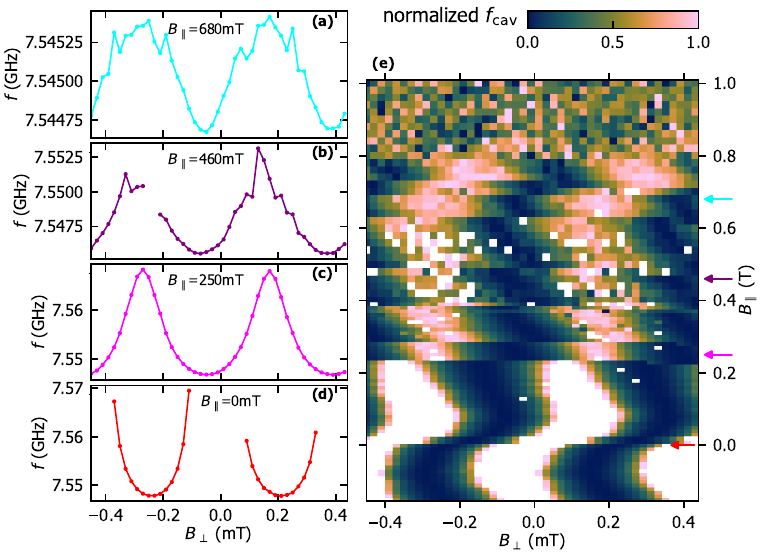}
  \caption{
    Flux instability at high in-plane magnetic fields. 
    \textbf{(a)}, \textbf{(b)}, \textbf{(c)} and \textbf{(d)} SQUID oscillations in the cavity frequency as a function of $\Bperp$.
    At zero field in \textbf{(d)}, the top sweet spot is very close to the cavity such that the top sweet spot is missing but at medium field in \textbf{(c)} we see a smooth dependence with well-defined top and bottom sweet spot.
    However, at higher fields, the cavity frequency shows strange jumps around the top sweet spot [see \textbf{(a)} and \textbf{(b)}].
    \textbf{(d)} Normalized $f_\mathrm{cav}$ as a function of $\Bperp$ and $\Bpar$.
    The fact that the offset of the oscillations does not shift much shows that $\Bpar$ is aligned well in the SQUID plane.
    However, around $\Bpar=0$, the offset shows a non-monotonous behavior possibly a magnet-inherent problem.
    At $\Bpar$ between \SI{0.4}{\tesla} and \SI{0.6}{\tesla}, there is a mix of jumps and missing data points (white spots) where there is no cavity resonance in the measured range - which could be due to jumps.
    We also generally found flux jumps in this range when measuring at the same field multiple times.
  }
  \label{fig:cavity_SQUID_alignment}
\end{figure}

As in Ref.~\cite{Krause22}, we observe that the SQUID oscillations with $\Bperp$ can become unstable at certain $\Bpar$.
This instability is the main reason why we can not measure the parity lifetime for $\Bpar>\SI{0.41}{\tesla}$.
Parity measurements require stability for several minutes for tuning-up the protocol.
However, the reduced coherence at high field would also have made measurements more challenging.
The SQUID instability is marked by frequency jumps in the flux dependence as well as instability in time, with jumps occurring so often that in unstable regions even quick time-domain measurements such as a $\Tone$ measurement are challenging.
In Ref.~\cite{Krause22} measurements of the SQUID device were impossible between \SI{0.4}{\tesla} and \SI{0.5}{\tesla}, but the SQUID became stable again at higher fields and we could take more data up to \SI{0.8}{\tesla}. 
Here, we did not find another region where the SQUID does not suffer from rapid flux jumps above \SI{0.5}{\tesla}.
The cavity frequency as a function of $\Bperp$ is displayed in \cref{fig:cavity_SQUID_alignment}, showing that cavity oscillations are clearly visible up to \SI{0.8}{\tesla}, where the transmon is effectively single-junction (cf. \cref{app:fraunhofer_and_critical_field}), and significantly weaker above.  
There is more noise especially between \SI{0.4}{\tesla} and \SI{0.6}{\tesla}.
We also see a decline in the cavity quality factor setting in around \SI{0.2}{\tesla} and showing a kink around \SI{0.4}{\tesla} [see \cref{fig:coherence_times_vs_bpar} \textbf{(c)}]. 
Currently, we do not have a good explanation for this phenomenology; possible culprits are spurious JJs, as already suggested in Ref.~\cite{Krause22}.


\begin{thebibliography}{72}%
\makeatletter
\providecommand \@ifxundefined [1]{%
 \@ifx{#1\undefined}
}%
\providecommand \@ifnum [1]{%
 \ifnum #1\expandafter \@firstoftwo
 \else \expandafter \@secondoftwo
 \fi
}%
\providecommand \@ifx [1]{%
 \ifx #1\expandafter \@firstoftwo
 \else \expandafter \@secondoftwo
 \fi
}%
\providecommand \natexlab [1]{#1}%
\providecommand \enquote  [1]{``#1''}%
\providecommand \bibnamefont  [1]{#1}%
\providecommand \bibfnamefont [1]{#1}%
\providecommand \citenamefont [1]{#1}%
\providecommand \href@noop [0]{\@secondoftwo}%
\providecommand \href [0]{\begingroup \@sanitize@url \@href}%
\providecommand \@href[1]{\@@startlink{#1}\@@href}%
\providecommand \@@href[1]{\endgroup#1\@@endlink}%
\providecommand \@sanitize@url [0]{\catcode `\\12\catcode `\$12\catcode
  `\&12\catcode `\#12\catcode `\^12\catcode `\_12\catcode `\%12\relax}%
\providecommand \@@startlink[1]{}%
\providecommand \@@endlink[0]{}%
\providecommand \url  [0]{\begingroup\@sanitize@url \@url }%
\providecommand \@url [1]{\endgroup\@href {#1}{\urlprefix }}%
\providecommand \urlprefix  [0]{URL }%
\providecommand \Eprint [0]{\href }%
\providecommand \doibase [0]{https://doi.org/}%
\providecommand \selectlanguage [0]{\@gobble}%
\providecommand \bibinfo  [0]{\@secondoftwo}%
\providecommand \bibfield  [0]{\@secondoftwo}%
\providecommand \translation [1]{[#1]}%
\providecommand \BibitemOpen [0]{}%
\providecommand \bibitemStop [0]{}%
\providecommand \bibitemNoStop [0]{.\EOS\space}%
\providecommand \EOS [0]{\spacefactor3000\relax}%
\providecommand \BibitemShut  [1]{\csname bibitem#1\endcsname}%
\let\auto@bib@innerbib\@empty
\bibitem [{\citenamefont {Kjaergaard}\ \emph {et~al.}(2020)\citenamefont
  {Kjaergaard}, \citenamefont {Schwartz}, \citenamefont {Braum\"{u}ller},
  \citenamefont {Krantz}, \citenamefont {Wang}, \citenamefont {Gustavsson},\
  and\ \citenamefont {Oliver}}]{Kjaergaard20}%
  \BibitemOpen
  \bibfield  {author} {\bibinfo {author} {\bibfnamefont {M.}~\bibnamefont
  {Kjaergaard}}, \bibinfo {author} {\bibfnamefont {M.~E.}\ \bibnamefont
  {Schwartz}}, \bibinfo {author} {\bibfnamefont {J.}~\bibnamefont
  {Braum\"{u}ller}}, \bibinfo {author} {\bibfnamefont {P.}~\bibnamefont
  {Krantz}}, \bibinfo {author} {\bibfnamefont {J.~I.-J.}\ \bibnamefont {Wang}},
  \bibinfo {author} {\bibfnamefont {S.}~\bibnamefont {Gustavsson}},\ and\
  \bibinfo {author} {\bibfnamefont {W.~D.}\ \bibnamefont {Oliver}},\ }\bibfield
   {title} {\bibinfo {title} {Superconducting qubits: Current state of play},\
  }\href {https://doi.org/10.1146/annurev-conmatphys-031119-050605} {\bibfield
  {journal} {\bibinfo  {journal} {Annu. Rev. Condens. Matter Phys.}\ }\textbf
  {\bibinfo {volume} {11}},\ \bibinfo {pages} {369} (\bibinfo {year}
  {2020})}\BibitemShut {NoStop}%
\bibitem [{\citenamefont {Koch}\ \emph {et~al.}(2007)\citenamefont {Koch},
  \citenamefont {Yu}, \citenamefont {Gambetta}, \citenamefont {Houck},
  \citenamefont {Schuster}, \citenamefont {Majer}, \citenamefont {Blais},
  \citenamefont {Devoret}, \citenamefont {Girvin},\ and\ \citenamefont
  {Schoelkopf}}]{Koch07}%
  \BibitemOpen
  \bibfield  {author} {\bibinfo {author} {\bibfnamefont {J.}~\bibnamefont
  {Koch}}, \bibinfo {author} {\bibfnamefont {T.~M.}\ \bibnamefont {Yu}},
  \bibinfo {author} {\bibfnamefont {J.}~\bibnamefont {Gambetta}}, \bibinfo
  {author} {\bibfnamefont {A.~A.}\ \bibnamefont {Houck}}, \bibinfo {author}
  {\bibfnamefont {D.~I.}\ \bibnamefont {Schuster}}, \bibinfo {author}
  {\bibfnamefont {J.}~\bibnamefont {Majer}}, \bibinfo {author} {\bibfnamefont
  {A.}~\bibnamefont {Blais}}, \bibinfo {author} {\bibfnamefont {M.~H.}\
  \bibnamefont {Devoret}}, \bibinfo {author} {\bibfnamefont {S.~M.}\
  \bibnamefont {Girvin}},\ and\ \bibinfo {author} {\bibfnamefont {R.~J.}\
  \bibnamefont {Schoelkopf}},\ }\bibfield  {title} {\bibinfo {title}
  {Charge-insensitive qubit design derived from the {Cooper} pair box},\ }\href
  {https://doi.org/10.1103/PhysRevA.76.042319} {\bibfield  {journal} {\bibinfo
  {journal} {Phys. Rev. A}\ }\textbf {\bibinfo {volume} {76}},\ \bibinfo
  {pages} {042319} (\bibinfo {year} {2007})}\BibitemShut {NoStop}%
\bibitem [{\citenamefont {Place}\ \emph {et~al.}(2021)\citenamefont {Place},
  \citenamefont {Rodgers}, \citenamefont {Mundada}, \citenamefont {Smitham},
  \citenamefont {Fitzpatrick}, \citenamefont {Leng}, \citenamefont {Premkumar},
  \citenamefont {Bryon}, \citenamefont {Vrajitoarea}, \citenamefont {Sussman}
  \emph {et~al.}}]{Place21}%
  \BibitemOpen
  \bibfield  {author} {\bibinfo {author} {\bibfnamefont {A.~P.}\ \bibnamefont
  {Place}}, \bibinfo {author} {\bibfnamefont {L.~V.}\ \bibnamefont {Rodgers}},
  \bibinfo {author} {\bibfnamefont {P.}~\bibnamefont {Mundada}}, \bibinfo
  {author} {\bibfnamefont {B.~M.}\ \bibnamefont {Smitham}}, \bibinfo {author}
  {\bibfnamefont {M.}~\bibnamefont {Fitzpatrick}}, \bibinfo {author}
  {\bibfnamefont {Z.}~\bibnamefont {Leng}}, \bibinfo {author} {\bibfnamefont
  {A.}~\bibnamefont {Premkumar}}, \bibinfo {author} {\bibfnamefont
  {J.}~\bibnamefont {Bryon}}, \bibinfo {author} {\bibfnamefont
  {A.}~\bibnamefont {Vrajitoarea}}, \bibinfo {author} {\bibfnamefont
  {S.}~\bibnamefont {Sussman}}, \emph {et~al.},\ }\bibfield  {title} {\bibinfo
  {title} {New material platform for superconducting transmon qubits with
  coherence times exceeding 0.3 milliseconds},\ }\href
  {https://doi.org/10.1038/s41467-021-22030-5} {\bibfield  {journal} {\bibinfo
  {journal} {Nat Commun}\ }\textbf {\bibinfo {volume} {12}},\ \bibinfo {pages}
  {1779} (\bibinfo {year} {2021})}\BibitemShut {NoStop}%
\bibitem [{\citenamefont {Serniak}\ \emph {et~al.}(2018)\citenamefont
  {Serniak}, \citenamefont {Hays}, \citenamefont {de~Lange}, \citenamefont
  {Diamond}, \citenamefont {Shankar}, \citenamefont {Burkhart}, \citenamefont
  {Frunzio}, \citenamefont {Houzet},\ and\ \citenamefont
  {Devoret}}]{Serniak18}%
  \BibitemOpen
  \bibfield  {author} {\bibinfo {author} {\bibfnamefont {K.}~\bibnamefont
  {Serniak}}, \bibinfo {author} {\bibfnamefont {M.}~\bibnamefont {Hays}},
  \bibinfo {author} {\bibfnamefont {G.}~\bibnamefont {de~Lange}}, \bibinfo
  {author} {\bibfnamefont {S.}~\bibnamefont {Diamond}}, \bibinfo {author}
  {\bibfnamefont {S.}~\bibnamefont {Shankar}}, \bibinfo {author} {\bibfnamefont
  {L.~D.}\ \bibnamefont {Burkhart}}, \bibinfo {author} {\bibfnamefont
  {L.}~\bibnamefont {Frunzio}}, \bibinfo {author} {\bibfnamefont
  {M.}~\bibnamefont {Houzet}},\ and\ \bibinfo {author} {\bibfnamefont {M.~H.}\
  \bibnamefont {Devoret}},\ }\bibfield  {title} {\bibinfo {title} {Hot
  nonequilibrium quasiparticles in transmon qubits},\ }\href
  {https://doi.org/10.1103/PhysRevLett.121.157701} {\bibfield  {journal}
  {\bibinfo  {journal} {Phys. Rev. Lett.}\ }\textbf {\bibinfo {volume} {121}},\
  \bibinfo {pages} {157701} (\bibinfo {year} {2018})}\BibitemShut {NoStop}%
\bibitem [{\citenamefont {McEwen}\ \emph {et~al.}(2022)\citenamefont {McEwen},
  \citenamefont {Faoro}, \citenamefont {Arya}, \citenamefont {Dunsworth},
  \citenamefont {Huang}, \citenamefont {Kim}, \citenamefont {Burkett},
  \citenamefont {Fowler}, \citenamefont {Arute}, \citenamefont {Bardin} \emph
  {et~al.}}]{McEwen22}%
  \BibitemOpen
  \bibfield  {author} {\bibinfo {author} {\bibfnamefont {M.}~\bibnamefont
  {McEwen}}, \bibinfo {author} {\bibfnamefont {L.}~\bibnamefont {Faoro}},
  \bibinfo {author} {\bibfnamefont {K.}~\bibnamefont {Arya}}, \bibinfo {author}
  {\bibfnamefont {A.}~\bibnamefont {Dunsworth}}, \bibinfo {author}
  {\bibfnamefont {T.}~\bibnamefont {Huang}}, \bibinfo {author} {\bibfnamefont
  {S.}~\bibnamefont {Kim}}, \bibinfo {author} {\bibfnamefont {B.}~\bibnamefont
  {Burkett}}, \bibinfo {author} {\bibfnamefont {A.}~\bibnamefont {Fowler}},
  \bibinfo {author} {\bibfnamefont {F.}~\bibnamefont {Arute}}, \bibinfo
  {author} {\bibfnamefont {J.~C.}\ \bibnamefont {Bardin}}, \emph {et~al.},\
  }\bibfield  {title} {\bibinfo {title} {Resolving catastrophic error bursts
  from cosmic rays in large arrays of superconducting qubits},\ }\href
  {https://doi.org/10.1038/s41567-021-01432-8} {\bibfield  {journal} {\bibinfo
  {journal} {Nat. Phys.}\ }\textbf {\bibinfo {volume} {18}},\ \bibinfo {pages}
  {107} (\bibinfo {year} {2022})}\BibitemShut {NoStop}%
\bibitem [{\citenamefont {Riwar}\ \emph {et~al.}(2016)\citenamefont {Riwar},
  \citenamefont {Hosseinkhani}, \citenamefont {Burkhart}, \citenamefont {Gao},
  \citenamefont {Schoelkopf}, \citenamefont {Glazman},\ and\ \citenamefont
  {Catelani}}]{Riwar.2016}%
  \BibitemOpen
  \bibfield  {author} {\bibinfo {author} {\bibfnamefont {R.-P.}\ \bibnamefont
  {Riwar}}, \bibinfo {author} {\bibfnamefont {A.}~\bibnamefont {Hosseinkhani}},
  \bibinfo {author} {\bibfnamefont {L.~D.}\ \bibnamefont {Burkhart}}, \bibinfo
  {author} {\bibfnamefont {Y.~Y.}\ \bibnamefont {Gao}}, \bibinfo {author}
  {\bibfnamefont {R.~J.}\ \bibnamefont {Schoelkopf}}, \bibinfo {author}
  {\bibfnamefont {L.~I.}\ \bibnamefont {Glazman}},\ and\ \bibinfo {author}
  {\bibfnamefont {G.}~\bibnamefont {Catelani}},\ }\bibfield  {title} {\bibinfo
  {title} {Normal-metal quasiparticle traps for superconducting qubits},\
  }\href {https://doi.org/10.1103/PhysRevB.94.104516} {\bibfield  {journal}
  {\bibinfo  {journal} {Phys. Rev. B}\ }\textbf {\bibinfo {volume} {94}},\
  \bibinfo {pages} {104516} (\bibinfo {year} {2016})}\BibitemShut {NoStop}%
\bibitem [{\citenamefont {Gustavsson}\ \emph {et~al.}(2016)\citenamefont
  {Gustavsson}, \citenamefont {Yan}, \citenamefont {Catelani}, \citenamefont
  {Bylander}, \citenamefont {Kamal}, \citenamefont {Birenbaum}, \citenamefont
  {Hover}, \citenamefont {Rosenberg}, \citenamefont {Samach}, \citenamefont
  {Sears}, \citenamefont {Weber}, \citenamefont {Yoder}, \citenamefont
  {Clarke}, \citenamefont {Kerman}, \citenamefont {Yoshihara}, \citenamefont
  {Nakamura}, \citenamefont {Orlando},\ and\ \citenamefont
  {Oliver}}]{Gustavsson16}%
  \BibitemOpen
  \bibfield  {author} {\bibinfo {author} {\bibfnamefont {S.}~\bibnamefont
  {Gustavsson}}, \bibinfo {author} {\bibfnamefont {F.}~\bibnamefont {Yan}},
  \bibinfo {author} {\bibfnamefont {G.}~\bibnamefont {Catelani}}, \bibinfo
  {author} {\bibfnamefont {J.}~\bibnamefont {Bylander}}, \bibinfo {author}
  {\bibfnamefont {A.}~\bibnamefont {Kamal}}, \bibinfo {author} {\bibfnamefont
  {J.}~\bibnamefont {Birenbaum}}, \bibinfo {author} {\bibfnamefont
  {D.}~\bibnamefont {Hover}}, \bibinfo {author} {\bibfnamefont
  {D.}~\bibnamefont {Rosenberg}}, \bibinfo {author} {\bibfnamefont
  {G.}~\bibnamefont {Samach}}, \bibinfo {author} {\bibfnamefont {A.~P.}\
  \bibnamefont {Sears}}, \bibinfo {author} {\bibfnamefont {S.~J.}\ \bibnamefont
  {Weber}}, \bibinfo {author} {\bibfnamefont {J.~L.}\ \bibnamefont {Yoder}},
  \bibinfo {author} {\bibfnamefont {J.}~\bibnamefont {Clarke}}, \bibinfo
  {author} {\bibfnamefont {A.~J.}\ \bibnamefont {Kerman}}, \bibinfo {author}
  {\bibfnamefont {F.}~\bibnamefont {Yoshihara}}, \bibinfo {author}
  {\bibfnamefont {Y.}~\bibnamefont {Nakamura}}, \bibinfo {author}
  {\bibfnamefont {T.~P.}\ \bibnamefont {Orlando}},\ and\ \bibinfo {author}
  {\bibfnamefont {W.~D.}\ \bibnamefont {Oliver}},\ }\bibfield  {title}
  {\bibinfo {title} {Suppressing relaxation in superconducting qubits by
  quasiparticle pumping},\ }\href {https://doi.org/10.1126/science.aah5844}
  {\bibfield  {journal} {\bibinfo  {journal} {Science}\ }\textbf {\bibinfo
  {volume} {354}},\ \bibinfo {pages} {1573} (\bibinfo {year}
  {2016})}\BibitemShut {NoStop}%
\bibitem [{\citenamefont {Riwar}\ and\ \citenamefont
  {Catelani}(2019)}]{Riwar.2019}%
  \BibitemOpen
  \bibfield  {author} {\bibinfo {author} {\bibfnamefont {R.-P.}\ \bibnamefont
  {Riwar}}\ and\ \bibinfo {author} {\bibfnamefont {G.}~\bibnamefont
  {Catelani}},\ }\bibfield  {title} {\bibinfo {title} {Efficient quasiparticle
  traps with low dissipation through gap engineering},\ }\href
  {https://doi.org/10.1103/PhysRevB.100.144514} {\bibfield  {journal} {\bibinfo
   {journal} {Phys. Rev. B}\ }\textbf {\bibinfo {volume} {100}},\ \bibinfo
  {pages} {144514} (\bibinfo {year} {2019})}\BibitemShut {NoStop}%
\bibitem [{\citenamefont {McEwen}\ \emph {et~al.}(2024)\citenamefont {McEwen},
  \citenamefont {Miao}, \citenamefont {Atalaya}, \citenamefont {Bilmes},
  \citenamefont {Crook}, \citenamefont {Bovaird}, \citenamefont {Kreikebaum},
  \citenamefont {Zobrist}, \citenamefont {Jeffrey}, \citenamefont {Ying},
  \citenamefont {Bengtsson}, \citenamefont {Chang}, \citenamefont {Dunsworth},
  \citenamefont {Kelly}, \citenamefont {Zhang}, \citenamefont {Forati},
  \citenamefont {Acharya}, \citenamefont {Iveland}, \citenamefont {Liu},
  \citenamefont {Kim}, \citenamefont {Burkett}, \citenamefont {Megrant},
  \citenamefont {Chen}, \citenamefont {Neill}, \citenamefont {Sank},
  \citenamefont {Devoret},\ and\ \citenamefont {Opremcak}}]{McEwen24}%
  \BibitemOpen
  \bibfield  {author} {\bibinfo {author} {\bibfnamefont {M.}~\bibnamefont
  {McEwen}}, \bibinfo {author} {\bibfnamefont {K.~C.}\ \bibnamefont {Miao}},
  \bibinfo {author} {\bibfnamefont {J.}~\bibnamefont {Atalaya}}, \bibinfo
  {author} {\bibfnamefont {A.}~\bibnamefont {Bilmes}}, \bibinfo {author}
  {\bibfnamefont {A.}~\bibnamefont {Crook}}, \bibinfo {author} {\bibfnamefont
  {J.}~\bibnamefont {Bovaird}}, \bibinfo {author} {\bibfnamefont {J.~M.}\
  \bibnamefont {Kreikebaum}}, \bibinfo {author} {\bibfnamefont
  {N.}~\bibnamefont {Zobrist}}, \bibinfo {author} {\bibfnamefont
  {E.}~\bibnamefont {Jeffrey}}, \bibinfo {author} {\bibfnamefont
  {B.}~\bibnamefont {Ying}}, \bibinfo {author} {\bibfnamefont {A.}~\bibnamefont
  {Bengtsson}}, \bibinfo {author} {\bibfnamefont {H.-S.}\ \bibnamefont
  {Chang}}, \bibinfo {author} {\bibfnamefont {A.}~\bibnamefont {Dunsworth}},
  \bibinfo {author} {\bibfnamefont {J.}~\bibnamefont {Kelly}}, \bibinfo
  {author} {\bibfnamefont {Y.}~\bibnamefont {Zhang}}, \bibinfo {author}
  {\bibfnamefont {E.}~\bibnamefont {Forati}}, \bibinfo {author} {\bibfnamefont
  {R.}~\bibnamefont {Acharya}}, \bibinfo {author} {\bibfnamefont
  {J.}~\bibnamefont {Iveland}}, \bibinfo {author} {\bibfnamefont
  {W.}~\bibnamefont {Liu}}, \bibinfo {author} {\bibfnamefont {S.}~\bibnamefont
  {Kim}}, \bibinfo {author} {\bibfnamefont {B.}~\bibnamefont {Burkett}},
  \bibinfo {author} {\bibfnamefont {A.}~\bibnamefont {Megrant}}, \bibinfo
  {author} {\bibfnamefont {Y.}~\bibnamefont {Chen}}, \bibinfo {author}
  {\bibfnamefont {C.}~\bibnamefont {Neill}}, \bibinfo {author} {\bibfnamefont
  {D.}~\bibnamefont {Sank}}, \bibinfo {author} {\bibfnamefont {M.}~\bibnamefont
  {Devoret}},\ and\ \bibinfo {author} {\bibfnamefont {A.}~\bibnamefont
  {Opremcak}},\ }\bibfield  {title} {\bibinfo {title} {Resisting high-energy
  impact events through gap engineering in superconducting qubit arrays},\
  }\href {https://doi.org/10.48550/arXiv.2402.15644} {\bibfield  {journal}
  {\bibinfo  {journal} {arXiv:2402.15644}\ } (\bibinfo {year}
  {2024})}\BibitemShut {NoStop}%
\bibitem [{\citenamefont {Henriques}\ \emph {et~al.}(2019)\citenamefont
  {Henriques}, \citenamefont {Valenti}, \citenamefont {Charpentier},
  \citenamefont {Lagoin}, \citenamefont {Gouriou}, \citenamefont {Martínez},
  \citenamefont {Cardani}, \citenamefont {Vignati}, \citenamefont {Grünhaupt},
  \citenamefont {Gusenkova}, \citenamefont {Ferrero}, \citenamefont {Skacel},
  \citenamefont {Wernsdorfer}, \citenamefont {Ustinov}, \citenamefont
  {Catelani}, \citenamefont {Sander},\ and\ \citenamefont {Pop}}]{Henriques19}%
  \BibitemOpen
  \bibfield  {author} {\bibinfo {author} {\bibfnamefont {F.}~\bibnamefont
  {Henriques}}, \bibinfo {author} {\bibfnamefont {F.}~\bibnamefont {Valenti}},
  \bibinfo {author} {\bibfnamefont {T.}~\bibnamefont {Charpentier}}, \bibinfo
  {author} {\bibfnamefont {M.}~\bibnamefont {Lagoin}}, \bibinfo {author}
  {\bibfnamefont {C.}~\bibnamefont {Gouriou}}, \bibinfo {author} {\bibfnamefont
  {M.}~\bibnamefont {Martínez}}, \bibinfo {author} {\bibfnamefont
  {L.}~\bibnamefont {Cardani}}, \bibinfo {author} {\bibfnamefont
  {M.}~\bibnamefont {Vignati}}, \bibinfo {author} {\bibfnamefont
  {L.}~\bibnamefont {Grünhaupt}}, \bibinfo {author} {\bibfnamefont
  {D.}~\bibnamefont {Gusenkova}}, \bibinfo {author} {\bibfnamefont
  {J.}~\bibnamefont {Ferrero}}, \bibinfo {author} {\bibfnamefont {S.~T.}\
  \bibnamefont {Skacel}}, \bibinfo {author} {\bibfnamefont {W.}~\bibnamefont
  {Wernsdorfer}}, \bibinfo {author} {\bibfnamefont {A.~V.}\ \bibnamefont
  {Ustinov}}, \bibinfo {author} {\bibfnamefont {G.}~\bibnamefont {Catelani}},
  \bibinfo {author} {\bibfnamefont {O.}~\bibnamefont {Sander}},\ and\ \bibinfo
  {author} {\bibfnamefont {I.~M.}\ \bibnamefont {Pop}},\ }\bibfield  {title}
  {\bibinfo {title} {Phonon traps reduce the quasiparticle density in
  superconducting circuits},\ }\href {https://doi.org/10.1063/1.5124967}
  {\bibfield  {journal} {\bibinfo  {journal} {Appl. Phys. Lett.}\ }\textbf
  {\bibinfo {volume} {115}},\ \bibinfo {pages} {212601} (\bibinfo {year}
  {2019})}\BibitemShut {NoStop}%
\bibitem [{\citenamefont {Iaia}\ \emph {et~al.}(2022)\citenamefont {Iaia},
  \citenamefont {Ku}, \citenamefont {Ballard}, \citenamefont {Larson},
  \citenamefont {Yelton}, \citenamefont {Liu}, \citenamefont {Patel},
  \citenamefont {McDermott},\ and\ \citenamefont {Plourde}}]{Iaia22}%
  \BibitemOpen
  \bibfield  {author} {\bibinfo {author} {\bibfnamefont {V.}~\bibnamefont
  {Iaia}}, \bibinfo {author} {\bibfnamefont {J.}~\bibnamefont {Ku}}, \bibinfo
  {author} {\bibfnamefont {A.}~\bibnamefont {Ballard}}, \bibinfo {author}
  {\bibfnamefont {C.~P.}\ \bibnamefont {Larson}}, \bibinfo {author}
  {\bibfnamefont {E.}~\bibnamefont {Yelton}}, \bibinfo {author} {\bibfnamefont
  {C.~H.}\ \bibnamefont {Liu}}, \bibinfo {author} {\bibfnamefont
  {S.}~\bibnamefont {Patel}}, \bibinfo {author} {\bibfnamefont
  {R.}~\bibnamefont {McDermott}},\ and\ \bibinfo {author} {\bibfnamefont
  {B.~L.~T.}\ \bibnamefont {Plourde}},\ }\bibfield  {title} {\bibinfo {title}
  {Phonon downconversion to suppress correlated errors in superconducting
  qubits},\ }\href {https://doi.org/10.1038/s41467-022-33997-0} {\bibfield
  {journal} {\bibinfo  {journal} {Nat Commun}\ }\textbf {\bibinfo {volume}
  {13}},\ \bibinfo {pages} {6425} (\bibinfo {year} {2022})}\BibitemShut
  {NoStop}%
\bibitem [{\citenamefont {Pan}\ \emph {et~al.}(2022)\citenamefont {Pan},
  \citenamefont {Zhou}, \citenamefont {Yuan}, \citenamefont {Nie},
  \citenamefont {Wei}, \citenamefont {Zhang}, \citenamefont {Li}, \citenamefont
  {Liu}, \citenamefont {Jiang}, \citenamefont {Catelani}, \citenamefont {Hu},
  \citenamefont {Yan},\ and\ \citenamefont {Yu}}]{Pan22}%
  \BibitemOpen
  \bibfield  {author} {\bibinfo {author} {\bibfnamefont {X.}~\bibnamefont
  {Pan}}, \bibinfo {author} {\bibfnamefont {Y.}~\bibnamefont {Zhou}}, \bibinfo
  {author} {\bibfnamefont {H.}~\bibnamefont {Yuan}}, \bibinfo {author}
  {\bibfnamefont {L.}~\bibnamefont {Nie}}, \bibinfo {author} {\bibfnamefont
  {W.}~\bibnamefont {Wei}}, \bibinfo {author} {\bibfnamefont {L.}~\bibnamefont
  {Zhang}}, \bibinfo {author} {\bibfnamefont {J.}~\bibnamefont {Li}}, \bibinfo
  {author} {\bibfnamefont {S.}~\bibnamefont {Liu}}, \bibinfo {author}
  {\bibfnamefont {Z.~H.}\ \bibnamefont {Jiang}}, \bibinfo {author}
  {\bibfnamefont {G.}~\bibnamefont {Catelani}}, \bibinfo {author}
  {\bibfnamefont {L.}~\bibnamefont {Hu}}, \bibinfo {author} {\bibfnamefont
  {F.}~\bibnamefont {Yan}},\ and\ \bibinfo {author} {\bibfnamefont
  {D.}~\bibnamefont {Yu}},\ }\bibfield  {title} {\bibinfo {title} {Engineering
  superconducting qubits to reduce quasiparticles and charge noise},\ }\href
  {https://doi.org/10.1038/s41467-022-34727-2} {\bibfield  {journal} {\bibinfo
  {journal} {Nat Commun}\ }\textbf {\bibinfo {volume} {13}},\ \bibinfo {pages}
  {7196} (\bibinfo {year} {2022})}\BibitemShut {NoStop}%
\bibitem [{\citenamefont {Gordon}\ \emph {et~al.}(2022)\citenamefont {Gordon},
  \citenamefont {Murray}, \citenamefont {Kurter}, \citenamefont {Sandberg},
  \citenamefont {Hall}, \citenamefont {Balakrishnan}, \citenamefont {Shelby},
  \citenamefont {Wacaser}, \citenamefont {Stabile}, \citenamefont {Sleight},
  \citenamefont {Brink}, \citenamefont {Rothwell}, \citenamefont {Rodbell},
  \citenamefont {Dial},\ and\ \citenamefont {Steffen}}]{Gordon22}%
  \BibitemOpen
  \bibfield  {author} {\bibinfo {author} {\bibfnamefont {R.~T.}\ \bibnamefont
  {Gordon}}, \bibinfo {author} {\bibfnamefont {C.~E.}\ \bibnamefont {Murray}},
  \bibinfo {author} {\bibfnamefont {C.}~\bibnamefont {Kurter}}, \bibinfo
  {author} {\bibfnamefont {M.}~\bibnamefont {Sandberg}}, \bibinfo {author}
  {\bibfnamefont {S.~A.}\ \bibnamefont {Hall}}, \bibinfo {author}
  {\bibfnamefont {K.}~\bibnamefont {Balakrishnan}}, \bibinfo {author}
  {\bibfnamefont {R.}~\bibnamefont {Shelby}}, \bibinfo {author} {\bibfnamefont
  {B.}~\bibnamefont {Wacaser}}, \bibinfo {author} {\bibfnamefont {A.~A.}\
  \bibnamefont {Stabile}}, \bibinfo {author} {\bibfnamefont {J.~W.}\
  \bibnamefont {Sleight}}, \bibinfo {author} {\bibfnamefont {M.}~\bibnamefont
  {Brink}}, \bibinfo {author} {\bibfnamefont {M.~B.}\ \bibnamefont {Rothwell}},
  \bibinfo {author} {\bibfnamefont {K.~P.}\ \bibnamefont {Rodbell}}, \bibinfo
  {author} {\bibfnamefont {O.}~\bibnamefont {Dial}},\ and\ \bibinfo {author}
  {\bibfnamefont {M.}~\bibnamefont {Steffen}},\ }\bibfield  {title} {\bibinfo
  {title} {{Environmental radiation impact on lifetimes and quasiparticle
  tunneling rates of fixed-frequency transmon qubits}},\ }\href
  {https://doi.org/10.1063/5.0078785} {\bibfield  {journal} {\bibinfo
  {journal} {Appl. Phys. Lett.}\ }\textbf {\bibinfo {volume} {120}},\ \bibinfo
  {pages} {074002} (\bibinfo {year} {2022})}\BibitemShut {NoStop}%
\bibitem [{\citenamefont {Veps{\"a}l{\"a}inen}\ \emph
  {et~al.}(2020)\citenamefont {Veps{\"a}l{\"a}inen}, \citenamefont {Karamlou},
  \citenamefont {Orrell}, \citenamefont {Dogra}, \citenamefont {Loer},
  \citenamefont {Vasconcelos}, \citenamefont {Kim}, \citenamefont {Melville},
  \citenamefont {Niedzielski}, \citenamefont {Yoder} \emph
  {et~al.}}]{Vepsalainen20}%
  \BibitemOpen
  \bibfield  {author} {\bibinfo {author} {\bibfnamefont {A.~P.}\ \bibnamefont
  {Veps{\"a}l{\"a}inen}}, \bibinfo {author} {\bibfnamefont {A.~H.}\
  \bibnamefont {Karamlou}}, \bibinfo {author} {\bibfnamefont {J.~L.}\
  \bibnamefont {Orrell}}, \bibinfo {author} {\bibfnamefont {A.~S.}\
  \bibnamefont {Dogra}}, \bibinfo {author} {\bibfnamefont {B.}~\bibnamefont
  {Loer}}, \bibinfo {author} {\bibfnamefont {F.}~\bibnamefont {Vasconcelos}},
  \bibinfo {author} {\bibfnamefont {D.~K.}\ \bibnamefont {Kim}}, \bibinfo
  {author} {\bibfnamefont {A.~J.}\ \bibnamefont {Melville}}, \bibinfo {author}
  {\bibfnamefont {B.~M.}\ \bibnamefont {Niedzielski}}, \bibinfo {author}
  {\bibfnamefont {J.~L.}\ \bibnamefont {Yoder}}, \emph {et~al.},\ }\bibfield
  {title} {\bibinfo {title} {Impact of ionizing radiation on superconducting
  qubit coherence},\ }\href {https://doi.org/10.1038/s41586-020-2619-8}
  {\bibfield  {journal} {\bibinfo  {journal} {Nature}\ }\textbf {\bibinfo
  {volume} {584}},\ \bibinfo {pages} {551} (\bibinfo {year}
  {2020})}\BibitemShut {NoStop}%
\bibitem [{\citenamefont {Cardani}\ \emph {et~al.}(2021)\citenamefont
  {Cardani}, \citenamefont {Valenti}, \citenamefont {Casali}, \citenamefont
  {Catelani}, \citenamefont {Charpentier}, \citenamefont {Clemenza},
  \citenamefont {Colantoni}, \citenamefont {Cruciani}, \citenamefont
  {D’Imperio}, \citenamefont {Gironi} \emph {et~al.}}]{Cardani21}%
  \BibitemOpen
  \bibfield  {author} {\bibinfo {author} {\bibfnamefont {L.}~\bibnamefont
  {Cardani}}, \bibinfo {author} {\bibfnamefont {F.}~\bibnamefont {Valenti}},
  \bibinfo {author} {\bibfnamefont {N.}~\bibnamefont {Casali}}, \bibinfo
  {author} {\bibfnamefont {G.}~\bibnamefont {Catelani}}, \bibinfo {author}
  {\bibfnamefont {T.}~\bibnamefont {Charpentier}}, \bibinfo {author}
  {\bibfnamefont {M.}~\bibnamefont {Clemenza}}, \bibinfo {author}
  {\bibfnamefont {I.}~\bibnamefont {Colantoni}}, \bibinfo {author}
  {\bibfnamefont {A.}~\bibnamefont {Cruciani}}, \bibinfo {author}
  {\bibfnamefont {G.}~\bibnamefont {D’Imperio}}, \bibinfo {author}
  {\bibfnamefont {L.}~\bibnamefont {Gironi}}, \emph {et~al.},\ }\bibfield
  {title} {\bibinfo {title} {Reducing the impact of radioactivity on quantum
  circuits in a deep-underground facility},\ }\href
  {https://doi.org/10.1038/s41467-021-23032-z} {\bibfield  {journal} {\bibinfo
  {journal} {Nat Commun}\ }\textbf {\bibinfo {volume} {12}},\ \bibinfo {pages}
  {2733} (\bibinfo {year} {2021})}\BibitemShut {NoStop}%
\bibitem [{\citenamefont {Krause}\ \emph {et~al.}(2022)\citenamefont {Krause},
  \citenamefont {Dickel}, \citenamefont {Vaal}, \citenamefont {Vielmetter},
  \citenamefont {Feng}, \citenamefont {Bounds}, \citenamefont {Catelani},
  \citenamefont {Fink},\ and\ \citenamefont {Ando}}]{Krause22}%
  \BibitemOpen
  \bibfield  {author} {\bibinfo {author} {\bibfnamefont {J.}~\bibnamefont
  {Krause}}, \bibinfo {author} {\bibfnamefont {C.}~\bibnamefont {Dickel}},
  \bibinfo {author} {\bibfnamefont {E.}~\bibnamefont {Vaal}}, \bibinfo {author}
  {\bibfnamefont {M.}~\bibnamefont {Vielmetter}}, \bibinfo {author}
  {\bibfnamefont {J.}~\bibnamefont {Feng}}, \bibinfo {author} {\bibfnamefont
  {R.}~\bibnamefont {Bounds}}, \bibinfo {author} {\bibfnamefont
  {G.}~\bibnamefont {Catelani}}, \bibinfo {author} {\bibfnamefont {J.~M.}\
  \bibnamefont {Fink}},\ and\ \bibinfo {author} {\bibfnamefont
  {Y.}~\bibnamefont {Ando}},\ }\bibfield  {title} {\bibinfo {title} {Magnetic
  field resilience of three-dimensional transmons with thin-film
  {${\text{Al/AlO}}_{x}/\text{Al}$} {Josephson} junctions approaching 1 {T}},\
  }\href {https://doi.org/10.1103/PhysRevApplied.17.034032} {\bibfield
  {journal} {\bibinfo  {journal} {Phys. Rev. Applied}\ }\textbf {\bibinfo
  {volume} {17}},\ \bibinfo {pages} {034032} (\bibinfo {year}
  {2022})}\BibitemShut {NoStop}%
\bibitem [{\citenamefont {Blais}\ \emph {et~al.}(2004)\citenamefont {Blais},
  \citenamefont {Huang}, \citenamefont {Wallraff}, \citenamefont {Girvin},\
  and\ \citenamefont {Schoelkopf}}]{Blais04}%
  \BibitemOpen
  \bibfield  {author} {\bibinfo {author} {\bibfnamefont {A.}~\bibnamefont
  {Blais}}, \bibinfo {author} {\bibfnamefont {R.-S.}\ \bibnamefont {Huang}},
  \bibinfo {author} {\bibfnamefont {A.}~\bibnamefont {Wallraff}}, \bibinfo
  {author} {\bibfnamefont {S.~M.}\ \bibnamefont {Girvin}},\ and\ \bibinfo
  {author} {\bibfnamefont {R.~J.}\ \bibnamefont {Schoelkopf}},\ }\bibfield
  {title} {\bibinfo {title} {Cavity quantum electrodynamics for superconducting
  electrical circuits: An architecture for quantum computation},\ }\href
  {https://doi.org/10.1103/PhysRevA.69.062320} {\bibfield  {journal} {\bibinfo
  {journal} {Phys. Rev. A}\ }\textbf {\bibinfo {volume} {69}},\ \bibinfo
  {pages} {062320} (\bibinfo {year} {2004})}\BibitemShut {NoStop}%
\bibitem [{\citenamefont {Rist\`e}\ \emph {et~al.}(2013)\citenamefont
  {Rist\`e}, \citenamefont {Bultink}, \citenamefont {Tiggelman}, \citenamefont
  {Schouten}, \citenamefont {Lehnert},\ and\ \citenamefont
  {DiCarlo}}]{Riste13}%
  \BibitemOpen
  \bibfield  {author} {\bibinfo {author} {\bibfnamefont {D.}~\bibnamefont
  {Rist\`e}}, \bibinfo {author} {\bibfnamefont {C.~C.}\ \bibnamefont
  {Bultink}}, \bibinfo {author} {\bibfnamefont {M.~J.}\ \bibnamefont
  {Tiggelman}}, \bibinfo {author} {\bibfnamefont {R.~N.}\ \bibnamefont
  {Schouten}}, \bibinfo {author} {\bibfnamefont {K.~W.}\ \bibnamefont
  {Lehnert}},\ and\ \bibinfo {author} {\bibfnamefont {L.}~\bibnamefont
  {DiCarlo}},\ }\bibfield  {title} {\bibinfo {title} {Millisecond charge-parity
  fluctuations and induced decoherence in a superconducting transmon qubit},\
  }\href {https://doi.org/10.1038/ncomms2936} {\bibfield  {journal} {\bibinfo
  {journal} {Nat.\ Commun.}\ }\textbf {\bibinfo {volume} {4}},\ \bibinfo
  {pages} {1913} (\bibinfo {year} {2013})}\BibitemShut {NoStop}%
\bibitem [{\citenamefont {Hassler}\ \emph {et~al.}(2011)\citenamefont
  {Hassler}, \citenamefont {Akhmerov},\ and\ \citenamefont
  {Beenakker}}]{Hassler11}%
  \BibitemOpen
  \bibfield  {author} {\bibinfo {author} {\bibfnamefont {F.}~\bibnamefont
  {Hassler}}, \bibinfo {author} {\bibfnamefont {A.~R.}\ \bibnamefont
  {Akhmerov}},\ and\ \bibinfo {author} {\bibfnamefont {C.~W.~J.}\ \bibnamefont
  {Beenakker}},\ }\bibfield  {title} {\bibinfo {title} {{The top-transmon: a
  hybrid superconducting qubit for parity-protected quantum computation}},\
  }\href {https://doi.org/10.1088/1367-2630/13/9/095004} {\bibfield  {journal}
  {\bibinfo  {journal} {New J.\ Phys.}\ }\textbf {\bibinfo {volume} {13}},\
  \bibinfo {pages} {095004} (\bibinfo {year} {2011})}\BibitemShut {NoStop}%
\bibitem [{\citenamefont {Hyart}\ \emph {et~al.}(2013)\citenamefont {Hyart},
  \citenamefont {van Heck}, \citenamefont {Fulga}, \citenamefont {Burrello},
  \citenamefont {Akhmerov},\ and\ \citenamefont {Beenakker}}]{Hyart13}%
  \BibitemOpen
  \bibfield  {author} {\bibinfo {author} {\bibfnamefont {T.}~\bibnamefont
  {Hyart}}, \bibinfo {author} {\bibfnamefont {B.}~\bibnamefont {van Heck}},
  \bibinfo {author} {\bibfnamefont {I.~C.}\ \bibnamefont {Fulga}}, \bibinfo
  {author} {\bibfnamefont {M.}~\bibnamefont {Burrello}}, \bibinfo {author}
  {\bibfnamefont {A.~R.}\ \bibnamefont {Akhmerov}},\ and\ \bibinfo {author}
  {\bibfnamefont {C.~W.~J.}\ \bibnamefont {Beenakker}},\ }\bibfield  {title}
  {\bibinfo {title} {Flux-controlled quantum computation with majorana
  fermions},\ }\href {https://doi.org/10.1103/PhysRevB.88.035121} {\bibfield
  {journal} {\bibinfo  {journal} {Phys. Rev. B}\ }\textbf {\bibinfo {volume}
  {88}},\ \bibinfo {pages} {035121} (\bibinfo {year} {2013})}\BibitemShut
  {NoStop}%
\bibitem [{\citenamefont {Luthi}\ \emph {et~al.}(2018)\citenamefont {Luthi},
  \citenamefont {Stavenga}, \citenamefont {Enzing}, \citenamefont {Bruno},
  \citenamefont {Dickel}, \citenamefont {Langford}, \citenamefont {Rol},
  \citenamefont {Jespersen}, \citenamefont {Nyg\aa{}rd}, \citenamefont
  {Krogstrup},\ and\ \citenamefont {DiCarlo}}]{Luthi18}%
  \BibitemOpen
  \bibfield  {author} {\bibinfo {author} {\bibfnamefont {F.}~\bibnamefont
  {Luthi}}, \bibinfo {author} {\bibfnamefont {T.}~\bibnamefont {Stavenga}},
  \bibinfo {author} {\bibfnamefont {O.~W.}\ \bibnamefont {Enzing}}, \bibinfo
  {author} {\bibfnamefont {A.}~\bibnamefont {Bruno}}, \bibinfo {author}
  {\bibfnamefont {C.}~\bibnamefont {Dickel}}, \bibinfo {author} {\bibfnamefont
  {N.~K.}\ \bibnamefont {Langford}}, \bibinfo {author} {\bibfnamefont {M.~A.}\
  \bibnamefont {Rol}}, \bibinfo {author} {\bibfnamefont {T.~S.}\ \bibnamefont
  {Jespersen}}, \bibinfo {author} {\bibfnamefont {J.}~\bibnamefont
  {Nyg\aa{}rd}}, \bibinfo {author} {\bibfnamefont {P.}~\bibnamefont
  {Krogstrup}},\ and\ \bibinfo {author} {\bibfnamefont {L.}~\bibnamefont
  {DiCarlo}},\ }\bibfield  {title} {\bibinfo {title} {Evolution of nanowire
  transmon qubits and their coherence in a magnetic field},\ }\href
  {https://doi.org/10.1103/PhysRevLett.120.100502} {\bibfield  {journal}
  {\bibinfo  {journal} {Phys. Rev. Lett.}\ }\textbf {\bibinfo {volume} {120}},\
  \bibinfo {pages} {100502} (\bibinfo {year} {2018})}\BibitemShut {NoStop}%
\bibitem [{\citenamefont {Kroll}\ \emph {et~al.}(2018)\citenamefont {Kroll},
  \citenamefont {Uilhoorn}, \citenamefont {van~der Enden}, \citenamefont
  {de~Jong}, \citenamefont {Watanabe}, \citenamefont {Taniguchi}, \citenamefont
  {Goswami}, \citenamefont {Cassidy},\ and\ \citenamefont
  {Kouwenhoven}}]{Kroll18}%
  \BibitemOpen
  \bibfield  {author} {\bibinfo {author} {\bibfnamefont {J.~G.}\ \bibnamefont
  {Kroll}}, \bibinfo {author} {\bibfnamefont {W.}~\bibnamefont {Uilhoorn}},
  \bibinfo {author} {\bibfnamefont {K.~L.}\ \bibnamefont {van~der Enden}},
  \bibinfo {author} {\bibfnamefont {D.}~\bibnamefont {de~Jong}}, \bibinfo
  {author} {\bibfnamefont {K.}~\bibnamefont {Watanabe}}, \bibinfo {author}
  {\bibfnamefont {T.}~\bibnamefont {Taniguchi}}, \bibinfo {author}
  {\bibfnamefont {S.}~\bibnamefont {Goswami}}, \bibinfo {author} {\bibfnamefont
  {M.~C.}\ \bibnamefont {Cassidy}},\ and\ \bibinfo {author} {\bibfnamefont
  {L.~P.}\ \bibnamefont {Kouwenhoven}},\ }\bibfield  {title} {\bibinfo {title}
  {Magnetic field compatible circuit quantum electrodynamics with graphene
  josephson junctions},\ }\href
  {https://www.nature.com/articles/s41467-018-07124-x} {\bibfield  {journal}
  {\bibinfo  {journal} {Nat Commun}\ }\textbf {\bibinfo {volume} {9}},\
  \bibinfo {pages} {4615} (\bibinfo {year} {2018})}\BibitemShut {NoStop}%
\bibitem [{\citenamefont {Kringh\o{}j}\ \emph {et~al.}(2021)\citenamefont
  {Kringh\o{}j}, \citenamefont {Larsen}, \citenamefont {Erlandsson},
  \citenamefont {Uilhoorn}, \citenamefont {Kroll}, \citenamefont {Hesselberg},
  \citenamefont {McNeil}, \citenamefont {Krogstrup}, \citenamefont {Casparis},
  \citenamefont {Marcus},\ and\ \citenamefont {Petersson}}]{Kringhoj21}%
  \BibitemOpen
  \bibfield  {author} {\bibinfo {author} {\bibfnamefont {A.}~\bibnamefont
  {Kringh\o{}j}}, \bibinfo {author} {\bibfnamefont {T.~W.}\ \bibnamefont
  {Larsen}}, \bibinfo {author} {\bibfnamefont {O.}~\bibnamefont {Erlandsson}},
  \bibinfo {author} {\bibfnamefont {W.}~\bibnamefont {Uilhoorn}}, \bibinfo
  {author} {\bibfnamefont {J.~G.}\ \bibnamefont {Kroll}}, \bibinfo {author}
  {\bibfnamefont {M.}~\bibnamefont {Hesselberg}}, \bibinfo {author}
  {\bibfnamefont {R.~P.~G.}\ \bibnamefont {McNeil}}, \bibinfo {author}
  {\bibfnamefont {P.}~\bibnamefont {Krogstrup}}, \bibinfo {author}
  {\bibfnamefont {L.}~\bibnamefont {Casparis}}, \bibinfo {author}
  {\bibfnamefont {C.~M.}\ \bibnamefont {Marcus}},\ and\ \bibinfo {author}
  {\bibfnamefont {K.~D.}\ \bibnamefont {Petersson}},\ }\bibfield  {title}
  {\bibinfo {title} {Magnetic-field-compatible superconducting transmon
  qubit},\ }\href {https://doi.org/10.1103/PhysRevApplied.15.054001} {\bibfield
   {journal} {\bibinfo  {journal} {Phys. Rev. Applied}\ }\textbf {\bibinfo
  {volume} {15}},\ \bibinfo {pages} {054001} (\bibinfo {year}
  {2021})}\BibitemShut {NoStop}%
\bibitem [{\citenamefont {Uilhoorn}\ \emph {et~al.}(2021)\citenamefont
  {Uilhoorn}, \citenamefont {Kroll}, \citenamefont {Bargerbos}, \citenamefont
  {Nabi}, \citenamefont {Yang}, \citenamefont {Krogstrup}, \citenamefont
  {Kouwenhoven}, \citenamefont {Kou},\ and\ \citenamefont
  {de~Lange}}]{Uilhoorn2021}%
  \BibitemOpen
  \bibfield  {author} {\bibinfo {author} {\bibfnamefont {W.}~\bibnamefont
  {Uilhoorn}}, \bibinfo {author} {\bibfnamefont {J.~G.}\ \bibnamefont {Kroll}},
  \bibinfo {author} {\bibfnamefont {A.}~\bibnamefont {Bargerbos}}, \bibinfo
  {author} {\bibfnamefont {S.~D.}\ \bibnamefont {Nabi}}, \bibinfo {author}
  {\bibfnamefont {C.-K.}\ \bibnamefont {Yang}}, \bibinfo {author}
  {\bibfnamefont {P.}~\bibnamefont {Krogstrup}}, \bibinfo {author}
  {\bibfnamefont {L.~P.}\ \bibnamefont {Kouwenhoven}}, \bibinfo {author}
  {\bibfnamefont {A.}~\bibnamefont {Kou}},\ and\ \bibinfo {author}
  {\bibfnamefont {G.}~\bibnamefont {de~Lange}},\ }\bibfield  {title} {\bibinfo
  {title} {Quasiparticle trapping by orbital effect in a hybrid
  superconducting-semiconducting circuit},\ }\href
  {https://arxiv.org/abs/2105.11038} {\bibfield  {journal} {\bibinfo  {journal}
  {arXiv:2105.11038}\ } (\bibinfo {year} {2021})}\BibitemShut {NoStop}%
\bibitem [{\citenamefont {Paik}\ \emph {et~al.}(2011)\citenamefont {Paik},
  \citenamefont {Schuster}, \citenamefont {Bishop}, \citenamefont {Kirchmair},
  \citenamefont {Catelani}, \citenamefont {Sears}, \citenamefont {Johnson},
  \citenamefont {Reagor}, \citenamefont {Frunzio}, \citenamefont {Glazman},
  \citenamefont {Girvin}, \citenamefont {Devoret},\ and\ \citenamefont
  {Schoelkopf}}]{Paik11}%
  \BibitemOpen
  \bibfield  {author} {\bibinfo {author} {\bibfnamefont {H.}~\bibnamefont
  {Paik}}, \bibinfo {author} {\bibfnamefont {D.~I.}\ \bibnamefont {Schuster}},
  \bibinfo {author} {\bibfnamefont {L.~S.}\ \bibnamefont {Bishop}}, \bibinfo
  {author} {\bibfnamefont {G.}~\bibnamefont {Kirchmair}}, \bibinfo {author}
  {\bibfnamefont {G.}~\bibnamefont {Catelani}}, \bibinfo {author}
  {\bibfnamefont {A.~P.}\ \bibnamefont {Sears}}, \bibinfo {author}
  {\bibfnamefont {B.~R.}\ \bibnamefont {Johnson}}, \bibinfo {author}
  {\bibfnamefont {M.~J.}\ \bibnamefont {Reagor}}, \bibinfo {author}
  {\bibfnamefont {L.}~\bibnamefont {Frunzio}}, \bibinfo {author} {\bibfnamefont
  {L.~I.}\ \bibnamefont {Glazman}}, \bibinfo {author} {\bibfnamefont {S.~M.}\
  \bibnamefont {Girvin}}, \bibinfo {author} {\bibfnamefont {M.~H.}\
  \bibnamefont {Devoret}},\ and\ \bibinfo {author} {\bibfnamefont {R.~J.}\
  \bibnamefont {Schoelkopf}},\ }\bibfield  {title} {\bibinfo {title}
  {Observation of high coherence in josephson junction qubits measured in a
  three-dimensional circuit qed architecture},\ }\href
  {https://doi.org/10.1103/PhysRevLett.107.240501} {\bibfield  {journal}
  {\bibinfo  {journal} {Phys. Rev. Lett.}\ }\textbf {\bibinfo {volume} {107}},\
  \bibinfo {pages} {240501} (\bibinfo {year} {2011})}\BibitemShut {NoStop}%
\bibitem [{\citenamefont {Kurter}\ \emph {et~al.}(2022)\citenamefont {Kurter},
  \citenamefont {Murray}, \citenamefont {Gordon}, \citenamefont {Wymore},
  \citenamefont {Sandberg}, \citenamefont {Shelby}, \citenamefont {Eddins},
  \citenamefont {Adiga}, \citenamefont {Finck}, \citenamefont {Rivera} \emph
  {et~al.}}]{Kurter22}%
  \BibitemOpen
  \bibfield  {author} {\bibinfo {author} {\bibfnamefont {C.}~\bibnamefont
  {Kurter}}, \bibinfo {author} {\bibfnamefont {C.}~\bibnamefont {Murray}},
  \bibinfo {author} {\bibfnamefont {R.}~\bibnamefont {Gordon}}, \bibinfo
  {author} {\bibfnamefont {B.}~\bibnamefont {Wymore}}, \bibinfo {author}
  {\bibfnamefont {M.}~\bibnamefont {Sandberg}}, \bibinfo {author}
  {\bibfnamefont {R.}~\bibnamefont {Shelby}}, \bibinfo {author} {\bibfnamefont
  {A.}~\bibnamefont {Eddins}}, \bibinfo {author} {\bibfnamefont
  {V.}~\bibnamefont {Adiga}}, \bibinfo {author} {\bibfnamefont
  {A.}~\bibnamefont {Finck}}, \bibinfo {author} {\bibfnamefont
  {E.}~\bibnamefont {Rivera}}, \emph {et~al.},\ }\bibfield  {title} {\bibinfo
  {title} {Quasiparticle tunneling as a probe of josephson junction barrier and
  capacitor material in superconducting qubits},\ }\href
  {https://doi.org/10.1038/s41534-022-00542-2} {\bibfield  {journal} {\bibinfo
  {journal} {npj Quantum Inf}\ }\textbf {\bibinfo {volume} {8}},\ \bibinfo
  {pages} {31} (\bibinfo {year} {2022})}\BibitemShut {NoStop}%
\bibitem [{\citenamefont {Tennant}\ \emph {et~al.}(2022)\citenamefont
  {Tennant}, \citenamefont {Martinez}, \citenamefont {Beck}, \citenamefont
  {O'Kelley}, \citenamefont {Wilen}, \citenamefont {McDermott}, \citenamefont
  {DuBois},\ and\ \citenamefont {Rosen}}]{Tennant22}%
  \BibitemOpen
  \bibfield  {author} {\bibinfo {author} {\bibfnamefont {D.~M.}\ \bibnamefont
  {Tennant}}, \bibinfo {author} {\bibfnamefont {L.~A.}\ \bibnamefont
  {Martinez}}, \bibinfo {author} {\bibfnamefont {K.~M.}\ \bibnamefont {Beck}},
  \bibinfo {author} {\bibfnamefont {S.~R.}\ \bibnamefont {O'Kelley}}, \bibinfo
  {author} {\bibfnamefont {C.~D.}\ \bibnamefont {Wilen}}, \bibinfo {author}
  {\bibfnamefont {R.}~\bibnamefont {McDermott}}, \bibinfo {author}
  {\bibfnamefont {J.~L.}\ \bibnamefont {DuBois}},\ and\ \bibinfo {author}
  {\bibfnamefont {Y.~J.}\ \bibnamefont {Rosen}},\ }\bibfield  {title} {\bibinfo
  {title} {Low-frequency correlated charge-noise measurements across multiple
  energy transitions in a tantalum transmon},\ }\href
  {https://doi.org/10.1103/PRXQuantum.3.030307} {\bibfield  {journal} {\bibinfo
   {journal} {PRX Quantum}\ }\textbf {\bibinfo {volume} {3}},\ \bibinfo {pages}
  {030307} (\bibinfo {year} {2022})}\BibitemShut {NoStop}%
\bibitem [{\citenamefont {Diamond}\ \emph {et~al.}(2022)\citenamefont
  {Diamond}, \citenamefont {Fatemi}, \citenamefont {Hays}, \citenamefont {Nho},
  \citenamefont {Kurilovich}, \citenamefont {Connolly}, \citenamefont {Joshi},
  \citenamefont {Serniak}, \citenamefont {Frunzio}, \citenamefont {Glazman},\
  and\ \citenamefont {Devoret}}]{Diamond22}%
  \BibitemOpen
  \bibfield  {author} {\bibinfo {author} {\bibfnamefont {S.}~\bibnamefont
  {Diamond}}, \bibinfo {author} {\bibfnamefont {V.}~\bibnamefont {Fatemi}},
  \bibinfo {author} {\bibfnamefont {M.}~\bibnamefont {Hays}}, \bibinfo {author}
  {\bibfnamefont {H.}~\bibnamefont {Nho}}, \bibinfo {author} {\bibfnamefont
  {P.~D.}\ \bibnamefont {Kurilovich}}, \bibinfo {author} {\bibfnamefont
  {T.}~\bibnamefont {Connolly}}, \bibinfo {author} {\bibfnamefont {V.~R.}\
  \bibnamefont {Joshi}}, \bibinfo {author} {\bibfnamefont {K.}~\bibnamefont
  {Serniak}}, \bibinfo {author} {\bibfnamefont {L.}~\bibnamefont {Frunzio}},
  \bibinfo {author} {\bibfnamefont {L.~I.}\ \bibnamefont {Glazman}},\ and\
  \bibinfo {author} {\bibfnamefont {M.~H.}\ \bibnamefont {Devoret}},\
  }\bibfield  {title} {\bibinfo {title} {Distinguishing parity-switching
  mechanisms in a superconducting qubit},\ }\href
  {https://doi.org/10.1103/PRXQuantum.3.040304} {\bibfield  {journal} {\bibinfo
   {journal} {PRX Quantum}\ }\textbf {\bibinfo {volume} {3}},\ \bibinfo {pages}
  {040304} (\bibinfo {year} {2022})}\BibitemShut {NoStop}%
\bibitem [{\citenamefont {Connolly}\ \emph {et~al.}(2023)\citenamefont
  {Connolly}, \citenamefont {Kurilovich}, \citenamefont {Diamond},
  \citenamefont {Nho}, \citenamefont {B{\o}ttcher}, \citenamefont {Glazman},
  \citenamefont {Fatemi},\ and\ \citenamefont
  {Devoret}}]{connolly2023coexistence}%
  \BibitemOpen
  \bibfield  {author} {\bibinfo {author} {\bibfnamefont {T.}~\bibnamefont
  {Connolly}}, \bibinfo {author} {\bibfnamefont {P.~D.}\ \bibnamefont
  {Kurilovich}}, \bibinfo {author} {\bibfnamefont {S.}~\bibnamefont {Diamond}},
  \bibinfo {author} {\bibfnamefont {H.}~\bibnamefont {Nho}}, \bibinfo {author}
  {\bibfnamefont {C.~G.}\ \bibnamefont {B{\o}ttcher}}, \bibinfo {author}
  {\bibfnamefont {L.~I.}\ \bibnamefont {Glazman}}, \bibinfo {author}
  {\bibfnamefont {V.}~\bibnamefont {Fatemi}},\ and\ \bibinfo {author}
  {\bibfnamefont {M.~H.}\ \bibnamefont {Devoret}},\ }\bibfield  {title}
  {\bibinfo {title} {Coexistence of nonequilibrium density and equilibrium
  energy distribution of quasiparticles in a superconducting qubit},\ }\href
  {https://doi.org/10.48550/arXiv.2302.12330} {\bibfield  {journal} {\bibinfo
  {journal} {arXiv:2302.12330}\ } (\bibinfo {year} {2023})}\BibitemShut
  {NoStop}%
\bibitem [{\citenamefont {Marchegiani}\ \emph {et~al.}(2022)\citenamefont
  {Marchegiani}, \citenamefont {Amico},\ and\ \citenamefont
  {Catelani}}]{Marchegiani22}%
  \BibitemOpen
  \bibfield  {author} {\bibinfo {author} {\bibfnamefont {G.}~\bibnamefont
  {Marchegiani}}, \bibinfo {author} {\bibfnamefont {L.}~\bibnamefont {Amico}},\
  and\ \bibinfo {author} {\bibfnamefont {G.}~\bibnamefont {Catelani}},\
  }\bibfield  {title} {\bibinfo {title} {Quasiparticles in superconducting
  qubits with asymmetric junctions},\ }\href
  {https://doi.org/10.1103/PRXQuantum.3.040338} {\bibfield  {journal} {\bibinfo
   {journal} {PRX Quantum}\ }\textbf {\bibinfo {volume} {3}},\ \bibinfo {pages}
  {040338} (\bibinfo {year} {2022})}\BibitemShut {NoStop}%
\bibitem [{\citenamefont {Serniak}\ \emph {et~al.}(2019)\citenamefont
  {Serniak}, \citenamefont {Diamond}, \citenamefont {Hays}, \citenamefont
  {Fatemi}, \citenamefont {Shankar}, \citenamefont {Frunzio}, \citenamefont
  {Schoelkopf},\ and\ \citenamefont {Devoret}}]{Serniak19}%
  \BibitemOpen
  \bibfield  {author} {\bibinfo {author} {\bibfnamefont {K.}~\bibnamefont
  {Serniak}}, \bibinfo {author} {\bibfnamefont {S.}~\bibnamefont {Diamond}},
  \bibinfo {author} {\bibfnamefont {M.}~\bibnamefont {Hays}}, \bibinfo {author}
  {\bibfnamefont {V.}~\bibnamefont {Fatemi}}, \bibinfo {author} {\bibfnamefont
  {S.}~\bibnamefont {Shankar}}, \bibinfo {author} {\bibfnamefont
  {L.}~\bibnamefont {Frunzio}}, \bibinfo {author} {\bibfnamefont
  {R.}~\bibnamefont {Schoelkopf}},\ and\ \bibinfo {author} {\bibfnamefont
  {M.}~\bibnamefont {Devoret}},\ }\bibfield  {title} {\bibinfo {title} {Direct
  dispersive monitoring of charge parity in offset-charge-sensitive
  transmons},\ }\href {https://doi.org/10.1103/PhysRevApplied.12.014052}
  {\bibfield  {journal} {\bibinfo  {journal} {Phys. Rev. Applied}\ }\textbf
  {\bibinfo {volume} {12}},\ \bibinfo {pages} {014052} (\bibinfo {year}
  {2019})}\BibitemShut {NoStop}%
\bibitem [{\citenamefont {Ginossar}\ and\ \citenamefont
  {Grosfeld}(2014)}]{Ginossar14}%
  \BibitemOpen
  \bibfield  {author} {\bibinfo {author} {\bibfnamefont {E.}~\bibnamefont
  {Ginossar}}\ and\ \bibinfo {author} {\bibfnamefont {E.}~\bibnamefont
  {Grosfeld}},\ }\bibfield  {title} {\bibinfo {title} {Microwave transitions as
  a signature of coherent parity mixing effects in the majorana-transmon
  qubit},\ }\href {https://doi.org/10.1038/ncomms5772} {\bibfield  {journal}
  {\bibinfo  {journal} {Nat.\ Commun.}\ }\textbf {\bibinfo {volume} {5}},\
  \bibinfo {pages} {4772} (\bibinfo {year} {2014})}\BibitemShut {NoStop}%
\bibitem [{\citenamefont {Joyez}\ \emph {et~al.}(1994)\citenamefont {Joyez},
  \citenamefont {Lafarge}, \citenamefont {Filipe}, \citenamefont {Esteve},\
  and\ \citenamefont {Devoret}}]{Joyez94}%
  \BibitemOpen
  \bibfield  {author} {\bibinfo {author} {\bibfnamefont {P.}~\bibnamefont
  {Joyez}}, \bibinfo {author} {\bibfnamefont {P.}~\bibnamefont {Lafarge}},
  \bibinfo {author} {\bibfnamefont {A.}~\bibnamefont {Filipe}}, \bibinfo
  {author} {\bibfnamefont {D.}~\bibnamefont {Esteve}},\ and\ \bibinfo {author}
  {\bibfnamefont {M.~H.}\ \bibnamefont {Devoret}},\ }\bibfield  {title}
  {\bibinfo {title} {Observation of parity-induced suppression of {J}osephson
  tunneling in the superconducting single electron transistor},\ }\href
  {https://doi.org/10.1103/PhysRevLett.72.2458} {\bibfield  {journal} {\bibinfo
   {journal} {Phys. Rev. Lett.}\ }\textbf {\bibinfo {volume} {72}},\ \bibinfo
  {pages} {2458} (\bibinfo {year} {1994})}\BibitemShut {NoStop}%
\bibitem [{\citenamefont {Aumentado}\ \emph {et~al.}(2004)\citenamefont
  {Aumentado}, \citenamefont {Keller}, \citenamefont {Martinis},\ and\
  \citenamefont {Devoret}}]{Aumentado04}%
  \BibitemOpen
  \bibfield  {author} {\bibinfo {author} {\bibfnamefont {J.}~\bibnamefont
  {Aumentado}}, \bibinfo {author} {\bibfnamefont {M.~W.}\ \bibnamefont
  {Keller}}, \bibinfo {author} {\bibfnamefont {J.~M.}\ \bibnamefont
  {Martinis}},\ and\ \bibinfo {author} {\bibfnamefont {M.~H.}\ \bibnamefont
  {Devoret}},\ }\bibfield  {title} {\bibinfo {title} {Nonequilibrium
  quasiparticles and $2e$ periodicity in single-{C}ooper-pair transistors},\
  }\href {https://doi.org/10.1103/PhysRevLett.92.066802} {\bibfield  {journal}
  {\bibinfo  {journal} {Phys. Rev. Lett.}\ }\textbf {\bibinfo {volume} {92}},\
  \bibinfo {pages} {066802} (\bibinfo {year} {2004})}\BibitemShut {NoStop}%
\bibitem [{\citenamefont {Lutchyn}\ \emph {et~al.}(2006)\citenamefont
  {Lutchyn}, \citenamefont {Glazman},\ and\ \citenamefont
  {Larkin}}]{Lutchyn06}%
  \BibitemOpen
  \bibfield  {author} {\bibinfo {author} {\bibfnamefont {R.~M.}\ \bibnamefont
  {Lutchyn}}, \bibinfo {author} {\bibfnamefont {L.~I.}\ \bibnamefont
  {Glazman}},\ and\ \bibinfo {author} {\bibfnamefont {A.~I.}\ \bibnamefont
  {Larkin}},\ }\bibfield  {title} {\bibinfo {title} {Kinetics of the
  superconducting charge qubit in the presence of a quasiparticle},\ }\href
  {https://doi.org/10.1103/PhysRevB.74.064515} {\bibfield  {journal} {\bibinfo
  {journal} {Phys. Rev. B}\ }\textbf {\bibinfo {volume} {74}},\ \bibinfo
  {pages} {064515} (\bibinfo {year} {2006})}\BibitemShut {NoStop}%
\bibitem [{\citenamefont {Houck}\ \emph {et~al.}(2008)\citenamefont {Houck},
  \citenamefont {Schreier}, \citenamefont {Johnson}, \citenamefont {Chow},
  \citenamefont {Koch}, \citenamefont {Gambetta}, \citenamefont {Schuster},
  \citenamefont {Frunzio}, \citenamefont {Devoret}, \citenamefont {Girvin},\
  and\ \citenamefont {Schoelkopf}}]{Houck08}%
  \BibitemOpen
  \bibfield  {author} {\bibinfo {author} {\bibfnamefont {A.~A.}\ \bibnamefont
  {Houck}}, \bibinfo {author} {\bibfnamefont {J.~A.}\ \bibnamefont {Schreier}},
  \bibinfo {author} {\bibfnamefont {B.~R.}\ \bibnamefont {Johnson}}, \bibinfo
  {author} {\bibfnamefont {J.~M.}\ \bibnamefont {Chow}}, \bibinfo {author}
  {\bibfnamefont {J.}~\bibnamefont {Koch}}, \bibinfo {author} {\bibfnamefont
  {J.~M.}\ \bibnamefont {Gambetta}}, \bibinfo {author} {\bibfnamefont {D.~I.}\
  \bibnamefont {Schuster}}, \bibinfo {author} {\bibfnamefont {L.}~\bibnamefont
  {Frunzio}}, \bibinfo {author} {\bibfnamefont {M.~H.}\ \bibnamefont
  {Devoret}}, \bibinfo {author} {\bibfnamefont {S.~M.}\ \bibnamefont
  {Girvin}},\ and\ \bibinfo {author} {\bibfnamefont {R.~J.}\ \bibnamefont
  {Schoelkopf}},\ }\bibfield  {title} {\bibinfo {title} {Controlling the
  spontaneous emission of a superconducting transmon qubit},\ }\href
  {https://doi.org/10.1103/PhysRevLett.101.080502} {\bibfield  {journal}
  {\bibinfo  {journal} {Phys. Rev. Lett.}\ }\textbf {\bibinfo {volume} {101}},\
  \bibinfo {pages} {080502} (\bibinfo {year} {2008})}\BibitemShut {NoStop}%
\bibitem [{\citenamefont {Christensen}\ \emph {et~al.}(2019)\citenamefont
  {Christensen}, \citenamefont {Wilen}, \citenamefont {Opremcak}, \citenamefont
  {Nelson}, \citenamefont {Schlenker}, \citenamefont {Zimonick}, \citenamefont
  {Faoro}, \citenamefont {Ioffe}, \citenamefont {Rosen}, \citenamefont
  {DuBois}, \citenamefont {Plourde},\ and\ \citenamefont
  {McDermott}}]{Christensen19}%
  \BibitemOpen
  \bibfield  {author} {\bibinfo {author} {\bibfnamefont {B.~G.}\ \bibnamefont
  {Christensen}}, \bibinfo {author} {\bibfnamefont {C.~D.}\ \bibnamefont
  {Wilen}}, \bibinfo {author} {\bibfnamefont {A.}~\bibnamefont {Opremcak}},
  \bibinfo {author} {\bibfnamefont {J.}~\bibnamefont {Nelson}}, \bibinfo
  {author} {\bibfnamefont {F.}~\bibnamefont {Schlenker}}, \bibinfo {author}
  {\bibfnamefont {C.~H.}\ \bibnamefont {Zimonick}}, \bibinfo {author}
  {\bibfnamefont {L.}~\bibnamefont {Faoro}}, \bibinfo {author} {\bibfnamefont
  {L.~B.}\ \bibnamefont {Ioffe}}, \bibinfo {author} {\bibfnamefont {Y.~J.}\
  \bibnamefont {Rosen}}, \bibinfo {author} {\bibfnamefont {J.~L.}\ \bibnamefont
  {DuBois}}, \bibinfo {author} {\bibfnamefont {B.~L.~T.}\ \bibnamefont
  {Plourde}},\ and\ \bibinfo {author} {\bibfnamefont {R.}~\bibnamefont
  {McDermott}},\ }\bibfield  {title} {\bibinfo {title} {Anomalous charge noise
  in superconducting qubits},\ }\href
  {https://doi.org/10.1103/PhysRevB.100.140503} {\bibfield  {journal} {\bibinfo
   {journal} {Phys. Rev. B}\ }\textbf {\bibinfo {volume} {100}},\ \bibinfo
  {pages} {140503} (\bibinfo {year} {2019})}\BibitemShut {NoStop}%
\bibitem [{\citenamefont {Fulde}(1973)}]{Fulde1973}%
  \BibitemOpen
  \bibfield  {author} {\bibinfo {author} {\bibfnamefont {P.}~\bibnamefont
  {Fulde}},\ }\bibfield  {title} {\bibinfo {title} {High field
  superconductivity in thin films},\ }\href
  {https://doi.org/10.1080/00018737300101369} {\bibfield  {journal} {\bibinfo
  {journal} {Adv. Phys.}\ }\textbf {\bibinfo {volume} {22}},\ \bibinfo {pages}
  {667} (\bibinfo {year} {1973})}\BibitemShut {NoStop}%
\bibitem [{\citenamefont {Meservey}\ and\ \citenamefont
  {Tedrow}(1994)}]{Meservey1994}%
  \BibitemOpen
  \bibfield  {author} {\bibinfo {author} {\bibfnamefont {R.}~\bibnamefont
  {Meservey}}\ and\ \bibinfo {author} {\bibfnamefont {P.}~\bibnamefont
  {Tedrow}},\ }\bibfield  {title} {\bibinfo {title} {Spin-polarized electron
  tunneling},\ }\href
  {https://doi.org/https://doi.org/10.1016/0370-1573(94)90105-8} {\bibfield
  {journal} {\bibinfo  {journal} {Phys. Rep.}\ }\textbf {\bibinfo {volume}
  {238}},\ \bibinfo {pages} {173} (\bibinfo {year} {1994})}\BibitemShut
  {NoStop}%
\bibitem [{\citenamefont {Chubov}\ \emph {et~al.}(1969)\citenamefont {Chubov},
  \citenamefont {Eremenko},\ and\ \citenamefont {Pilipenko}}]{Chubov1969}%
  \BibitemOpen
  \bibfield  {author} {\bibinfo {author} {\bibfnamefont {P.}~\bibnamefont
  {Chubov}}, \bibinfo {author} {\bibfnamefont {V.}~\bibnamefont {Eremenko}},\
  and\ \bibinfo {author} {\bibfnamefont {Y.~A.}\ \bibnamefont {Pilipenko}},\
  }\bibfield  {title} {\bibinfo {title} {Dependence of the critical temperature
  and energy gap on the thickness of superconducting aluminum films},\ }\href
  {http://jetp.ras.ru/cgi-bin/e/index/e/28/3/p389?a=list} {\bibfield  {journal}
  {\bibinfo  {journal} {Sov. Phys. JETP}\ }\textbf {\bibinfo {volume} {28}},\
  \bibinfo {pages} {389} (\bibinfo {year} {1969})}\BibitemShut {NoStop}%
\bibitem [{\citenamefont {Meservey}\ and\ \citenamefont
  {Tedrow}(1971)}]{Meservey1971}%
  \BibitemOpen
  \bibfield  {author} {\bibinfo {author} {\bibfnamefont {R.}~\bibnamefont
  {Meservey}}\ and\ \bibinfo {author} {\bibfnamefont {P.~M.}\ \bibnamefont
  {Tedrow}},\ }\bibfield  {title} {\bibinfo {title} {Properties of very thin
  aluminum films},\ }\href {https://doi.org/10.1063/1.1659648} {\bibfield
  {journal} {\bibinfo  {journal} {J. Appl. Phys.}\ }\textbf {\bibinfo {volume}
  {42}},\ \bibinfo {pages} {51} (\bibinfo {year} {1971})}\BibitemShut {NoStop}%
\bibitem [{SOM()}]{SOM}%
  \BibitemOpen
  \href@noop {} {}\bibinfo {howpublished} {The supplemental material at [URL
  will be inserted by publisher] provides experimental details and additional
  data and analysis supporting the claims in the main text.}\BibitemShut
  {Stop}%
\bibitem [{\citenamefont {Tinkham}(2004)}]{Tinkham04}%
  \BibitemOpen
  \bibfield  {author} {\bibinfo {author} {\bibfnamefont {M.}~\bibnamefont
  {Tinkham}},\ }\href@noop {} {\emph {\bibinfo {title} {Introduction to
  Superconductivity}}},\ \bibinfo {edition} {2nd}\ ed.\ (\bibinfo  {publisher}
  {Dover Publications},\ \bibinfo {year} {2004})\BibitemShut {NoStop}%
\bibitem [{Note1()}]{Note1}%
  \BibitemOpen
  \bibinfo {note} {More generally, given the penetration depths $\lambda
  _T,\lambda _B$ and the film thicknesses of the two electrodes $t_{T}$,
  $t_{R}$, the effective thickness determining the Fraunhofer critical field
  amounts to $\lambda _T \tanh (t_T/2\lambda _T)+\lambda _B \tanh (t_B/2\lambda
  _B)+t_{\protect \rm AlOx}$, where $t_{\protect \rm AlOx}$ is the thickness of
  the insulating oxide-barrier of the Josephson junction.}\BibitemShut {Stop}%
\bibitem [{\citenamefont {Willsch}\ \emph {et~al.}(2024)\citenamefont
  {Willsch}, \citenamefont {Rieger}, \citenamefont {Winkel}, \citenamefont
  {Willsch}, \citenamefont {Dickel}, \citenamefont {Krause}, \citenamefont
  {Ando}, \citenamefont {Lescanne}, \citenamefont {Leghtas}, \citenamefont
  {Bronn}, \citenamefont {Deb}, \citenamefont {Lanes}, \citenamefont {Minev},
  \citenamefont {Dennig}, \citenamefont {Geisert}, \citenamefont {Günzler},
  \citenamefont {Ihssen}, \citenamefont {Paluch}, \citenamefont {Reisinger},
  \citenamefont {Hanna}, \citenamefont {Bae}, \citenamefont {Schüffelgen},
  \citenamefont {Grützmacher}, \citenamefont {Buimaga-Iarinca}, \citenamefont
  {Morari}, \citenamefont {Wernsdorfer}, \citenamefont {DiVincenzo},
  \citenamefont {Michielsen}, \citenamefont {Catelani},\ and\ \citenamefont
  {Pop}}]{Willsch23}%
  \BibitemOpen
  \bibfield  {author} {\bibinfo {author} {\bibfnamefont {D.}~\bibnamefont
  {Willsch}}, \bibinfo {author} {\bibfnamefont {D.}~\bibnamefont {Rieger}},
  \bibinfo {author} {\bibfnamefont {P.}~\bibnamefont {Winkel}}, \bibinfo
  {author} {\bibfnamefont {M.}~\bibnamefont {Willsch}}, \bibinfo {author}
  {\bibfnamefont {C.}~\bibnamefont {Dickel}}, \bibinfo {author} {\bibfnamefont
  {J.}~\bibnamefont {Krause}}, \bibinfo {author} {\bibfnamefont
  {Y.}~\bibnamefont {Ando}}, \bibinfo {author} {\bibfnamefont {R.}~\bibnamefont
  {Lescanne}}, \bibinfo {author} {\bibfnamefont {Z.}~\bibnamefont {Leghtas}},
  \bibinfo {author} {\bibfnamefont {N.~T.}\ \bibnamefont {Bronn}}, \bibinfo
  {author} {\bibfnamefont {P.}~\bibnamefont {Deb}}, \bibinfo {author}
  {\bibfnamefont {O.}~\bibnamefont {Lanes}}, \bibinfo {author} {\bibfnamefont
  {Z.~K.}\ \bibnamefont {Minev}}, \bibinfo {author} {\bibfnamefont
  {B.}~\bibnamefont {Dennig}}, \bibinfo {author} {\bibfnamefont
  {S.}~\bibnamefont {Geisert}}, \bibinfo {author} {\bibfnamefont
  {S.}~\bibnamefont {Günzler}}, \bibinfo {author} {\bibfnamefont
  {S.}~\bibnamefont {Ihssen}}, \bibinfo {author} {\bibfnamefont
  {P.}~\bibnamefont {Paluch}}, \bibinfo {author} {\bibfnamefont
  {T.}~\bibnamefont {Reisinger}}, \bibinfo {author} {\bibfnamefont
  {R.}~\bibnamefont {Hanna}}, \bibinfo {author} {\bibfnamefont {J.~H.}\
  \bibnamefont {Bae}}, \bibinfo {author} {\bibfnamefont {P.}~\bibnamefont
  {Schüffelgen}}, \bibinfo {author} {\bibfnamefont {D.}~\bibnamefont
  {Grützmacher}}, \bibinfo {author} {\bibfnamefont {L.}~\bibnamefont
  {Buimaga-Iarinca}}, \bibinfo {author} {\bibfnamefont {C.}~\bibnamefont
  {Morari}}, \bibinfo {author} {\bibfnamefont {W.}~\bibnamefont {Wernsdorfer}},
  \bibinfo {author} {\bibfnamefont {D.~P.}\ \bibnamefont {DiVincenzo}},
  \bibinfo {author} {\bibfnamefont {K.}~\bibnamefont {Michielsen}}, \bibinfo
  {author} {\bibfnamefont {G.}~\bibnamefont {Catelani}},\ and\ \bibinfo
  {author} {\bibfnamefont {I.~M.}\ \bibnamefont {Pop}},\ }\bibfield  {title}
  {\bibinfo {title} {Observation of {Josephson} harmonics in tunnel
  junctions},\ }\href {https://doi.org/10.1038/s41567-024-02400-8} {\bibfield
  {journal} {\bibinfo  {journal} {Nat. Phys.}\ } (\bibinfo {year}
  {2024})}\BibitemShut {NoStop}%
\bibitem [{\citenamefont {Catelani}(2014)}]{Catelani14}%
  \BibitemOpen
  \bibfield  {author} {\bibinfo {author} {\bibfnamefont {G.}~\bibnamefont
  {Catelani}},\ }\bibfield  {title} {\bibinfo {title} {Parity switching and
  decoherence by quasiparticles in single-junction transmons},\ }\href
  {https://doi.org/10.1103/PhysRevB.89.094522} {\bibfield  {journal} {\bibinfo
  {journal} {Phys. Rev. B}\ }\textbf {\bibinfo {volume} {89}},\ \bibinfo
  {pages} {094522} (\bibinfo {year} {2014})}\BibitemShut {NoStop}%
\bibitem [{\citenamefont {Houzet}\ \emph {et~al.}(2019)\citenamefont {Houzet},
  \citenamefont {Serniak}, \citenamefont {Catelani}, \citenamefont {Devoret},\
  and\ \citenamefont {Glazman}}]{Houzet19}%
  \BibitemOpen
  \bibfield  {author} {\bibinfo {author} {\bibfnamefont {M.}~\bibnamefont
  {Houzet}}, \bibinfo {author} {\bibfnamefont {K.}~\bibnamefont {Serniak}},
  \bibinfo {author} {\bibfnamefont {G.}~\bibnamefont {Catelani}}, \bibinfo
  {author} {\bibfnamefont {M.~H.}\ \bibnamefont {Devoret}},\ and\ \bibinfo
  {author} {\bibfnamefont {L.~I.}\ \bibnamefont {Glazman}},\ }\bibfield
  {title} {\bibinfo {title} {Photon-assisted charge-parity jumps in a
  superconducting qubit},\ }\href
  {https://doi.org/10.1103/PhysRevLett.123.107704} {\bibfield  {journal}
  {\bibinfo  {journal} {Phys. Rev. Lett.}\ }\textbf {\bibinfo {volume} {123}},\
  \bibinfo {pages} {107704} (\bibinfo {year} {2019})}\BibitemShut {NoStop}%
\bibitem [{\citenamefont {Fischer}\ and\ \citenamefont
  {Catelani}(2024)}]{Fischer24}%
  \BibitemOpen
  \bibfield  {author} {\bibinfo {author} {\bibfnamefont {P.}~\bibnamefont
  {Fischer}}\ and\ \bibinfo {author} {\bibfnamefont {G.}~\bibnamefont
  {Catelani}},\ }\bibfield  {title} {\bibinfo {title} {Nonequilibrium
  quasiparticle distribution in superconducting resonators: effect of
  pair-breaking photons},\ }\href {https://arxiv.org/abs/2401.12607} {\bibfield
   {journal} {\bibinfo  {journal} {arXiv:2401.12607}\ } (\bibinfo {year}
  {2024})}\BibitemShut {NoStop}%
\bibitem [{Note2()}]{Note2}%
  \BibitemOpen
  \bibinfo {note} {The gap suppression with temperature can be safely neglected
  in this range, temperature being much smaller than the critical temperatures
  of the two electrodes.}\BibitemShut {Stop}%
\bibitem [{\citenamefont {Wang}\ \emph {et~al.}(2014)\citenamefont {Wang},
  \citenamefont {Gao}, \citenamefont {Pop}, \citenamefont {Vool}, \citenamefont
  {Axline}, \citenamefont {Brecht}, \citenamefont {Heeres}, \citenamefont
  {Frunzio}, \citenamefont {Devoret}, \citenamefont {Catelani} \emph
  {et~al.}}]{Wang14}%
  \BibitemOpen
  \bibfield  {author} {\bibinfo {author} {\bibfnamefont {C.}~\bibnamefont
  {Wang}}, \bibinfo {author} {\bibfnamefont {Y.~Y.}\ \bibnamefont {Gao}},
  \bibinfo {author} {\bibfnamefont {I.~M.}\ \bibnamefont {Pop}}, \bibinfo
  {author} {\bibfnamefont {U.}~\bibnamefont {Vool}}, \bibinfo {author}
  {\bibfnamefont {C.}~\bibnamefont {Axline}}, \bibinfo {author} {\bibfnamefont
  {T.}~\bibnamefont {Brecht}}, \bibinfo {author} {\bibfnamefont {R.~W.}\
  \bibnamefont {Heeres}}, \bibinfo {author} {\bibfnamefont {L.}~\bibnamefont
  {Frunzio}}, \bibinfo {author} {\bibfnamefont {M.~H.}\ \bibnamefont
  {Devoret}}, \bibinfo {author} {\bibfnamefont {G.}~\bibnamefont {Catelani}},
  \emph {et~al.},\ }\bibfield  {title} {\bibinfo {title} {Measurement and
  control of quasiparticle dynamics in a superconducting qubit},\ }\href
  {https://doi.org/10.1038/ncomms6836} {\bibfield  {journal} {\bibinfo
  {journal} {Nat Commun}\ }\textbf {\bibinfo {volume} {5}},\ \bibinfo {pages}
  {5836} (\bibinfo {year} {2014})}\BibitemShut {NoStop}%
\bibitem [{\citenamefont {Van~Woerkom}\ \emph {et~al.}(2015)\citenamefont
  {Van~Woerkom}, \citenamefont {Geresdi},\ and\ \citenamefont
  {Kouwenhoven}}]{vanWoerkom15}%
  \BibitemOpen
  \bibfield  {author} {\bibinfo {author} {\bibfnamefont {D.~J.}\ \bibnamefont
  {Van~Woerkom}}, \bibinfo {author} {\bibfnamefont {A.}~\bibnamefont
  {Geresdi}},\ and\ \bibinfo {author} {\bibfnamefont {L.~P.}\ \bibnamefont
  {Kouwenhoven}},\ }\bibfield  {title} {\bibinfo {title} {One minute parity
  lifetime of a nbtin cooper-pair transistor},\ }\href
  {https://doi.org/10.1038/nphys3342} {\bibfield  {journal} {\bibinfo
  {journal} {Nat. Phys.}\ }\textbf {\bibinfo {volume} {11}},\ \bibinfo {pages}
  {547} (\bibinfo {year} {2015})}\BibitemShut {NoStop}%
\bibitem [{\citenamefont {Rafferty}\ \emph {et~al.}(2021)\citenamefont
  {Rafferty}, \citenamefont {Patel}, \citenamefont {Liu}, \citenamefont
  {Abdullah}, \citenamefont {Wilen}, \citenamefont {Harrison},\ and\
  \citenamefont {McDermott}}]{Rafferty21}%
  \BibitemOpen
  \bibfield  {author} {\bibinfo {author} {\bibfnamefont {O.}~\bibnamefont
  {Rafferty}}, \bibinfo {author} {\bibfnamefont {S.}~\bibnamefont {Patel}},
  \bibinfo {author} {\bibfnamefont {C.}~\bibnamefont {Liu}}, \bibinfo {author}
  {\bibfnamefont {S.}~\bibnamefont {Abdullah}}, \bibinfo {author}
  {\bibfnamefont {C.}~\bibnamefont {Wilen}}, \bibinfo {author} {\bibfnamefont
  {D.}~\bibnamefont {Harrison}},\ and\ \bibinfo {author} {\bibfnamefont
  {R.}~\bibnamefont {McDermott}},\ }\bibfield  {title} {\bibinfo {title}
  {Spurious antenna modes of the transmon qubit},\ }\href
  {https://arxiv.org/pdf/2103.06803.pdf} {\bibfield  {journal} {\bibinfo
  {journal} {arXiv:2103.06803}\ } (\bibinfo {year} {2021})}\BibitemShut
  {NoStop}%
\bibitem [{\citenamefont {Liu}\ \emph {et~al.}(2024)\citenamefont {Liu},
  \citenamefont {Harrison}, \citenamefont {Patel}, \citenamefont {Wilen},
  \citenamefont {Rafferty}, \citenamefont {Shearrow}, \citenamefont {Ballard},
  \citenamefont {Iaia}, \citenamefont {Ku}, \citenamefont {Plourde},\ and\
  \citenamefont {McDermott}}]{Liu22}%
  \BibitemOpen
  \bibfield  {author} {\bibinfo {author} {\bibfnamefont {C.~H.}\ \bibnamefont
  {Liu}}, \bibinfo {author} {\bibfnamefont {D.~C.}\ \bibnamefont {Harrison}},
  \bibinfo {author} {\bibfnamefont {S.}~\bibnamefont {Patel}}, \bibinfo
  {author} {\bibfnamefont {C.~D.}\ \bibnamefont {Wilen}}, \bibinfo {author}
  {\bibfnamefont {O.}~\bibnamefont {Rafferty}}, \bibinfo {author}
  {\bibfnamefont {A.}~\bibnamefont {Shearrow}}, \bibinfo {author}
  {\bibfnamefont {A.}~\bibnamefont {Ballard}}, \bibinfo {author} {\bibfnamefont
  {V.}~\bibnamefont {Iaia}}, \bibinfo {author} {\bibfnamefont {J.}~\bibnamefont
  {Ku}}, \bibinfo {author} {\bibfnamefont {B.~L.~T.}\ \bibnamefont {Plourde}},\
  and\ \bibinfo {author} {\bibfnamefont {R.}~\bibnamefont {McDermott}},\
  }\bibfield  {title} {\bibinfo {title} {Quasiparticle poisoning of
  superconducting qubits from resonant absorption of pair-breaking photons},\
  }\href {https://doi.org/10.1103/PhysRevLett.132.017001} {\bibfield  {journal}
  {\bibinfo  {journal} {Phys. Rev. Lett.}\ }\textbf {\bibinfo {volume} {132}},\
  \bibinfo {pages} {017001} (\bibinfo {year} {2024})}\BibitemShut {NoStop}%
\bibitem [{\citenamefont {Nielsen}\ \emph {et~al.}(2024)\citenamefont
  {Nielsen}, \citenamefont {Astafev}, \citenamefont {Nielsen}, \citenamefont
  {Vogel}, \citenamefont {lakhotiaharshit}, \citenamefont {Johnson},
  \citenamefont {Hardal}, \citenamefont {Akshita}, \citenamefont {sohail
  chatoor}, \citenamefont {Bonabi}, \citenamefont {Liang}, \citenamefont
  {Ungaretti}, \citenamefont {Pauka}, \citenamefont {Morgan}, \citenamefont
  {Adriaan}, \citenamefont {Eendebak}, \citenamefont {Nijholt}, \citenamefont
  {qSaevar}, \citenamefont {Eendebak}, \citenamefont {Droege}, \citenamefont
  {Samantha}, \citenamefont {Darulova}, \citenamefont {van Gulik},
  \citenamefont {Pearson}, \citenamefont {Larsen},\ and\ \citenamefont
  {Corna}}]{Qcodes}%
  \BibitemOpen
  \bibfield  {author} {\bibinfo {author} {\bibfnamefont {J.~H.}\ \bibnamefont
  {Nielsen}}, \bibinfo {author} {\bibfnamefont {M.}~\bibnamefont {Astafev}},
  \bibinfo {author} {\bibfnamefont {W.~H.}\ \bibnamefont {Nielsen}}, \bibinfo
  {author} {\bibfnamefont {D.}~\bibnamefont {Vogel}}, \bibinfo {author}
  {\bibnamefont {lakhotiaharshit}}, \bibinfo {author} {\bibfnamefont
  {A.}~\bibnamefont {Johnson}}, \bibinfo {author} {\bibfnamefont
  {A.}~\bibnamefont {Hardal}}, \bibinfo {author} {\bibnamefont {Akshita}},
  \bibinfo {author} {\bibnamefont {sohail chatoor}}, \bibinfo {author}
  {\bibfnamefont {F.}~\bibnamefont {Bonabi}}, \bibinfo {author} {\bibnamefont
  {Liang}}, \bibinfo {author} {\bibfnamefont {G.}~\bibnamefont {Ungaretti}},
  \bibinfo {author} {\bibfnamefont {S.}~\bibnamefont {Pauka}}, \bibinfo
  {author} {\bibfnamefont {T.}~\bibnamefont {Morgan}}, \bibinfo {author}
  {\bibnamefont {Adriaan}}, \bibinfo {author} {\bibfnamefont {P.}~\bibnamefont
  {Eendebak}}, \bibinfo {author} {\bibfnamefont {B.}~\bibnamefont {Nijholt}},
  \bibinfo {author} {\bibnamefont {qSaevar}}, \bibinfo {author} {\bibfnamefont
  {P.}~\bibnamefont {Eendebak}}, \bibinfo {author} {\bibfnamefont
  {S.}~\bibnamefont {Droege}}, \bibinfo {author} {\bibnamefont {Samantha}},
  \bibinfo {author} {\bibfnamefont {J.}~\bibnamefont {Darulova}}, \bibinfo
  {author} {\bibfnamefont {R.}~\bibnamefont {van Gulik}}, \bibinfo {author}
  {\bibfnamefont {N.}~\bibnamefont {Pearson}}, \bibinfo {author} {\bibfnamefont
  {T.}~\bibnamefont {Larsen}},\ and\ \bibinfo {author} {\bibfnamefont
  {A.}~\bibnamefont {Corna}},\ }\href {https://doi.org/10.5281/zenodo.10459033}
  {\bibinfo {title} {Qcodes/qcodes: Qcodes 0.43.0}} (\bibinfo {year}
  {2024})\BibitemShut {NoStop}%
\bibitem [{\citenamefont {Rol}\ \emph {et~al.}(2021)\citenamefont {Rol},
  \citenamefont {Attryde}, \citenamefont {van Oven}, \citenamefont {Loh},
  \citenamefont {Gloudemans}, \citenamefont {Neg{\^\i}rneac}, \citenamefont
  {Last},\ and\ \citenamefont {Bultink}}]{rol2021quantify}%
  \BibitemOpen
  \bibfield  {author} {\bibinfo {author} {\bibfnamefont {M.~A.}\ \bibnamefont
  {Rol}}, \bibinfo {author} {\bibfnamefont {C.}~\bibnamefont {Attryde}},
  \bibinfo {author} {\bibfnamefont {J.~C.}\ \bibnamefont {van Oven}}, \bibinfo
  {author} {\bibfnamefont {K.}~\bibnamefont {Loh}}, \bibinfo {author}
  {\bibfnamefont {J.}~\bibnamefont {Gloudemans}}, \bibinfo {author}
  {\bibfnamefont {V.}~\bibnamefont {Neg{\^\i}rneac}}, \bibinfo {author}
  {\bibfnamefont {T.}~\bibnamefont {Last}},\ and\ \bibinfo {author}
  {\bibfnamefont {C.~C.}\ \bibnamefont {Bultink}},\ }\bibfield  {title}
  {\bibinfo {title} {Quantify: An open-source framework for operating quantum
  computers in the nisq era},\ }in\ \href
  {https://ui.adsabs.harvard.edu/abs/2021APS..MARM34001R/abstract} {\emph
  {\bibinfo {booktitle} {APS March Meeting Abstracts}}},\ Vol.\ \bibinfo
  {volume} {2021}\ (\bibinfo {year} {2021})\ pp.\ \bibinfo {pages}
  {M34--001}\BibitemShut {NoStop}%
\bibitem [{\citenamefont {Crielaard}\ \emph {et~al.}(2022)\citenamefont
  {Crielaard}, \citenamefont {De~Jong}, \citenamefont {Gloudemans},
  \citenamefont {Vyas}, \citenamefont {Negirneac}, \citenamefont {Valada},
  \citenamefont {Sindile}, \citenamefont {Attryde}, \citenamefont {Lawrence},
  \citenamefont {Reynders} \emph {et~al.}}]{crielaard2022quantify}%
  \BibitemOpen
  \bibfield  {author} {\bibinfo {author} {\bibfnamefont {D.}~\bibnamefont
  {Crielaard}}, \bibinfo {author} {\bibfnamefont {D.}~\bibnamefont {De~Jong}},
  \bibinfo {author} {\bibfnamefont {J.}~\bibnamefont {Gloudemans}}, \bibinfo
  {author} {\bibfnamefont {R.}~\bibnamefont {Vyas}}, \bibinfo {author}
  {\bibfnamefont {V.}~\bibnamefont {Negirneac}}, \bibinfo {author}
  {\bibfnamefont {D.}~\bibnamefont {Valada}}, \bibinfo {author} {\bibfnamefont
  {C.}~\bibnamefont {Sindile}}, \bibinfo {author} {\bibfnamefont
  {C.}~\bibnamefont {Attryde}}, \bibinfo {author} {\bibfnamefont
  {A.}~\bibnamefont {Lawrence}}, \bibinfo {author} {\bibfnamefont
  {T.}~\bibnamefont {Reynders}}, \emph {et~al.},\ }\bibfield  {title} {\bibinfo
  {title} {Quantify-scheduler: An open-source hybrid compiler for operating
  quantum computers in the nisq era},\ }in\ \href
  {https://ui.adsabs.harvard.edu/abs/2022APS..MARQ36010C/abstract} {\emph
  {\bibinfo {booktitle} {APS March Meeting Abstracts}}},\ Vol.\ \bibinfo
  {volume} {2022}\ (\bibinfo {year} {2022})\ pp.\ \bibinfo {pages}
  {Q36--010}\BibitemShut {NoStop}%
\bibitem [{\citenamefont {Johansson}\ \emph {et~al.}(2013)\citenamefont
  {Johansson}, \citenamefont {Nation},\ and\ \citenamefont
  {Nori}}]{Johansson2013}%
  \BibitemOpen
  \bibfield  {author} {\bibinfo {author} {\bibfnamefont {J.}~\bibnamefont
  {Johansson}}, \bibinfo {author} {\bibfnamefont {P.}~\bibnamefont {Nation}},\
  and\ \bibinfo {author} {\bibfnamefont {F.}~\bibnamefont {Nori}},\ }\bibfield
  {title} {\bibinfo {title} {{QuTiP} 2: A python framework for the dynamics of
  open quantum systems},\ }\href {https://doi.org/10.1016/j.cpc.2012.11.019}
  {\bibfield  {journal} {\bibinfo  {journal} {Comput. Phys. Commun.}\ }\textbf
  {\bibinfo {volume} {184}},\ \bibinfo {pages} {1234} (\bibinfo {year}
  {2013})}\BibitemShut {NoStop}%
\bibitem [{\citenamefont {Weiss}\ \emph {et~al.}()\citenamefont {Weiss},
  \citenamefont {Du}, \citenamefont {Grobler}, \citenamefont {Cournapeau},
  \citenamefont {Pedregosa}, \citenamefont {Varoquaux}, \citenamefont
  {Mueller}, \citenamefont {Thirion}, \citenamefont {Nouri}, \citenamefont
  {G.}, \citenamefont {J.}, \citenamefont {Benediktsson}, \citenamefont {L.},
  \citenamefont {Korobov}, \citenamefont {R.}, \citenamefont {Lattarini},
  \citenamefont {Niculae}, \citenamefont {csytracy}, \citenamefont {Gramfort},
  \citenamefont {Lebedev}, \citenamefont {Huppenkothen}, \citenamefont
  {Farrow}, \citenamefont {Yanenko}, \citenamefont {Lee}, \citenamefont
  {Danielson},\ and\ \citenamefont {Rockhill}}]{hmmlearn}%
  \BibitemOpen
  \bibfield  {author} {\bibinfo {author} {\bibfnamefont {R.}~\bibnamefont
  {Weiss}}, \bibinfo {author} {\bibfnamefont {S.}~\bibnamefont {Du}}, \bibinfo
  {author} {\bibfnamefont {J.}~\bibnamefont {Grobler}}, \bibinfo {author}
  {\bibfnamefont {D.}~\bibnamefont {Cournapeau}}, \bibinfo {author}
  {\bibfnamefont {F.}~\bibnamefont {Pedregosa}}, \bibinfo {author}
  {\bibfnamefont {G.}~\bibnamefont {Varoquaux}}, \bibinfo {author}
  {\bibfnamefont {A.}~\bibnamefont {Mueller}}, \bibinfo {author} {\bibfnamefont
  {B.}~\bibnamefont {Thirion}}, \bibinfo {author} {\bibfnamefont
  {D.}~\bibnamefont {Nouri}}, \bibinfo {author} {\bibfnamefont
  {L.}~\bibnamefont {G.}}, \bibinfo {author} {\bibfnamefont {V.}~\bibnamefont
  {J.}}, \bibinfo {author} {\bibfnamefont {J.}~\bibnamefont {Benediktsson}},
  \bibinfo {author} {\bibfnamefont {B.}~\bibnamefont {L.}}, \bibinfo {author}
  {\bibfnamefont {M.}~\bibnamefont {Korobov}}, \bibinfo {author} {\bibfnamefont
  {M.}~\bibnamefont {R.}}, \bibinfo {author} {\bibfnamefont {S.}~\bibnamefont
  {Lattarini}}, \bibinfo {author} {\bibfnamefont {V.}~\bibnamefont {Niculae}},
  \bibinfo {author} {\bibnamefont {csytracy}}, \bibinfo {author} {\bibfnamefont
  {A.}~\bibnamefont {Gramfort}}, \bibinfo {author} {\bibfnamefont
  {S.}~\bibnamefont {Lebedev}}, \bibinfo {author} {\bibfnamefont
  {D.}~\bibnamefont {Huppenkothen}}, \bibinfo {author} {\bibfnamefont
  {C.}~\bibnamefont {Farrow}}, \bibinfo {author} {\bibfnamefont
  {A.}~\bibnamefont {Yanenko}}, \bibinfo {author} {\bibfnamefont
  {A.}~\bibnamefont {Lee}}, \bibinfo {author} {\bibfnamefont {M.}~\bibnamefont
  {Danielson}},\ and\ \bibinfo {author} {\bibfnamefont {A.}~\bibnamefont
  {Rockhill}},\ }\href@noop {} {\bibinfo {title} {hmmlearn python package -
  hidden markov models in python, with scikit-learn like api}},\ \bibinfo
  {howpublished} {\url{https://github.com/hmmlearn/hmmlearn}}\BibitemShut
  {NoStop}%
\bibitem [{\citenamefont {Motzoi}\ \emph {et~al.}(2009)\citenamefont {Motzoi},
  \citenamefont {Gambetta}, \citenamefont {Rebentrost},\ and\ \citenamefont
  {Wilhelm}}]{Motzoi09}%
  \BibitemOpen
  \bibfield  {author} {\bibinfo {author} {\bibfnamefont {F.}~\bibnamefont
  {Motzoi}}, \bibinfo {author} {\bibfnamefont {J.~M.}\ \bibnamefont
  {Gambetta}}, \bibinfo {author} {\bibfnamefont {P.}~\bibnamefont
  {Rebentrost}},\ and\ \bibinfo {author} {\bibfnamefont {F.~K.}\ \bibnamefont
  {Wilhelm}},\ }\bibfield  {title} {\bibinfo {title} {Simple pulses for
  elimination of leakage in weakly nonlinear qubits},\ }\href
  {https://doi.org/10.1103/PhysRevLett.103.110501} {\bibfield  {journal}
  {\bibinfo  {journal} {Phys. Rev. Lett.}\ }\textbf {\bibinfo {volume} {103}},\
  \bibinfo {pages} {110501} (\bibinfo {year} {2009})}\BibitemShut {NoStop}%
\bibitem [{\citenamefont {Chow}\ \emph {et~al.}(2010)\citenamefont {Chow},
  \citenamefont {DiCarlo}, \citenamefont {Gambetta}, \citenamefont {Motzoi},
  \citenamefont {Frunzio}, \citenamefont {Girvin},\ and\ \citenamefont
  {Schoelkopf}}]{Chow10b}%
  \BibitemOpen
  \bibfield  {author} {\bibinfo {author} {\bibfnamefont {J.~M.}\ \bibnamefont
  {Chow}}, \bibinfo {author} {\bibfnamefont {L.}~\bibnamefont {DiCarlo}},
  \bibinfo {author} {\bibfnamefont {J.~M.}\ \bibnamefont {Gambetta}}, \bibinfo
  {author} {\bibfnamefont {F.}~\bibnamefont {Motzoi}}, \bibinfo {author}
  {\bibfnamefont {L.}~\bibnamefont {Frunzio}}, \bibinfo {author} {\bibfnamefont
  {S.~M.}\ \bibnamefont {Girvin}},\ and\ \bibinfo {author} {\bibfnamefont
  {R.~J.}\ \bibnamefont {Schoelkopf}},\ }\bibfield  {title} {\bibinfo {title}
  {Optimized driving of superconducting artificial atoms for improved
  single-qubit gates},\ }\href {https://doi.org/10.1103/PhysRevA.82.040305}
  {\bibfield  {journal} {\bibinfo  {journal} {Phys. Rev. A}\ }\textbf {\bibinfo
  {volume} {82}},\ \bibinfo {pages} {040305} (\bibinfo {year}
  {2010})}\BibitemShut {NoStop}%
\bibitem [{\citenamefont {Reed}(2013)}]{ReedPhD13}%
  \BibitemOpen
  \bibfield  {author} {\bibinfo {author} {\bibfnamefont {M.}~\bibnamefont
  {Reed}},\ }\emph {\bibinfo {title} {Entanglement and quantum error correction
  with superconducting qubits}},\ \href@noop {} {\bibinfo {type} {Ph{D}
  {D}issertation}},\ \bibinfo  {school} {Yale University} (\bibinfo {year}
  {2013})\BibitemShut {NoStop}%
\bibitem [{\citenamefont {Cherney}\ and\ \citenamefont
  {Shewchun}(1969)}]{Cherney1969}%
  \BibitemOpen
  \bibfield  {author} {\bibinfo {author} {\bibfnamefont {O.~A.~E.}\
  \bibnamefont {Cherney}}\ and\ \bibinfo {author} {\bibfnamefont
  {J.}~\bibnamefont {Shewchun}},\ }\bibfield  {title} {\bibinfo {title}
  {Enhancement of superconductivity in thin aluminium films},\ }\href
  {https://doi.org/10.1139/p69-138} {\bibfield  {journal} {\bibinfo  {journal}
  {Can. J. Phys.}\ }\textbf {\bibinfo {volume} {47}},\ \bibinfo {pages} {1101}
  (\bibinfo {year} {1969})}\BibitemShut {NoStop}%
\bibitem [{\citenamefont {Barone}\ and\ \citenamefont
  {Patern{\`o}}(1982)}]{Barone}%
  \BibitemOpen
  \bibfield  {author} {\bibinfo {author} {\bibfnamefont {A.}~\bibnamefont
  {Barone}}\ and\ \bibinfo {author} {\bibfnamefont {G.}~\bibnamefont
  {Patern{\`o}}},\ }\href@noop {} {\emph {\bibinfo {title} {Physics and
  applications of the Josephson effect}}}\ (\bibinfo  {publisher} {Wiley, New
  York},\ \bibinfo {year} {1982})\BibitemShut {NoStop}%
\bibitem [{\citenamefont {Janssen}\ \emph {et~al.}(2024)\citenamefont
  {Janssen}, \citenamefont {Butseraen}, \citenamefont {Krause}, \citenamefont
  {Coissard}, \citenamefont {Planat}, \citenamefont {Roch}, \citenamefont
  {Catelani}, \citenamefont {Ando},\ and\ \citenamefont
  {Dickel}}]{Janssen2024}%
  \BibitemOpen
  \bibfield  {author} {\bibinfo {author} {\bibfnamefont {L.~M.}\ \bibnamefont
  {Janssen}}, \bibinfo {author} {\bibfnamefont {G.}~\bibnamefont {Butseraen}},
  \bibinfo {author} {\bibfnamefont {J.}~\bibnamefont {Krause}}, \bibinfo
  {author} {\bibfnamefont {A.}~\bibnamefont {Coissard}}, \bibinfo {author}
  {\bibfnamefont {L.}~\bibnamefont {Planat}}, \bibinfo {author} {\bibfnamefont
  {N.}~\bibnamefont {Roch}}, \bibinfo {author} {\bibfnamefont {G.}~\bibnamefont
  {Catelani}}, \bibinfo {author} {\bibfnamefont {Y.}~\bibnamefont {Ando}},\
  and\ \bibinfo {author} {\bibfnamefont {C.}~\bibnamefont {Dickel}},\
  }\bibfield  {title} {\bibinfo {title} {Magnetic-field dependence of a
  josephson traveling-wave parametric amplifier and integration into a
  high-field setup},\ }\href {http://arxiv.org/abs/2402.19398} {\bibfield
  {journal} {\bibinfo  {journal} {arXiv:2402.19398}\ } (\bibinfo {year}
  {2024})}\BibitemShut {NoStop}%
\bibitem [{\citenamefont {Rist\`{e}}\ \emph {et~al.}(2015)\citenamefont
  {Rist\`{e}}, \citenamefont {Poletto}, \citenamefont {Huang}, \citenamefont
  {Bruno}, \citenamefont {Vesterinen}, \citenamefont {Saira},\ and\
  \citenamefont {DiCarlo}}]{Riste15}%
  \BibitemOpen
  \bibfield  {author} {\bibinfo {author} {\bibfnamefont {D.}~\bibnamefont
  {Rist\`{e}}}, \bibinfo {author} {\bibfnamefont {S.}~\bibnamefont {Poletto}},
  \bibinfo {author} {\bibfnamefont {M.~Z.}\ \bibnamefont {Huang}}, \bibinfo
  {author} {\bibfnamefont {A.}~\bibnamefont {Bruno}}, \bibinfo {author}
  {\bibfnamefont {V.}~\bibnamefont {Vesterinen}}, \bibinfo {author}
  {\bibfnamefont {O.~P.}\ \bibnamefont {Saira}},\ and\ \bibinfo {author}
  {\bibfnamefont {L.}~\bibnamefont {DiCarlo}},\ }\bibfield  {title} {\bibinfo
  {title} {Detecting bit-flip errors in a logical qubit using stabilizer
  measurements},\ }\href {https://doi.org/10.1038/ncomms7983} {\bibfield
  {journal} {\bibinfo  {journal} {Nat Commun}\ }\textbf {\bibinfo {volume}
  {6}},\ \bibinfo {pages} {6983} (\bibinfo {year} {2015})}\BibitemShut
  {NoStop}%
\bibitem [{\citenamefont {Vool}\ \emph {et~al.}(2014)\citenamefont {Vool},
  \citenamefont {Pop}, \citenamefont {Sliwa}, \citenamefont {Abdo},
  \citenamefont {Wang}, \citenamefont {Brecht}, \citenamefont {Gao},
  \citenamefont {Shankar}, \citenamefont {Hatridge}, \citenamefont {Catelani},
  \citenamefont {Mirrahimi}, \citenamefont {Frunzio}, \citenamefont
  {Schoelkopf}, \citenamefont {Glazman},\ and\ \citenamefont
  {Devoret}}]{Vool14}%
  \BibitemOpen
  \bibfield  {author} {\bibinfo {author} {\bibfnamefont {U.}~\bibnamefont
  {Vool}}, \bibinfo {author} {\bibfnamefont {I.~M.}\ \bibnamefont {Pop}},
  \bibinfo {author} {\bibfnamefont {K.}~\bibnamefont {Sliwa}}, \bibinfo
  {author} {\bibfnamefont {B.}~\bibnamefont {Abdo}}, \bibinfo {author}
  {\bibfnamefont {C.}~\bibnamefont {Wang}}, \bibinfo {author} {\bibfnamefont
  {T.}~\bibnamefont {Brecht}}, \bibinfo {author} {\bibfnamefont {Y.~Y.}\
  \bibnamefont {Gao}}, \bibinfo {author} {\bibfnamefont {S.}~\bibnamefont
  {Shankar}}, \bibinfo {author} {\bibfnamefont {M.}~\bibnamefont {Hatridge}},
  \bibinfo {author} {\bibfnamefont {G.}~\bibnamefont {Catelani}}, \bibinfo
  {author} {\bibfnamefont {M.}~\bibnamefont {Mirrahimi}}, \bibinfo {author}
  {\bibfnamefont {L.}~\bibnamefont {Frunzio}}, \bibinfo {author} {\bibfnamefont
  {R.~J.}\ \bibnamefont {Schoelkopf}}, \bibinfo {author} {\bibfnamefont
  {L.~I.}\ \bibnamefont {Glazman}},\ and\ \bibinfo {author} {\bibfnamefont
  {M.~H.}\ \bibnamefont {Devoret}},\ }\bibfield  {title} {\bibinfo {title}
  {Non-poissonian quantum jumps of a fluxonium qubit due to quasiparticle
  excitations},\ }\href {https://doi.org/10.1103/PhysRevLett.113.247001}
  {\bibfield  {journal} {\bibinfo  {journal} {Phys. Rev. Lett.}\ }\textbf
  {\bibinfo {volume} {113}},\ \bibinfo {pages} {247001} (\bibinfo {year}
  {2014})}\BibitemShut {NoStop}%
\bibitem [{\citenamefont {Glazman}\ and\ \citenamefont
  {Catelani}(2021)}]{glazman21}%
  \BibitemOpen
  \bibfield  {author} {\bibinfo {author} {\bibfnamefont {L.}~\bibnamefont
  {Glazman}}\ and\ \bibinfo {author} {\bibfnamefont {G.}~\bibnamefont
  {Catelani}},\ }\bibfield  {title} {\bibinfo {title} {Bogoliubov
  quasiparticles in superconducting qubits},\ }\href
  {https://doi.org/10.21468/SciPostPhysLectNotes.31} {\bibfield  {journal}
  {\bibinfo  {journal} {SciPost Phys. Lect. Notes}\ }\textbf {\bibinfo {volume}
  {31}} (\bibinfo {year} {2021})}\BibitemShut {NoStop}%
\bibitem [{\citenamefont {Catelani}\ \emph {et~al.}(2011)\citenamefont
  {Catelani}, \citenamefont {Schoelkopf}, \citenamefont {Devoret},\ and\
  \citenamefont {Glazman}}]{Catelani11}%
  \BibitemOpen
  \bibfield  {author} {\bibinfo {author} {\bibfnamefont {G.}~\bibnamefont
  {Catelani}}, \bibinfo {author} {\bibfnamefont {R.~J.}\ \bibnamefont
  {Schoelkopf}}, \bibinfo {author} {\bibfnamefont {M.~H.}\ \bibnamefont
  {Devoret}},\ and\ \bibinfo {author} {\bibfnamefont {L.~I.}\ \bibnamefont
  {Glazman}},\ }\bibfield  {title} {\bibinfo {title} {Relaxation and frequency
  shifts induced by quasiparticles in superconducting qubits},\ }\href
  {https://journals.aps.org/prb/abstract/10.1103/PhysRevB.84.064517} {\bibfield
   {journal} {\bibinfo  {journal} {Phys. Rev. B}\ }\textbf {\bibinfo {volume}
  {84}},\ \bibinfo {pages} {064517} (\bibinfo {year} {2011})}\BibitemShut
  {NoStop}%
\bibitem [{\citenamefont {Peterer}\ \emph {et~al.}(2015)\citenamefont
  {Peterer}, \citenamefont {Bader}, \citenamefont {Jin}, \citenamefont {Yan},
  \citenamefont {Kamal}, \citenamefont {Gudmundsen}, \citenamefont {Leek},
  \citenamefont {Orlando}, \citenamefont {Oliver},\ and\ \citenamefont
  {Gustavsson}}]{peterer12}%
  \BibitemOpen
  \bibfield  {author} {\bibinfo {author} {\bibfnamefont {M.~J.}\ \bibnamefont
  {Peterer}}, \bibinfo {author} {\bibfnamefont {S.~J.}\ \bibnamefont {Bader}},
  \bibinfo {author} {\bibfnamefont {X.}~\bibnamefont {Jin}}, \bibinfo {author}
  {\bibfnamefont {F.}~\bibnamefont {Yan}}, \bibinfo {author} {\bibfnamefont
  {A.}~\bibnamefont {Kamal}}, \bibinfo {author} {\bibfnamefont {T.~J.}\
  \bibnamefont {Gudmundsen}}, \bibinfo {author} {\bibfnamefont {P.~J.}\
  \bibnamefont {Leek}}, \bibinfo {author} {\bibfnamefont {T.~P.}\ \bibnamefont
  {Orlando}}, \bibinfo {author} {\bibfnamefont {W.~D.}\ \bibnamefont
  {Oliver}},\ and\ \bibinfo {author} {\bibfnamefont {S.}~\bibnamefont
  {Gustavsson}},\ }\bibfield  {title} {\bibinfo {title} {Coherence and decay of
  higher energy levels of a superconducting transmon qubit},\ }\href
  {https://doi.org/10.1103/PhysRevLett.114.010501} {\bibfield  {journal}
  {\bibinfo  {journal} {Phys. Rev. Lett.}\ }\textbf {\bibinfo {volume} {114}},\
  \bibinfo {pages} {010501} (\bibinfo {year} {2015})}\BibitemShut {NoStop}%
\bibitem [{\citenamefont {Morvan}\ \emph {et~al.}(2021)\citenamefont {Morvan},
  \citenamefont {Ramasesh}, \citenamefont {Blok}, \citenamefont {Kreikebaum},
  \citenamefont {O'Brien}, \citenamefont {Chen}, \citenamefont {Mitchell},
  \citenamefont {Naik}, \citenamefont {Santiago},\ and\ \citenamefont
  {Siddiqi}}]{Morvan2021}%
  \BibitemOpen
  \bibfield  {author} {\bibinfo {author} {\bibfnamefont {A.}~\bibnamefont
  {Morvan}}, \bibinfo {author} {\bibfnamefont {V.~V.}\ \bibnamefont
  {Ramasesh}}, \bibinfo {author} {\bibfnamefont {M.~S.}\ \bibnamefont {Blok}},
  \bibinfo {author} {\bibfnamefont {J.~M.}\ \bibnamefont {Kreikebaum}},
  \bibinfo {author} {\bibfnamefont {K.}~\bibnamefont {O'Brien}}, \bibinfo
  {author} {\bibfnamefont {L.}~\bibnamefont {Chen}}, \bibinfo {author}
  {\bibfnamefont {B.~K.}\ \bibnamefont {Mitchell}}, \bibinfo {author}
  {\bibfnamefont {R.~K.}\ \bibnamefont {Naik}}, \bibinfo {author}
  {\bibfnamefont {D.~I.}\ \bibnamefont {Santiago}},\ and\ \bibinfo {author}
  {\bibfnamefont {I.}~\bibnamefont {Siddiqi}},\ }\bibfield  {title} {\bibinfo
  {title} {Qutrit randomized benchmarking},\ }\href
  {https://doi.org/10.1103/PhysRevLett.126.210504} {\bibfield  {journal}
  {\bibinfo  {journal} {Phys. Rev. Lett.}\ }\textbf {\bibinfo {volume} {126}},\
  \bibinfo {pages} {210504} (\bibinfo {year} {2021})}\BibitemShut {NoStop}%
\bibitem [{\citenamefont {Mörstedt}\ \emph {et~al.}(2024)\citenamefont
  {Mörstedt}, \citenamefont {Teixeira}, \citenamefont {Viitanen},
  \citenamefont {Kivijärvi}, \citenamefont {Tiiri}, \citenamefont {Rasola},
  \citenamefont {Gunyho}, \citenamefont {Kundu}, \citenamefont {Lattier},
  \citenamefont {Vadimov}, \citenamefont {Catelani}, \citenamefont {Sevriuk},
  \citenamefont {Heinsoo}, \citenamefont {Räbinä}, \citenamefont
  {Ankerhold},\ and\ \citenamefont {Möttönen}}]{morstedt2024}%
  \BibitemOpen
  \bibfield  {author} {\bibinfo {author} {\bibfnamefont {T.~F.}\ \bibnamefont
  {Mörstedt}}, \bibinfo {author} {\bibfnamefont {W.~S.}\ \bibnamefont
  {Teixeira}}, \bibinfo {author} {\bibfnamefont {A.}~\bibnamefont {Viitanen}},
  \bibinfo {author} {\bibfnamefont {H.}~\bibnamefont {Kivijärvi}}, \bibinfo
  {author} {\bibfnamefont {M.}~\bibnamefont {Tiiri}}, \bibinfo {author}
  {\bibfnamefont {M.}~\bibnamefont {Rasola}}, \bibinfo {author} {\bibfnamefont
  {A.~M.}\ \bibnamefont {Gunyho}}, \bibinfo {author} {\bibfnamefont
  {S.}~\bibnamefont {Kundu}}, \bibinfo {author} {\bibfnamefont
  {L.}~\bibnamefont {Lattier}}, \bibinfo {author} {\bibfnamefont
  {V.}~\bibnamefont {Vadimov}}, \bibinfo {author} {\bibfnamefont
  {G.}~\bibnamefont {Catelani}}, \bibinfo {author} {\bibfnamefont
  {V.}~\bibnamefont {Sevriuk}}, \bibinfo {author} {\bibfnamefont
  {J.}~\bibnamefont {Heinsoo}}, \bibinfo {author} {\bibfnamefont
  {J.}~\bibnamefont {Räbinä}}, \bibinfo {author} {\bibfnamefont
  {J.}~\bibnamefont {Ankerhold}},\ and\ \bibinfo {author} {\bibfnamefont
  {M.}~\bibnamefont {Möttönen}},\ }\bibfield  {title} {\bibinfo {title}
  {Rapid on-demand generation of thermal states in superconducting quantum
  circuits},\ }\href {http://arxiv.org/abs/2402.09594} {\bibfield  {journal}
  {\bibinfo  {journal} {arXiv:2402.09594}\ } (\bibinfo {year}
  {2024})}\BibitemShut {NoStop}%
\bibitem [{\citenamefont {Sultanov}\ \emph {et~al.}(2021)\citenamefont
  {Sultanov}, \citenamefont {Kuzmanović}, \citenamefont {Lebedev},\ and\
  \citenamefont {Paraoanu}}]{Sultanov21}%
  \BibitemOpen
  \bibfield  {author} {\bibinfo {author} {\bibfnamefont {A.}~\bibnamefont
  {Sultanov}}, \bibinfo {author} {\bibfnamefont {M.}~\bibnamefont
  {Kuzmanović}}, \bibinfo {author} {\bibfnamefont {A.~V.}\ \bibnamefont
  {Lebedev}},\ and\ \bibinfo {author} {\bibfnamefont {G.~S.}\ \bibnamefont
  {Paraoanu}},\ }\bibfield  {title} {\bibinfo {title} {{Protocol for
  temperature sensing using a three-level transmon circuit}},\ }\href
  {https://doi.org/10.1063/5.0065224} {\bibfield  {journal} {\bibinfo
  {journal} {Appl. Phys. Lett.}\ }\textbf {\bibinfo {volume} {119}},\ \bibinfo
  {pages} {144002} (\bibinfo {year} {2021})}\BibitemShut {NoStop}%
\end{thebibliography}
\end{document}